\documentclass[11pt,a4paper]{article}
\usepackage{jheppub}

\usepackage[normalem]{ulem}
\usepackage{amsmath}
\usepackage{amsfonts}
\usepackage{amsthm}
\usepackage{slashed}
\usepackage{graphicx}
\usepackage{mathrsfs,bbm}
\usepackage{subfig}
\usepackage{booktabs}
\usepackage[numbers,sort&compress]{natbib}
\usepackage{lscape}
\usepackage{breqn}
\usepackage{amsmath,amssymb,textcomp,enumerate,alltt,xspace,multirow}

\usepackage{multicol}
\usepackage{boxedminipage}
\usepackage[dvipsnames]{xcolor}
\usepackage{comment}
\usepackage{tabularx}
\usepackage{adjustbox}
\usepackage{cleveref}

\usepackage{tikz}
\usetikzlibrary{decorations.pathmorphing}
\usetikzlibrary{decorations.markings}
\usetikzlibrary{calc}
\usetikzlibrary{fit, shapes, tikzmark}

\tikzset{
    vector/.style={decorate, decoration={snake}, draw},
	provector/.style={decorate, decoration={snake,amplitude=2.5pt}, draw},
	antivector/.style={decorate, decoration={snake,amplitude=-2.5pt}, draw},
    fermion/.style={draw=black, postaction={decorate},
        decoration={markings,mark=at position .55 with {\arrow[draw=black]{>}}}},
    fermionx/.style={draw=black, postaction={decorate},
        decoration={markings,mark=at position .55 with {\arrow[draw=black]{<}}}},        
    fermionbar/.style={draw=black, postaction={decorate},
        decoration={markings,mark=at position .55 with {\arrow[draw=black]{<}}}},
    fermionnoarrow/.style={draw=black},
    higgs/.style={draw=blue},
    gluon/.style={decorate, draw=black,
        decoration={coil, amplitude=1.5pt, segment length=3pt}},
    photon/.style={decorate, draw=black,
        decoration={coil, aspect=0,amplitude=1.5pt, segment length=4.5pt}},        
    scalar/.style={dashed,draw=black, postaction={decorate}},
    phi/.style={dashed,draw=red, postaction={decorate}},    
    scalarbar/.style={dashed,draw=black, postaction={decorate},
        decoration={markings,mark=at position .55 with {\arrow[draw=black]{<}}}},
    scalarnoarrow/.style={dashed,draw=black},
    electron/.style={draw=black, postaction={decorate},
        decoration={markings,mark=at position .7 with {\arrow[draw=black]{>}}}},
    positron/.style={draw=black, postaction={decorate},
        decoration={markings,mark=at position .35 with {\arrow[draw=black]{<}}}},        
	bigvector/.style={decorate, decoration={snake,amplitude=4pt}, draw},
}

\usepackage[colorlinks=true]{hyperref}
\hypersetup{
colorlinks=true,
linkcolor=blue,
linktocpage=true,
citecolor=violet
}

\usepackage{multirow}
\allowdisplaybreaks

\makeatletter
\g@addto@macro\bfseries{\boldmath}
\makeatother

\theoremstyle{remark}

\DeclareMathOperator{\td}{d}
\newcommand{\dlog}{\mathrm{dlog}}
\newcommand{\iu}{\mathrm{i}}

\newcommand{\TOP}[1]{\textsf{#1}}

\newcommand{\liverpool}{Department of Mathematical Sciences, University of Liverpool, \\
Liverpool L69 3BX, U.K.
}
\newcommand{\salamanca}{Departamento de Física Fundamental and IUFFyM, Universidad de Salamanca, \\
Salamanca E-37008, Spain}

\title{
First look at the evaluation of two-loop Feynman integrals for radiative return processes
}

\author[a,b]{Mattia Pozzoli}
\emailAdd{mattia.pozzoli@kit.edu}
\author[c,d]{and William J. Torres~Bobadilla}
\emailAdd{william.torres@usal.es}

\affiliation[a]{Dipartimento di Fisica e Astronomia, Università di Bologna \\
INFN, Sezione di Bologna, \\
via Irnerio 46, I-40126 Bologna, Italy}
\affiliation[b]{Institute for Theoretical Physics, Karlsruhe Institute of Technology (KIT),\\ 
Wolfgang-Gaede-Str. 1, 76131 Karlsruhe, Germany}
\affiliation[c]{\liverpool}
\affiliation[d]{\salamanca}

\abstract{
Precision studies of radiative return processes at low-energy electron--positron colliders require next-to-next-to-leading order QED predictions retaining full dependence on the electron mass. We present the calculation of planar two-loop four-point Feynman integrals relevant for initial-state radiation contributions to these processes. The calculation presents considerable analytical complexity, due to the presence of a nested square root and of integrals associated with elliptic geometries. We construct differential equations for the Feynman integrals which are polynomial in the dimensional regulator, and are suitable for numerical integration. We demonstrate stable numerical evaluations throughout the physical region relevant for low-energy experiments, despite the presence of large hierarchies of scales. Our results provide essential building blocks for NNLO predictions for radiative return processes.
}

\begin{document}

\maketitle

\section{Introduction}
\label{sec:introduction}
Precision measurements at low-energy electron--positron colliders continue to play a central role in testing the Standard Model. Experiments based on radiative return processes, such as BaBar~\cite{BaBar:2012bdw}, Belle II~\cite{Belle:2007ebm}, BESIII~\cite{BESIII:2015equ}, and KLOE~\cite{KLOE:2008fmq,KLOE:2010qei,KLOE:2012anl,KLOE-2:2017fda}, 
provide essential input for the data-driven determination of the hadronic vacuum polarisation (HVP) contribution to the anomalous magnetic moment of the muon $(g-2)_\mu$. 
The resulting Standard Model prediction exhibits a tension with experimental measurements when the leading-order HVP contribution is evaluated using experimental $e^+e^-\to\text{hadrons}$ data. However, lattice-QCD determinations have yielded results that reduce or remove this tension, highlighting the importance of understanding the differences between these approaches~\cite{Aliberti:2025beg}.
Improving the precision of both the experimental measurements and the corresponding theoretical calculations remains a pressing task. 
In this context, achieving accuracy beyond next-to-leading order (NLO) in QED for radiative return processes such as $e^+e^- \to \pi^+\pi^-\gamma$ and $e^+e^- \to \mu^+\mu^-\gamma$ becomes particularly important for reducing the theoretical uncertainties in the data-driven approach~\cite{Aliberti:2024fpq,Budassi:2026lmr,PetitRosas:2026iuq,CarloniCalame:2026hhy,Dave:2026pvq}.

The structure of radiative return amplitudes naturally allows for a decomposition into contributions where the energetic photon is emitted either from the initial electron--positron pair (initial-state radiation) or from the final state (final-state radiation)~\cite{Aliberti:2024fpq}. These contributions are individually gauge-invariant, and they can be computed independently. Therefore, focusing on the initial-state radiation component provides a well-defined starting point for the construction of NNLO QED predictions, as it reduces the problem to the evaluation of genuine four-point two-loop amplitudes. These amplitudes are closely related to the process $e^+e^- \to \gamma\gamma^*$, whose two-loop calculations, in the massless-electron approximation, have been studied in the literature~\cite{Badger:2023xtl,Fadin:2023phc}. 

The main bottleneck in the computation of the two-loop amplitudes retaining full dependence on the electron mass is the evaluation of the Feynman integrals. Firstly, the additional scale and the presence of massive virtual particle increase the algebraic complexity of the expressions appearing in the \emph{integration-by-parts identities} (IBPs)~\cite{Tkachov:1981wb,Chetyrkin:1981qh,Laporta:2000dsw} relating the Feynman integrals, and in the \emph{differential equations} (DEs)~\cite{Barucchi:1973zm,Kotikov:1990kg,Kotikov:1991hm,Gehrmann:1999as,Bern:1993kr} satisfied by the \emph{master integrals} (MIs). 
Secondly, Feynman integrals with massive propagators are known to involve special functions stemming from higher-genus geometries~\cite{Bourjaily:2022bwx,Bargiela:2025vwl}, such as elliptic curves or, more generally, Calabi-Yau manifolds. While the first challenge can be efficiently overcome by exploiting \emph{finite-field} techniques~\cite{vonManteuffel:2014ixa,Peraro:2016wsq}, methods to tackle the second one are still an object of extensive study by the community~\cite{Adams:2017tga,Adams:2018yfj,Frellesvig:2021hkr,Dlapa:2022wdu,Pogel:2022ken,Pogel:2022vat,Frellesvig:2023iwr,Driesse:2024feo,Duhr:2024uid,Duhr:2025lbz,Chen:2025hzq}. These efforts have lead to the development of general techniques~\cite{Gorges:2023zgv,e-collaboration:2025frv,Bree:2025tug} and lately also to the interpretation of the problem in terms of leading singularities~\cite{Chaubey:2025adn,Forner:2026vby}. In particular, canonical DEs~\cite{Henn:2013pwa} for various integral families associated with elliptic curves with four-point kinematics have become available~\cite{Adams:2018yfj,Gorges:2023zgv,Delto:2023kqv,Chen:2025hzq,Becchetti:2025rrz,Becchetti:2025oyb,Coro:2025vgn}. Despite this progress, the task of obtaining canonical DEs in the elliptic (or in general higher-genus) case remains substantially more complicated than in the polylogarithmic case.

As a first step towards the computation of the two-loop amplitude for $e^+e^- \to \gamma\gamma^*$, in this work we compute the planar integral families contributing to the amplitude. These integral families include integrals associated with five non-isomorphic elliptic curves, two of which are associated with the elliptic sunrise~\cite{Remiddi:2003ci,Laporta:2004rb,Pozzorini:2005ff}, while the remaining three appear in genuine four-point topologies.
Rather than attempting to construct a canonical basis for the corresponding integrals, we tackle the problem of finding a suitable basis of MIs following the method that was developed in~\cite{Badger:2024fgb} and successfully applied also in~\cite{Becchetti:2025qlu}. Namely, we construct DEs that are canonical for the polylogarithmic MIs and polynomial in the dimensional regulator $\varepsilon$ for the elliptic integrals. This strategy allows us to simplify the DEs compared to an arbitrary basis, without introducing transcendental functions.

In addition to elliptic curves, these integrals contain also a \emph{nested square root}. This kind of analytic structure has already been observed in the literature~\cite{FebresCordero:2023pww,Badger:2024fgb,Becchetti:2025oyb,Aliaj:2026iny,Li:2026emp}, and, while it is possible to cast the corresponding DEs in $\varepsilon$-factorised form through algebraic transformations, it is not clear whether they can be expressed in terms of logarithmic one-forms~\cite{FebresCordero:2023pww}. Moreover, due to their intricate branch structure, the solution of DEs involving nested square roots would be more complicated. For this reason, we prefer to avoid introducing these structures in the DEs.

As for the solution of the DEs, we rely on numerical methods. This strategy has been successfully applied in many calculations~\cite{Boughezal:2007ny,Czakon:2008zk,Mandal:2018cdj,Czakon:2020vql,Czakon:2021yub,Calisto:2023vmm,Haisch:2024nzv,PetitRosas:2025xhm,Badger:2025ilt,Badger:2025ljy,Czakon:2026tog}, and we expect it to be suitable for phenomenological applications. Practically, we obtain boundary conditions for the MIs with \textsc{AMFlow}~\cite{Liu:2017jxz,Liu:2022chg}, and we integrate the DEs employing the strategy developed in~\cite{PetitRosas:2025xhm} with an in-house implementation in the programming language {\sc Julia}, which we benchmark against \textsc{DiffExp}~\cite{Hidding:2020ytt}.

~

The paper is organised as follows. In \cref{sec:kinematics}, we define the kinematic setup of the process and  integral families we consider in this work. In \cref{sec:description}, we describe the problem of finding DEs that are suitable for numerical integration. In \cref{sec:elliptic}, we discuss the details of the sectors involving elliptic geometries and the nested square root. We discuss the structure and numerical integration of the DEs  in \cref{sec:benchmarks}. Finally, in \cref{sec:conclusion}, we draw our conclusions.
The supplemental material of this manuscript, containing all the results of this work in
{\sc Mathematica} format, is available on Zenodo~\cite{zenodo}, and we refer to the file \verb"README.md" for their description.

\section{Two-loop planar integrals}
\label{sec:kinematics}
We study the two-loop four-point Feynman integrals required for the scattering process
\begin{align}
e^+(p_1)\,e^-(p_2)\to \gamma(p_3)\gamma^*(p_4)\,,
\label{eq:ampl}
\end{align}
by retaining full dependence on the electron mass.
We consider all momenta as incoming, satisfying momentum conservation, $p_1+p_2+p_3+p_4=0$, 
and use the kinematic configuration:
\begin{align}
& s = (p_1+p_2)^2\,, && p_1^2=p_2^2=m^2\\ 
& t = (p_2+p_3)^2\,, && p_3^2=0\,,\\
& u = (p_1+p_3)^2\,, && p_4^2=q^2\,,
\end{align}
with the relation, $s+t+u=2m^2+q^2$.

The kinematics of events relevant for the process in \cref{eq:ampl} corresponds to the $s$-scattering channel. The corresponding physical region is defined through constraints on the kinematic invariants, 
\begin{subequations}
\begin{align}
p_1^2 >0\,,&& p_2^2 >0\,, && p_4^2 >0\,, && p_1\cdot p_2>0\,,
\notag\\
p_1\cdot p_3<0\,, && p_1\cdot p_4<0\,, &&  p_2\cdot p_3<0\,, && p_2\cdot p_4<0\,,
\end{align}
and further constrained by conditions arising from Gram determinants,
\begin{align}
G(p_i,p_j)<0\,, && G(p_i,p_j,p_k)>0\,, \end{align}
with $i,j,k\in\{1,\hdots,4\}$ and $i\ne j\ne k$.
The Gram determinants are defined as
\begin{align}
G(p_{i_1},\hdots, p_{i_n})
=
\text{det}\left(\begin{array}{ccc}
p_{i_{1}}^{2} & \cdots & p_{i_{1}}\cdot p_{i_{n}}\\
\vdots & \ddots & \vdots\\
p_{i_{n}}\cdot p_{i_{1}} & \cdots & p_{i_{n}}^{2}
\end{array}\right)
\,.
\end{align}
\label{eq:physical_region}
\end{subequations}

\begin{figure}[t]
\centering
\subfloat[\TOP{PL1}]{
\begin{tikzpicture}
\coordinate (v1) at (0,0);
\coordinate (v2) at (0,1.5);
\coordinate (v5) at (1.5,1.5);
\coordinate (v6) at (1.5,0.0);
\coordinate (p1) at (-1,-0.5);
\coordinate (p2) at (-1,2);
\coordinate (p3) at (2.5,2);
\coordinate (p4) at (2.5,-0.5);
\coordinate (v3) at (0.75,1.5);
\coordinate (v4) at (0.75,0);
\draw[fermionnoarrow,  thin] (v1) -- (v2);
\draw[fermionnoarrow,  thin] (v3) -- (v4);
\draw[fermionnoarrow,  ultra thick] (v2) -- (v5);
\draw[fermionnoarrow,  ultra thick] (v5) -- (v6);
\draw[fermionnoarrow,  ultra thick] (v6) -- (v1);
\draw[fermionnoarrow, ultra thick] (v1) -- (p1);
\draw[fermionnoarrow, ultra thick] (v2) -- (p2);
\draw[fermionnoarrow, thin] (v5) -- (p3);
\draw[fermionnoarrow, ultra thick, blue] (v6) -- (p4);
\end{tikzpicture}
\label{fig:PL1}
}
\subfloat[\TOP{PL2}]{
\begin{tikzpicture}
\draw[fermionnoarrow,  ultra thick] (v1) -- (v2);
\draw[fermionnoarrow,  ultra thick] (v2) -- (v3);
\draw[fermionnoarrow,  ultra thick] (v3) -- (v4);
\draw[fermionnoarrow,  thin] (v3) -- (v5);
\draw[fermionnoarrow,  ultra thick] (v5) -- (v6);
\draw[fermionnoarrow,  ultra thick] (v4) -- (v6);
\draw[fermionnoarrow,  thin] (v4) -- (v1);
\draw[fermionnoarrow, ultra thick] (v1) -- (p1);
\draw[fermionnoarrow, thin] (v2) -- (p2);
\draw[fermionnoarrow, ultra thick] (v5) -- (p3);
\draw[fermionnoarrow, ultra thick, blue] (v6) -- (p4);
\end{tikzpicture}
\label{fig:PL2}
}
\subfloat[\TOP{PL3}]{
\begin{tikzpicture}
\draw[fermionnoarrow,  ultra thick] (v1) -- (v2);
\draw[fermionnoarrow,  ultra thick] (v3) -- (v4);
\draw[fermionnoarrow,  thin] (v2) -- (v3);
\draw[fermionnoarrow,  ultra thick] (v3) -- (v5);
\draw[fermionnoarrow,  ultra thick] (v5) -- (v6);
\draw[fermionnoarrow,  thin] (v4) -- (v1);
\draw[fermionnoarrow,  ultra thick] (v4) -- (v6);
\draw[fermionnoarrow, ultra thick] (v1) -- (p1);
\draw[fermionnoarrow, ultra thick] (v2) -- (p2);
\draw[fermionnoarrow, thin] (v5) -- (p3);
\draw[fermionnoarrow, ultra thick, blue] (v6) -- (p4);
\end{tikzpicture}
\label{fig:PL3}
}\\
\subfloat[\TOP{NP1}]{
\begin{tikzpicture}
\draw[fermionnoarrow,  thin] (v1) -- (v3);
\draw[fermionnoarrow,  thin] (v2) -- (v4);
\draw[fermionnoarrow,  ultra thick] (v2) -- (v5);
\draw[fermionnoarrow,  ultra thick] (v5) -- (v6);
\draw[fermionnoarrow,  ultra thick] (v6) -- (v1);
\draw[fermionnoarrow, ultra thick] (v1) -- (p1);
\draw[fermionnoarrow, ultra thick] (v2) -- (p2);
\draw[fermionnoarrow, thin] (v5) -- (p3);
\draw[fermionnoarrow, ultra thick, blue] (v6) -- (p4);
\end{tikzpicture}
\label{fig:NP1}
}
\subfloat[\TOP{NP2}]{
\begin{tikzpicture}
\draw[fermionnoarrow,  ultra thick] (v1) -- (v2);
\draw[fermionnoarrow,  ultra thick] (v2) -- (v3);
\draw[fermionnoarrow,  ultra thick] (v3) -- (v6);
\draw[fermionnoarrow,  ultra thick] (v4) -- (v5);
\draw[fermionnoarrow,  thin] (v3) -- (v5);
\draw[fermionnoarrow,  ultra thick] (v4) -- (v6);
\draw[fermionnoarrow,  thin] (v4) -- (v1);
\draw[fermionnoarrow, ultra thick] (v1) -- (p1);
\draw[fermionnoarrow, thin] (v2) -- (p2);
\draw[fermionnoarrow, ultra thick] (v5) -- (p3);
\draw[fermionnoarrow, ultra thick, blue] (v6) -- (p4);
\end{tikzpicture}
\label{fig:NP2}
}
\subfloat[\TOP{NP3}]{
\begin{tikzpicture}
\draw[fermionnoarrow,  ultra thick] (v1) -- (v2);
\draw[fermionnoarrow,  ultra thick] (v3) -- (v6);
\draw[fermionnoarrow,  thin] (v2) -- (v3);
\draw[fermionnoarrow,  ultra thick] (v3) -- (v5);
\draw[fermionnoarrow,  ultra thick] (v5) -- (v4);
\draw[fermionnoarrow,  thin] (v4) -- (v1);
\draw[fermionnoarrow,  ultra thick] (v4) -- (v6);
\draw[fermionnoarrow, ultra thick] (v1) -- (p1);
\draw[fermionnoarrow, ultra thick] (v2) -- (p2);
\draw[fermionnoarrow, thin] (v5) -- (p3);
\draw[fermionnoarrow, ultra thick, blue] (v6) -- (p4);
\end{tikzpicture}
\label{fig:NP5}
}
\subfloat[\TOP{NP4}]{
\begin{tikzpicture}
\draw[fermionnoarrow,  ultra thick] (v1) -- (v3);
\draw[fermionnoarrow,  ultra thick] (v2) -- (v4);
\draw[fermionnoarrow,  ultra thick] (v2) -- (v3);
\draw[fermionnoarrow,  thin] (v3) -- (v5);
\draw[fermionnoarrow,  ultra thick] (v5) -- (v6);
\draw[fermionnoarrow,  ultra thick] (v4) -- (v6);
\draw[fermionnoarrow,  thin] (v4) -- (v1);
\draw[fermionnoarrow, ultra thick] (v1) -- (p1);
\draw[fermionnoarrow, thin] (v2) -- (p2);
\draw[fermionnoarrow, ultra thick] (v5) -- (p3);
\draw[fermionnoarrow, ultra thick, blue] (v6) -- (p4);
\end{tikzpicture}
\label{fig:NP4}
}
\caption{Two-loop integral families. 
Thin lines correspond to massless external  momenta and propagators, 
thick black lines account for propagators and momenta with mass $m^2$,
and blue ones for the off-shell external momentum with $p_4^2 = q^2$.}
\label{fig:families}
\end{figure}
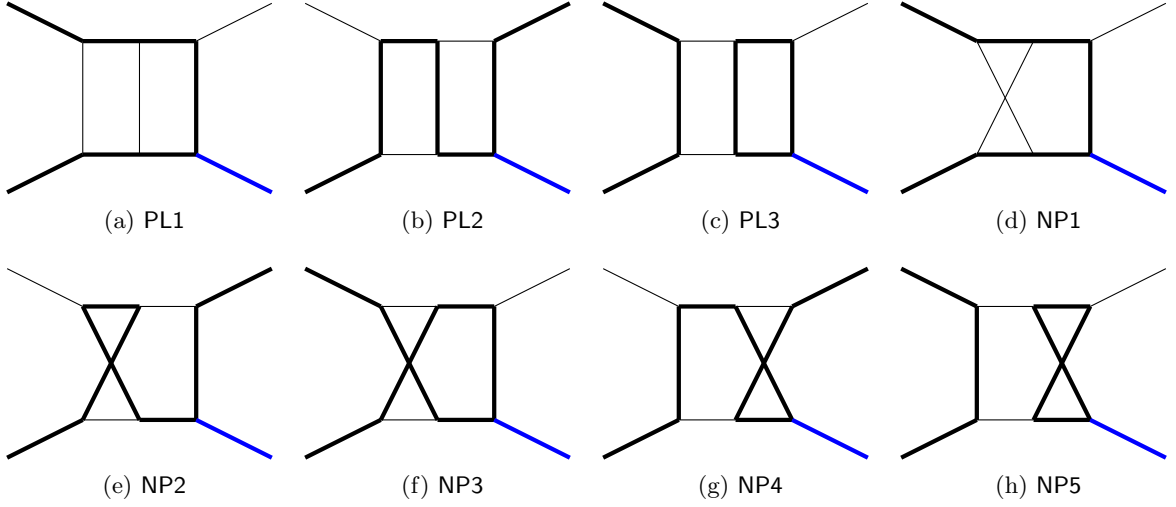

We regularise Feynman integrals in $D=4-2\varepsilon$ dimensions and adopt the normalisation,
\begin{align}
I^{(\TOP{X})}_{a_{1}a_{2}\dots a_{r}} &= 
e^{L\varepsilon\gamma_{\text{E}}}\,
\left({m^{2}}\right)^{L\varepsilon-a}\,
\int
\prod_{k=1}^L 
\frac{\mathrm{d}^{D}k_{L}}{\iu\pi^{D/2}}
\prod_{i=1}^{r}\frac{1}{D_{i}^{a_{i}}}\,.
\label{eq:Feyn_int}
\end{align}
Here, $D_i$ denote loop and auxiliary propagators, whose powers $a_i$ can take positive and negative integer values, respectively, and $a=\sum_{i=1}^r a_i$. This choice expresses all scalar products involving loop and external momenta in terms of the denominators $D_i$. 
The amplitude for the process in \cref{eq:ampl} is expressed as linear combinations of scalar integrals belonging to the integral families in \cref{fig:families}, up to permutations of the external momenta.

\begin{table}[t]
\centering
\begin{tabular}{c@{\qquad}l@{\qquad}l@{\qquad}l}
\toprule
Denominator & \TOP{PL1} & \TOP{PL2} & \TOP{PL3} \tabularnewline
\midrule
$D_{1}$ & $\left(k_{1}-p_{1}\right)^{2}-m^{2}$ & $\left(k_{1}-p_{1}\right)^{2}$ & $\left(k_{1}-p_{1}\right)^{2}$  \\
$D_{2}$ & $k_{1}^{2}$ & $k_{1}^{2}-m^{2}$ & $k_{1}^{2}-m^{2}$ \\
$D_{3}$ & $\left(k_{1}+p_{2}\right)^{2}-m^{2}$ & $\left(k_{1}+p_{3}\right)^{2}-m^{2}$ & $\left(k_{1}+p_{2}\right)^{2}$ \\
$D_{4}$ & $\left(k_{2}-p_{1}\right)^{2}-m^{2}$ & $\left(k_{2}-p_{1}\right)^{2}-m^{2}$ & $\left(k_{2}-p_{1}\right)^{2}-m^{2}$ \\
$D_{5}$ & $\left(k_{2}+p_{2}\right)^{2}-m^{2}$ & $\left(k_{2}+p_{3}\right)^{2}$ & $\left(k_{2}+p_{2}\right)^{2}-m^{2}$ \\
$D_{6}$ & $\left(k_{2}+p_{23}\right)^{2}-m^{2}$ & $\left(k_{2}+p_{23}\right)^{2}-m^{2}$ & $\left(k_{2}+p_{23}\right)^{2}-m^{2}$\\
$D_{7}$ & $\left(k_{1}-k_{2}\right)^{2}$ & $\left(k_{1}-k_{2}\right)^{2}-m^{2}$ & $\left(k_{1}-k_{2}\right)^{2}-m^{2}$ \\
$D_{8}$ & $\left(k_{1}+p_{23}\right)^{2}-m^{2}$ & $\left(k_{1}+p_{23}\right)^{2}-m^{2}$ & $\left(k_{1}+p_{23}\right)^{2}$  \\
$D_{9}$ & $k_{2}^{2}$ & $k_{2}^{2}$ & $k_{2}^{2}$ \\
\midrule
MIs & 68 & 74 & 70 \\
Elliptic sectors (MIs) & 2 (6) & 7 (26) & 6 (24)\\
\bottomrule
\end{tabular}
\caption{Definition of the planar families \TOP{PL1}, \TOP{PL2}, and \TOP{PL3}. 
The Feynman prescription $+ \iu\delta$ is understood for each propagator and not written explicitly. The last two rows report the total number of MIs of each family and the number of MIs associated to elliptic curves.
}
\label{tab:families}
\end{table}

In this work, we begin by considering the set of planar families \TOP{PL1}, \TOP{PL2}, and \TOP{PL3}.\footnote{
The extension to the corresponding non-planar families is currently in progress and will directly benefit from the analytic and numerical framework developed here.
}
Their definition is given in \cref{tab:families}, where the first seven denominators $D_i$ correspond to loop propagators, while the remaining two are auxiliary propagators.
To compute these integrals, we follow the standard approach used in the analytic evaluation of multi-loop Feynman integrals. By employing integration-by-parts identities (IBPs), we first reduce all integrals to a minimal set $\vec{\mathcal{I}}^{\TOP{X}}$ of master integrals (MIs).
We use automated tools for this reduction. In particular, we employ {\sc LiteRed}~\cite{Lee:2012cn} to generate identities between integrals  and {\sc FiniteFlow}~\cite{Peraro:2019svx} to construct and solve the resulting linear systems of equations over finite fields. During the numerical reduction, we identify additional relations by exploiting connections between graph polynomials of Feynman integrals~\cite{Pak:2011xt}. Including these relations leads to a number of master integrals consistent with independent reductions obtained using {\sc Kira-3}~\cite{Lange:2025fba} and {\sc NeatIBP}~\cite{Wu:2023upw}. The number of MIs is reported in \cref{tab:families}. 
We then construct a system of linear differential equations (DEs) for the MIs
\begin{equation}
    \mathrm{d} \vec{\mathcal{I}}^{\TOP{X}} (\vec{x};\varepsilon) = \mathrm{d} A^{(\TOP{X})}(\vec{x};\varepsilon) \cdot \vec{\mathcal{I}}^{\TOP{X}} (\vec{x};\varepsilon)\,,
    \label{eq:DEs_MIs}
\end{equation}
whose solution allows us to evaluate the MIs.
It is known that the complexity of the DEs in \cref{eq:DEs_MIs} strongly depends on the choice of bases of MIs. In the following section, we discuss the strategy we adopt to select a suitable set of MIs for our purposes.

\section{Differential equations for Feynman integrals}
\label{sec:description}
There is a natural choice of MIs: those that satisfy DEs in \emph{canonical form}~\cite{Henn:2013pwa}. These are characterised by the factorisation of the $\varepsilon$-dependence of the connection matrix $\mathrm{d}A$ in \cref{eq:DEs_MIs}, which is then written in terms of one-forms with locally at most simple poles. This form of the DEs is very convenient for the goal of solving them, as it separates the dependence on the kinematics and on the dimensional regulator, and is devoid of spurious poles. The simplest case is that in which the canonical MIs are associated with the geometry of the Riemann sphere, and the connection matrix is expressed in terms of $\dlog$-forms. We find that several sectors in the integral families \TOP{PL1}, \TOP{PL2}, and \TOP{PL3} fall into this case. For these sectors, we use a combination of techniques~\cite{Argeri:2014qva,Dlapa:2021qsl,Flieger:2022xyq} and public tools~\cite{Henn:2020lye,Meyer:2017joq,Flieger:2022xyq} to construct a basis of MIs that brings the differential equations into canonical~form.

Due to the presence of massive virtual particles, we expect that some of the integrals in the families of \cref{fig:families} will involve also more complicated geometries, such as elliptic curves or even Calabi-Yau manifolds. As discussed in \cref{sec:introduction}, a lot of effort has been put into the development of techniques to construct canonical integrals in these cases. However, as we motivate in this section, for our scope of numerically solving the DEs obtaining a canonical form is not a necessity. In the following, we thus explain the strategy that we followed to select the basis of MIs for the families \TOP{PL1}, \TOP{PL2} and \TOP{PL3}.

\subsection*{Differential equations}
Motivated by earlier works~\cite{Badger:2024fgb,Becchetti:2025qlu}, we aim to obtain DEs whose connection matrix can be expressed as
\begin{equation}
    \td\! A^{(\TOP{X})}(\vec{x};\varepsilon) = \sum_{k=0}^2 \varepsilon^k \left[ \sum_\alpha c_{k \alpha}^{(\TOP{X})} \dlog \left( W_\alpha (\vec{x})\right) + \sum_\beta d_{k \beta}^{(\TOP{X})} \omega_\beta (\vec{x}) \right]\,,
\label{eq:deq}
\end{equation}
where $c_{k \alpha}^{(\TOP{X})}$ and $d_{k \beta}^{(\TOP{X})}$ are matrices of rational numbers. The arguments $W_\alpha (\vec{x})$ of the $\dlog$-forms are algebraic functions of the kinematics, called letters, while the non-$\dlog$ one-forms $\omega_\beta (\vec{x})$ are expressed as:
\begin{equation}
    \omega_\beta (\vec{x}) = \omega_\beta^{(s)} (\vec{x}) \mathrm{d} s + \omega_\beta^{(t)} (\vec{x}) \mathrm{d} t +\omega_\beta^{(m^2)}  (\vec{x}) \mathrm{d} m^2 + \omega_\beta^{(q^2)} (\vec{x}) \mathrm{d} q^2,
\end{equation}
where the coefficients $\omega_\beta^{(x_i)} (\vec{x}) \mathrm{d} x_i$ are algebraic functions of the kinematics.

The polynomial dependence on $\varepsilon$ of \cref{eq:deq} is very convenient for the numerical solution of the DEs~\cite{PetitRosas:2025xhm,Badger:2025ilt,Badger:2025ljy}. Moreover, the form in \cref{eq:deq} can be achieved without introducing in the DEs transcendental functions, such as periods of elliptic curves, and the DEs contain only algebraic functions of the kinematics. 

Canonical integrals associated with $\dlog$-forms naturally satisfy DEs of the form of \cref{eq:deq}. As mentioned in \cref{sec:kinematics}, the problem of finding such canonical integrals is well studied in the literature and many public tools are available to this end. From now on, we will therefore focus on the sectors that involve more complicated geometries. For the families at hand, we do not find anything more complicated than elliptic curves. For the sectors associated to these curves, we follow the approach of~\cite{Gorges:2023zgv}. By analysing their analytic structure, either in momentum space or in the Baikov representation, we relate the corresponding Feynman integrals to elliptic differentials of the first, second, and third kind. This naturally leads to DEs of the form of \cref{eq:deq}.

\subsection*{Toy example: Three-point two-mass triangle}

\begin{figure}
\centering
\subfloat[\label{fig:sec_t1}]{\begin{tikzpicture}
\coordinate (v1) at (0,0);
\coordinate (v2) at (1,0.5);
\coordinate (v3) at (1.0,-0.5);
\coordinate (v4) at (2.,0.5);
\coordinate (v5) at (2.,-0.5);
\coordinate (p1) at (-0.75,0);
\coordinate (p2) at (2.75,0.5);
\coordinate (p3) at (2.75,-0.5);
\draw[fermionnoarrow,  thin] (v1) -- (v2);
\draw[fermionnoarrow,  ultra thick] (v1) -- (v3);
\draw[fermionnoarrow,  ultra thick] (v2) -- (v3);
\draw[fermionnoarrow,  ultra thick] (v4) -- (v2);
\draw[fermionnoarrow,  ultra thick] (v4) -- (v5);
\draw[fermionnoarrow,  thin] (v3) -- (v5);
\draw[higgs, ultra thick] (v1) -- (p1);
\draw[fermionnoarrow, thin] (v4) -- (p2);
\draw[fermionnoarrow, ultra thick] (v5) -- (p3);
\end{tikzpicture}}\qquad
\subfloat[\label{fig:sec_t2}]{\begin{tikzpicture}
\coordinate (v3) at (2.0,0.);
\coordinate (p1) at (-0.75,0);
\coordinate (p2) at (2.75,0.5);
\coordinate (p3) at (2.75,-0.5);
\draw[fermionnoarrow,  ultra thick] (v1) -- (v3);
\draw[fermionnoarrow,  ultra thick] (v1) -- (v4);
\draw[fermionnoarrow,  ultra thick] (v1) -- (v5);
\draw[fermionnoarrow,  ultra thick] (v4) -- (v3);
\draw[fermionnoarrow,  thin] (v3) -- (v5);
\draw[higgs, ultra thick] (v1) -- (p1);
\draw[fermionnoarrow, thin] (v4) -- (p2);
\draw[fermionnoarrow, ultra thick] (v5) -- (p3);
\end{tikzpicture}}
\caption{Two-loop two-mass triangle integral families.}
\label{fig:3pt_2m}
\end{figure}

As a warm-up exercise, we study the differential equations for the integral families depicted in \cref{fig:3pt_2m}. The families require a total of 19 master integrals, of which only two (belonging to a single sector) exhibit elliptic geometry. This sector corresponds to the well-known two-loop equal-mass sunrise, which has been extensively studied in the literature~\cite{Remiddi:2003ci,Laporta:2004rb,Pozzorini:2005ff}.
We recover the general structure of \cref{eq:deq}, with logarithmic forms given by the letters of the alphabet
\begin{subequations}
\begin{align}
W_\alpha (\vec{x}) \in  \{m^2\,,s\,,m^2-s\,,9m^2-s,m^2+s\}\,.
\label{eq:alpha_sun}
\end{align}
As expected, all integrals satisfy canonical differential equations on the maximal cut, except the two equal-mass sunrises. 
By analysing the Baikov representation of the latter in $D=2$, we construct the master integrals:
\begin{align}
\mathcal{I}_1 &= 
s\,
\parbox{35mm}{\begin{tikzpicture}
\coordinate (v1) at (0,0);
\coordinate (v2) at (1,0.5);
\coordinate (v3) at (1.0,-0.5);
\coordinate (v4) at (2.,0.0);
\coordinate (p1) at (-0.75,0);
\coordinate (p2) at (2.75,0);
\draw[fermionnoarrow,  ultra thick] (v1) -- (v4);
\draw[fermionnoarrow,  ultra thick] (v1) -- (v4);
\draw[higgs, ultra thick] (v1) -- (p1);
\draw[higgs, ultra thick] (v4) -- (p2);
\draw[fermionnoarrow, ultra thick](1,0) ellipse (1 and 0.5);
\end{tikzpicture}}
\sim\int\frac{\td\!z}{y}\,,
\end{align}
where 
\begin{equation}
    y^2 = z (z-4 m^2) \left(z^2 -2(t+m^2)+(m^2-t)^2 \right) \,,
    \label{eq:elliptic_curve_equal_mass_sunrise}
\end{equation}
defines an elliptic curve in the $(z,y)$-plane. Hence the scalar integral of the equal-mass sunrise in $D=2$ is associated with an elliptic differential of the first kind.

A second MI, related to the differential of the second kind, can be obtained taking the derivative of the first one. In this case, this is essentially equivalent to dotting a propagator. This leads us to the second MI:
\begin{align}
\mathcal{I}_2 &= 
\frac{m^2(s-9 m^2) (s-m^2)}{s}\,
\parbox{35mm}{\begin{tikzpicture}
\coordinate (v1) at (0,0);
\coordinate (v2) at (1,0.5);
\coordinate (v3) at (1.0,-0.5);
\coordinate (v4) at (2.,0.0);
\coordinate (p1) at (-0.75,0);
\coordinate (p2) at (2.75,0);
\draw[fermionnoarrow,  ultra thick] (v1) -- (v4);
\draw[fermionnoarrow,  ultra thick] (v1) -- (v4);
\draw[higgs, ultra thick] (v1) -- (p1);
\draw[higgs, ultra thick] (v4) -- (p2);
\draw[fermionnoarrow, ultra thick](1,0) ellipse (1 and 0.5);
\draw[fill=black] (v2) circle (.07cm);
\end{tikzpicture}}\ .
\end{align}
\label{eq:y_sunrise}
\end{subequations}
These two integrals satisfy coupled differential equations,
\begin{align}
\td\left(\begin{array}{c}
\mathcal{I}_{1}\\
\mathcal{I}_{2}
\end{array}\right)=&\Bigg[\epsilon\left(\begin{array}{cc}
-2L_{3}-2L_{4}-L_{1}+3L_{2} & 0\\
0 & -2L_{2}
\end{array}\right)
\notag\\
&+\left(\begin{array}{cc}
0 & (2\epsilon+1)(3\epsilon+1)\omega_{1}\\
\frac{3}{8}L_{3}-\frac{27}{8}L_{4}+3L_{1} & 0
\end{array}\right)\Bigg] \cdot \left(\begin{array}{c}
\mathcal{I}_{1}\\
\mathcal{I}_{2}
\end{array}\right)\,,
\end{align}
with $L_i=\dlog (W_{\alpha_{i}})$, and the one form is
\begin{align}
\omega_1 = \mathrm{d} \left[\frac{m^{2}}{s}\left(1-\frac{3}{2}\frac{m^{2}}{s}\right)\right]\,.
\end{align}
In this case, the one-form is the total differential of an algebraic function, i.e.~it is an exact one-form. When we discuss the final form of our DEs in \cref{sec:benchmarks}, we shall see that this is in general not the case for our choice of MIs.

This elliptic sector appears as sub-sector of those discussed in the following sections. Due to its simplicity, we do not analyse it further here and instead focus on the treatment of elliptic sectors arising in genuine four-point integral families with five to seven propagators. The systematic numerical integration of the corresponding differential equations is discussed in more detail in \cref{sec:benchmarks}, where we compare the precision of {\sc DiffExp}~\cite{Hidding:2020ytt} and our proof-of-concept implementation.

\section{Elliptic sectors}
\label{sec:elliptic}
In this section, we discuss our approach to choose a suitable set of MIs for the elliptic sectors appearing in the families \TOP{PL1}\,, \TOP{PL2}\,, and \TOP{PL3}\,, by highlighting the features of the differential equations they satisfy. We follow the strategy of~\cite{Gorges:2023zgv} to identify suitable candidates on the maximal cut of each sector as starting point. 
We begin with the simpler elliptic sectors containing three master integrals for \TOP{PL1} (see Sec.~\ref{sec:simple_sectors_PL1}) and analogously for \TOP{PL2} and \TOP{PL3} (see Sec.~\ref{sec:simple_sectors}). We then address the most challenging sectors encountered in this work, namely the four-point kite sectors, in \cref{sec:fourpt_kite}. For the latter, we build on insights from~\cite{Bargiela:2025vwl} (and the explicit calculation of \TOP{PL3} with $q^2=0$ in Refs.~\cite{Adams:2018bsn,Adams:2018kez}), which show that the corner integrals in these sectors admit algebraic leading singularities, while elliptic behaviour arises in integrals with dots ($a_i>1$) or numerators ($a_i<0$).
Finally, we turn to the top sectors of the families \TOP{PL2} and \TOP{PL3} in Sec.~\ref{sec:top_sectors}.

\subsection{The sectors $\{ 0,1,1,1,1,1,1,0,0 \}$ and $\{ 1,1,0,1,1,1,1,0,0 \}$ of \TOP{PL1}}
\label{sec:simple_sectors_PL1}

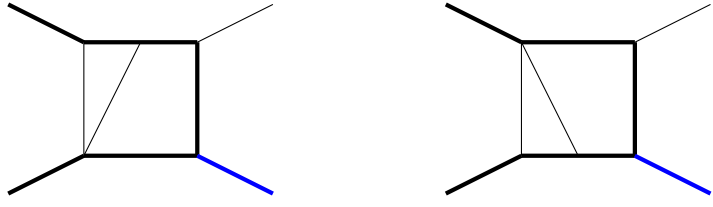
\begin{figure}[t]
    \centering
\subfloat[Sector $\{ 0,1,1,1,1,1,1,0,0 \}$:~3~MIs]{
\begin{tikzpicture}
\coordinate (v1) at (0,0);
\coordinate (v2) at (0,1.5);
\coordinate (v5) at (1.5,1.5);
\coordinate (v6) at (1.5,0.0);
\coordinate (p1) at (-1,-0.5);
\coordinate (p2) at (-1,2);
\coordinate (p3) at (2.5,2);
\coordinate (p4) at (2.5,-0.5);
\coordinate (v3) at (0.75,1.5);
\coordinate (v4) at (0.75,0);
\draw[fermionnoarrow,  thin] (v1) -- (v2);
\draw[fermionnoarrow,  thin] (v1) -- (v3);
\draw[fermionnoarrow,  ultra thick] (v2) -- (v5);
\draw[fermionnoarrow,  ultra thick] (v5) -- (v6);
\draw[fermionnoarrow,  ultra thick] (v6) -- (v1);
\draw[fermionnoarrow, ultra thick] (v1) -- (p1);
\draw[fermionnoarrow, ultra thick] (v2) -- (p2);
\draw[fermionnoarrow, thin] (v5) -- (p3);
\draw[fermionnoarrow, ultra thick, blue] (v6) -- (p4);
\node  at (-1.4,-0.5) {\phantom{$p_1$}}; 
\node  at (3,-0.5) {\phantom{$p_4$}}; 
\end{tikzpicture}
\label{fig:PL1_011111100}
}\quad 
\subfloat[Sector $\{ 1,1,0,1,1,1,1,0,0 \}$:~3~MIs]{
\begin{tikzpicture}
\draw[fermionnoarrow,  thin] (v1) -- (v2);
\draw[fermionnoarrow,  thin] (v2) -- (v4);
\draw[fermionnoarrow,  ultra thick] (v2) -- (v5);
\draw[fermionnoarrow,  ultra thick] (v5) -- (v6);
\draw[fermionnoarrow,  ultra thick] (v6) -- (v1);
\draw[fermionnoarrow, ultra thick] (v1) -- (p1);
\draw[fermionnoarrow, ultra thick] (v2) -- (p2);
\draw[fermionnoarrow, thin] (v5) -- (p3);
\draw[fermionnoarrow, ultra thick, blue] (v6) -- (p4);
\node  at (-1.4,-0.5) {\phantom{$p_1$}}; 
\node  at (3,-0.5) {\phantom{$p_4$}}; 
\end{tikzpicture}
\label{fig:PL1_110111100}
}
\caption{Elliptic sectors $\{ 0,1,1,1,1,1,1,0,0 \}$ and $\{ 1,1,0,1,1,1,1,0,0 \}$ of \TOP{PL1}.}
\label{fig:6-prop_elliptic_PL1}
\end{figure}

We present the two sectors in \cref{fig:6-prop_elliptic_PL1} together, as they are associated to the same elliptic curve, and their treatment is essentially identical. Without loss of generality, we therefore work with the sector in \cref{fig:PL1_011111100}. Employing \texttt{BaikovPackage}~\cite{Frellesvig:2024ymq}, we investigate the loop-by-loop Baikov representation~\cite{Baikov:1996iu, Baikov:1996rk,Frellesvig:2017aai} of the corner integral of this sector, working first on the maximal cut, and up to $\varepsilon$ corrections. Starting from the $k_1$-loop, the only residual auxiliary propagator (or irreducible scalar product) is $D_9$, which we relabel $z$ in the following. Up to a prefactor the integral is
\begin{equation}
\begin{split}
    I_{011111100}^{(\TOP{PL1})} \bigl|^{\rm MC}_{\varepsilon=0} &\propto \int \frac{\mathrm{d} z}{(q^2-s)\sqrt{ \mathcal{P}^{(\TOP{PL1})}_4(z)}} \,,\\
    \mathcal{P}^{(\TOP{PL1})}_4(z) &= z (z -4 m^2)\mathcal{P}^{(\TOP{PL1})}_{2}(z) \,,\\
    \mathcal{P}^{(\TOP{PL1})}_{2}(z) &= z^2 +2 \frac{s t + m^2 (s-2 q^2)}{q^2-s} z + \frac{s(s-4 m^2)(m^2-t)^2}{(q^2-s)^2}\,,
\end{split}
    \label{eq:maxcut_PL1_011111100}
\end{equation}
where the roots of degree two polynomial $\mathcal{P}^{(\TOP{PL1})}_{2}(z)$ are distinct and different from $z=0, 4 m^2$. Therefore, the quartic polynomial in \cref{eq:maxcut_PL1_011111100} defines an elliptic curve, and the maximal cut of this integral cannot be expressed as a product of one-forms of logarithmic type. For instance, \cref{eq:maxcut_PL1_011111100} shows that the corner integral of this sector is associated with an elliptic differential of the first kind. This immediately tells us that a first master integral for this sector should be
\begin{equation}
    \mathcal{I}^{\TOP{PL1}}_5 = 
    \varepsilon^4 m^2 (q^2-s) I_{011111100}^{(\TOP{PL1})}\,,
\end{equation}
where the factor $(q^2-s)$ is chosen to normalise the $z^4$ monomial in \cref{eq:maxcut_PL1_011111100} to one, and the $m^2$ normalisation is included to make the integral dimensionless in $D=4$.

As our sector has three MIs, we need to identify two more. 
We choose the second one as a derivative of the first one:
\begin{equation}
    \mathcal{I}^{\TOP{PL1}}_6 = \varepsilon^4 m^4 \partial_t \bigg( (q^2-s) I_{011111100}^{(\TOP{PL1})} \bigg),
\end{equation}
where we chose the derivative w.r.t. $t$ because it yields simpler DEs, and we multiplied by an additional power of $m^2$ to make the integral dimensionless in four dimensions. We want the last integral to be associated with an elliptic differential of the third kind, which complements the elliptic curve with a simple pole. The easiest way to achieve this, is to consider the integral
\begin{equation}
    \mathcal{I}^{\TOP{PL1}}_4 = \varepsilon^4 (q^2-s) I_{01111110-1}^{(\TOP{PL1})}\,,
\end{equation}
whose maximal cut, up to a prefactor, is
\begin{equation}
    I_{01111110-1}^{(\TOP{PL1})} \bigl|^{\rm MC}_{\varepsilon=0} \propto \int \frac{z \ \mathrm{d} z}{\sqrt{z (z -4 m^2)\mathcal{P}^{(\TOP{PL1})}_{2}(z)}} \,,
\end{equation}
which has a pole at infinity. This can be seen by performing the change of variables $z \to 1/\tilde{z}$, and computing the Laurent expansion of the integrand around $\tilde{z}=0$, which corresponds to $z \to \infty$. The leading term is a simple pole with unit residue.

On the maximal cut, the DEs for this set of integrals take the following form
\begin{equation}
   \td \begin{pmatrix}
       \mathcal{I}^{\TOP{PL1}}_4\\
       \mathcal{I}^{\TOP{PL1}}_5\\
       \mathcal{I}^{\TOP{PL1}}_6
   \end{pmatrix} = \begin{pmatrix}
        * \varepsilon & * + * \varepsilon & *\\
        * \varepsilon & * + * \varepsilon & *\\
        * \varepsilon(* + * \varepsilon) & * + * \varepsilon + * \varepsilon^2 & * + * \varepsilon
    \end{pmatrix} \cdot \begin{pmatrix}
       \mathcal{I}^{\TOP{PL1}}_4\\
       \mathcal{I}^{\TOP{PL1}}_5\\
       \mathcal{I}^{\TOP{PL1}}_6
   \end{pmatrix}\,,
   \label{eq:PL1_simple_maxcut_DEs}
\end{equation}
which clearly fulfils the properties that we are looking for. We remark that, despite the fact that they are $\varepsilon$-factorised, the first two entries of the first column cannot be written in terms of logarithmic one-forms. This is expected, since all the MIs of the sector are associated with elliptic differentials. Releasing the cuts, we verify that the DEs exhibit the same $\varepsilon$-dependence also beyond the maximal cut: the entries involving $\mathcal{I}^{\TOP{PL1}}_4$ and $\mathcal{I}^{\TOP{PL1}}_5$ are at worst linear in $\varepsilon$, while the couplings of $\mathcal{I}^{\TOP{PL1}}_6$ to the sub-sectors are at worst quadratic in $\varepsilon$. We can further factorise some entries of the DEs in $\varepsilon$ by including sub-sector contributions to $\mathcal{I}_4^{\TOP{PL1}}$, the integral associated with the differential of the third kind. These can be determined from the DEs, integrating the $\varepsilon^0$ term.

~

As mentioned at the beginning of this subsection, the sector in \cref{fig:PL1_110111100} is associated with the same elliptic curve of the one we just discussed, and we thus determine a basis for it following the same strategy. Summarising, the set of MIs,
\begin{equation}
\begin{split}
\mathcal{I}_{4}^{\TOP{PL1}} & =\varepsilon^{4}\left(q^{2}-s\right)\left[I_{01111110-1}^{(\TOP{PL1})}-I_{011101100}^{(\TOP{PL1})}\right]\,,\\
\mathcal{I}_{5}^{\TOP{PL1}} & =\varepsilon^{4} m^2 \left(q^{2}-s\right)I_{011111100}^{(\TOP{PL1})}\,,\\
\mathcal{I}_{6}^{\TOP{PL1}} & =\varepsilon^{4} m^{4} \partial_{t}\bigg(\left(q^{2}-s\right)I_{011111100}^{(\TOP{PL1})}\bigg)\,,\\
\mathcal{I}_{9}^{\TOP{PL1}} & =\varepsilon^{4}\left[\left(q^{2}-s\right)I_{11011110-1}^{(\TOP{PL1})}+\left(m^{2}-t\right)I_{110011100}^{(\TOP{PL1})}\right]\,,\\
\mathcal{I}_{10}^{\TOP{PL1}} & =\varepsilon^{4} m^2 \left(q^{2}-s\right)I_{110111100}^{(\TOP{PL1})}\,,\\
\mathcal{I}_{11}^{\TOP{PL1}} & =\varepsilon^{4}m^{4}\partial_{t}\bigg(\left(q^{2}-s\right)I_{110111100}^{(\TOP{PL1})}\bigg)\,,
\end{split}
\label{eq:basis_PL1}
\end{equation}
satisfies a system of DEs which is at most quadratic in $\varepsilon$ and does not have any poles in $\varepsilon$, according to the structure of \cref{eq:deq}.

\subsection{The other simple elliptic sectors}
\label{sec:simple_sectors}

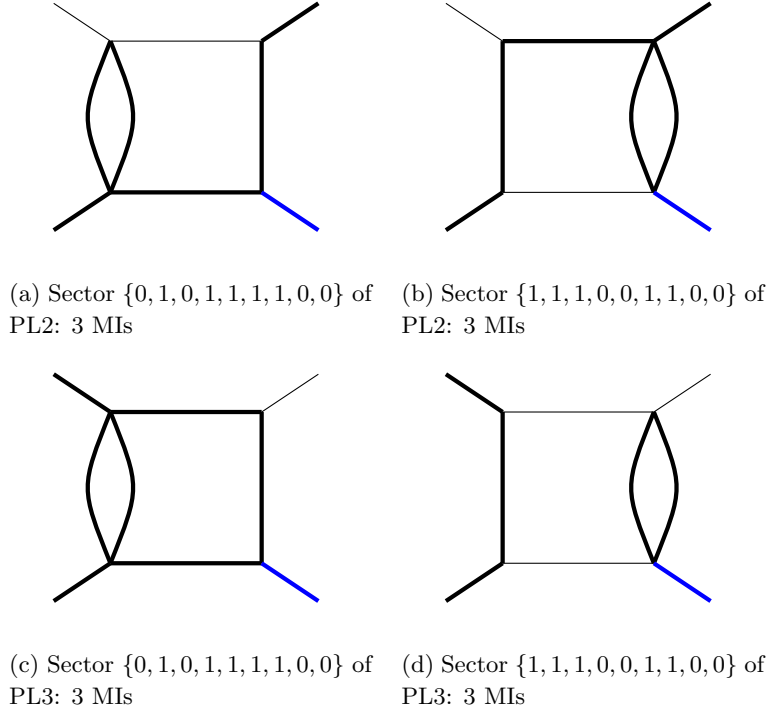
\begin{figure}[h!]
    \centering
\subfloat[Sector $\{ 0,1,0,1,1,1,1,0,0 \}$ of PL2: 3 MIs]{
\begin{tikzpicture}
\coordinate (v1) at (0,0);
\coordinate (v2) at (0,2);
\coordinate (v3) at (2,0);
\coordinate (v4) at (2,2);
\coordinate (p1) at (-0.75,-0.5);
\coordinate (p2) at (-0.75,2.5);
\coordinate (p3) at (2.75,-0.5);
\coordinate (p4) at (2.75,2.5);
\draw[fermionnoarrow,  ultra thick] (v3) -- (v4);
\draw[fermionnoarrow,  ultra thick] (v1) .. controls (0.4,1) .. (v2);
\draw[fermionnoarrow,  ultra thick] (v1) .. controls (-0.4,1) .. (v2);
\draw[fermionnoarrow,  ultra thick] (v1) -- (v3);
\draw[fermionnoarrow,  thin] (v2) -- (v4);
\draw[fermionnoarrow,  ultra thick] (v1) -- (p1);
\draw[fermionnoarrow,  ultra thick, blue] (v3) -- (p3);
\draw[fermionnoarrow,  thin] (v2) -- (p2);
\draw[fermionnoarrow,  ultra thick] (v4) -- (p4);
\node  at (-0.9,-0.75) {\phantom{$p_1$}}; 
\node  at (2.9,-0.75) {\phantom{$p_4$}}; 
\end{tikzpicture}
\label{fig:PL2_010111100}
}\quad 
\subfloat[Sector $\{ 1,1,1,0,0,1,1,0,0 \}$ of PL2: 3 MIs]{
\begin{tikzpicture}
\coordinate (v1) at (0,0);
\coordinate (v2) at (0,2);
\coordinate (v3) at (2,0);
\coordinate (v4) at (2,2);
\coordinate (p1) at (-0.75,-0.5);
\coordinate (p2) at (-0.75,2.5);
\coordinate (p3) at (2.75,-0.5);
\coordinate (p4) at (2.75,2.5);
\draw[fermionnoarrow,  ultra thick] (v1) -- (v2);
\draw[fermionnoarrow,  ultra thick] (v3) .. controls (2.4,1) .. (v4);
\draw[fermionnoarrow,  ultra thick] (v3) .. controls (1.6,1) .. (v4);
\draw[fermionnoarrow,  ultra thick] (v2) -- (v4);
\draw[fermionnoarrow,  thin] (v1) -- (v3);
\draw[fermionnoarrow,  ultra thick] (v1) -- (p1);
\draw[fermionnoarrow,  thin] (v2) -- (p2);
\draw[fermionnoarrow,  ultra thick, blue] (v3) -- (p3);
\draw[fermionnoarrow,  ultra thick] (v4) -- (p4);
\node  at (-0.9,-0.75) {\phantom{$p_1$}}; 
\node  at (2.9,-0.75) {\phantom{$p_4$}}; 
\end{tikzpicture}
\label{fig:PL2_111001100}
}\\
\subfloat[Sector $\{ 0,1,0,1,1,1,1,0,0 \}$ of PL3: 3 MIs]{
\begin{tikzpicture}
\coordinate (v1) at (0,0);
\coordinate (v2) at (0,2);
\coordinate (v3) at (2,0);
\coordinate (v4) at (2,2);
\coordinate (p1) at (-0.75,-0.5);
\coordinate (p2) at (-0.75,2.5);
\coordinate (p3) at (2.75,-0.5);
\coordinate (p4) at (2.75,2.5);
\draw[fermionnoarrow,  ultra thick] (v3) -- (v4);
\draw[fermionnoarrow,  ultra thick] (v1) .. controls (0.4,1) .. (v2);
\draw[fermionnoarrow,  ultra thick] (v1) .. controls (-0.4,1) .. (v2);
\draw[fermionnoarrow,  ultra thick] (v2) -- (v4);
\draw[fermionnoarrow,  ultra thick] (v1) -- (v3);
\draw[fermionnoarrow,  thin] (v4) -- (p4);
\draw[fermionnoarrow,  ultra thick, blue] (v3) -- (p3);
\draw[fermionnoarrow,  ultra thick] (v2) -- (p2);
\draw[fermionnoarrow,  ultra thick] (v1) -- (p1);
\node  at (-0.9,-0.75) {\phantom{$p_1$}}; 
\node  at (2.9,-0.75) {\phantom{$p_4$}}; 
\end{tikzpicture}
\label{fig:PL3_010111100}
}\quad 
\subfloat[Sector $\{ 1,1,1,0,0,1,1,0,0 \}$ of PL3: 3 MIs]{
\begin{tikzpicture}
\coordinate (v1) at (0,0);
\coordinate (v2) at (0,2);
\coordinate (v3) at (2,0);
\coordinate (v4) at (2,2);
\coordinate (p1) at (-0.75,-0.5);
\coordinate (p2) at (-0.75,2.5);
\coordinate (p3) at (2.75,-0.5);
\coordinate (p4) at (2.75,2.5);
\draw[fermionnoarrow,  ultra thick] (v1) -- (v2);
\draw[fermionnoarrow,  ultra thick] (v3) .. controls (2.4,1) .. (v4);
\draw[fermionnoarrow,  ultra thick] (v3) .. controls (1.6,1) .. (v4);
\draw[fermionnoarrow,  thin] (v2) -- (v4);
\draw[fermionnoarrow,  thin] (v1) -- (v3);
\draw[fermionnoarrow,  ultra thick] (v1) -- (p1);
\draw[fermionnoarrow,  ultra thick] (v2) -- (p2);
\draw[fermionnoarrow,  ultra thick, blue] (v3) -- (p3);
\draw[fermionnoarrow,  thin] (v4) -- (p4);
\node  at (-0.9,-0.75) {\phantom{$p_1$}}; 
\node  at (2.9,-0.75) {\phantom{$p_4$}}; 
\end{tikzpicture}
\label{fig:PL3_111001100}
}
\caption{The graphs of the box-bubble elliptic sectors of PL2 and PL3}
\label{fig:box_bubble_elliptic_PL2_PL3}
\end{figure}

The two elliptic sectors that we have just discussed in \cref{sec:simple_sectors_PL1} are the only two elliptic sectors of \TOP{PL1}. In \TOP{PL2} and \TOP{PL3}\,, apart from the elliptic sunrises that we already presented in \cref{sec:description}\,, there are four sectors that we classify as simple, i.e.~that can be treated analogously to the two previously presented. There are two such sectors for each of the two families, all of box-bubble type (see \cref{fig:box_bubble_elliptic_PL2_PL3}). In all cases, analysing the integral with a dot on the seventh propagator in the loop-by-loop Baikov representation, starting from the loop momentum associated with the bubble, we unveil an elliptic differential of the first~kind.

The sector in \cref{fig:PL2_010111100} is associated with the elliptic curve similar to the one in \cref{eq:maxcut_PL1_011111100}\,, but with
\begin{equation}
    \mathcal{P}^{(\TOP{PL2})}_{2}(z) = z^2 +2 \frac{(u-m^2)((m^2-t)^2+s(m^2+t)-2 m^2 q^2)}{\lambda_{\text{K}}\left(q^2,u,m^{2}\right)} z + \frac{(u-m^2)^2(m^2-t)^2}{\lambda_{\text{K}}\left(q^2,u,m^{2}\right)}\,,
    \label{eq:elliptic_curve_010111100_PL2}
\end{equation}
where $\lambda_{\text{K}}(a,b,c)=a^2+b^2+c^2-2ab-2ac-2bc$.

The elliptic curve associated with the other box-bubble sector of \TOP{PL2} is isomorphic to the one in \cref{eq:elliptic_curve_010111100_PL2}\,, as can be checked by computing their $j$-invariants~\cite{Lang1987-zb}. 
Analogously, the curve associated with the sector in \cref{fig:PL3_010111100} is isomorphic to the one appearing in \cref{eq:maxcut_PL1_011111100}. Finally, the sector in \cref{fig:PL3_111001100} is associated with the curve
\begin{equation}
    \mathcal{P}^{(\TOP{PL3)}}_{2}(z) = z^2 +2 \frac{s(m^2(s-2 q^2)+s t)}{\lambda_{\text{K}}\left(s,m^2,m^{2}\right)} z + \frac{s^2(m^2-t)^2}{\lambda_{\text{K}}\left(s,m^2,m^{2}\right)}\,.
    \label{eq:elliptic_curve_111001100_PL3}
\end{equation}
All these sectors contain three master integrals, which we choose in the same fashion as we chose those in \cref{sec:simple_sectors_PL1}. 

For the two sectors of \TOP{PL2} we choose the integrals
\begin{equation}
\begin{split}
    \mathcal{I}^{\TOP{PL2}}_{20} &= \varepsilon^3 m^2 \sqrt{\lambda_{\text{K}}\left(q^2,u,m^{2}\right)} I_{010111200}^{(\TOP{PL2})}\,,\\
    \mathcal{I}^{\TOP{PL2}}_{21} &= \varepsilon^3 \sqrt{\lambda_{\text{K}}\left(q^2,u,m^{2}\right)} I_{01011120-1}^{(\TOP{PL2})}\,,\\
    \mathcal{I}^{\TOP{PL2}}_{22} &= \varepsilon^3 m^4 \partial_{q^2} \bigg( \sqrt{\lambda_{\text{K}}\left(q^2,u,m^{2}\right)} I_{010111200}^{(\TOP{PL2})} \bigg)\,,\\
    \mathcal{I}^{\TOP{PL2}}_{35} &= \varepsilon^3 m^2 (u-m^2) I_{111001200}^{(\TOP{PL2})}\,,\\
    \mathcal{I}^{\TOP{PL2}}_{36} &= \varepsilon^3 (u-m^2) I_{1110012-10}^{(\TOP{PL2})}\,,\\
    \mathcal{I}^{\TOP{PL2}}_{37} &= \varepsilon^3 m^4 \partial_{q^2} \bigg( (u-m^2) I_{111001200}^{(\TOP{PL2})} \bigg)\,,
\end{split}
\label{eq:basis_simple_PL2}
\end{equation}
where we normalised by $\varepsilon^3$ because of the dotted propagator.

Similarly, for \TOP{PL3} we pick the MIs:
\begin{equation}
\begin{split}
    \mathcal{I}^{\TOP{PL3}}_{22} &= \varepsilon^3 \sqrt{\lambda_{\text{K}}\left(s,m^2,m^{2}\right)} I_{1110012-10}^{(\TOP{PL3})}\,,\\
    \mathcal{I}^{\TOP{PL3}}_{23} &= \varepsilon^3 m^2 \sqrt{\lambda_{\text{K}}\left(s,m^2,m^{2}\right)} I_{111001200}^{(\TOP{PL3})}\,,\\
    \mathcal{I}^{\TOP{PL3}}_{24} &= \varepsilon^3 m^4 \partial_{t} \bigg( \sqrt{\lambda_{\text{K}}\left(s,m^2,m^{2}\right)} I_{111001200}^{(\TOP{PL3})} \bigg)\,,\\
    \mathcal{I}^{\TOP{PL3}}_{25} &= \varepsilon^3 (q^2-s) I_{01011120-1}^{(\TOP{PL3})}\,,\\
    \mathcal{I}^{\TOP{PL3}}_{26} &= \varepsilon^3 m^2 (q^2-s) I_{010111200}^{(\TOP{PL3})}\,,\\
    \mathcal{I}^{\TOP{PL3}}_{27} &= \varepsilon^3 m^4 \partial_{t} \bigg( (q^2-s) I_{010111200}^{(\TOP{PL3})} \bigg)\,.
\end{split}
\label{eq:basis_simple_PL3}
\end{equation}
Unsurprisingly, on the maximal cut these MIs satisfy DEs with the same $\varepsilon$-dependence as in \cref{eq:PL1_simple_maxcut_DEs}. Releasing the cuts, we see that no sub-sector contribution is missing from the MIs associated with the differential forms of the third kind.

\subsection{The four-point kite sectors}
\label{sec:fourpt_kite}

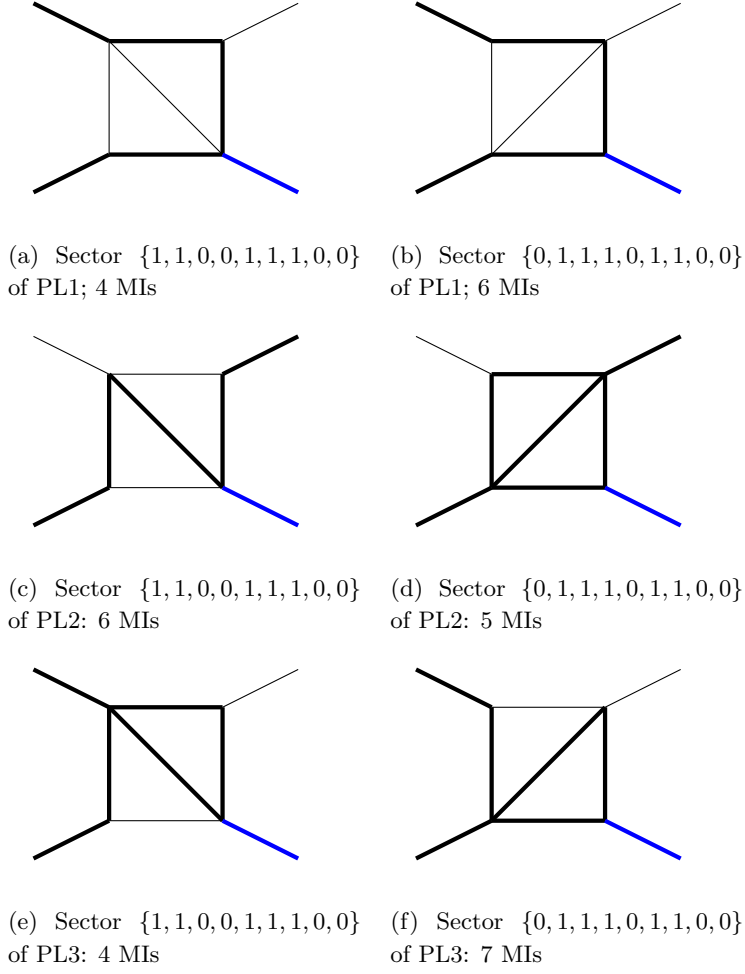
\begin{figure}[h]
    \centering
\subfloat[Sector $\{1,1,0,0,1,1,1,0,0\}$ of PL1; 4 MIs]{\label{fig:PL1_110011100}
\begin{tikzpicture}
\coordinate (v1) at (0,0);
\coordinate (v2) at (0,1.5);
\coordinate (v5) at (1.5,1.5);
\coordinate (v6) at (1.5,0.0);
\coordinate (v3) at (0.75,1.5);
\coordinate (v4) at (0.75,0);
\coordinate (p1) at (-1,-0.5);
\coordinate (p2) at (-1,2);
\coordinate (p3) at (2.5,2);
\coordinate (p4) at (2.5,-0.5);
\draw[fermionnoarrow,  thin] (v1) -- (v2);
\draw[fermionnoarrow,  thin] (v2) -- (v6);
\draw[fermionnoarrow,  ultra thick] (v2) -- (v5);
\draw[fermionnoarrow,  ultra thick] (v5) -- (v6);
\draw[fermionnoarrow,  ultra thick] (v6) -- (v1);
\draw[fermionnoarrow, ultra thick] (v1) -- (p1);
\draw[fermionnoarrow, ultra thick] (v2) -- (p2);
\draw[fermionnoarrow, thin] (v5) -- (p3);
\draw[fermionnoarrow, ultra thick, blue] (v6) -- (p4);
\node  at (-0.9,-0.75) {\phantom{$p_1$}}; 
\node  at (2.9,-0.75) {\phantom{$p_4$}}; 
\end{tikzpicture}
}\quad
\subfloat[Sector $\{0,1,1,1,0,1,1,0,0\}$ of PL1; 6 MIs]{\label{fig:PL1_011101100}
\begin{tikzpicture}
\coordinate (v1) at (0,0);
\coordinate (v2) at (0,1.5);
\coordinate (v5) at (1.5,1.5);
\coordinate (v6) at (1.5,0.0);
\coordinate (v3) at (0.75,1.5);
\coordinate (v4) at (0.75,0);
\coordinate (p1) at (-1,-0.5);
\coordinate (p2) at (-1,2);
\coordinate (p3) at (2.5,2);
\coordinate (p4) at (2.5,-0.5);
\draw[fermionnoarrow,  thin] (v1) -- (v2);
\draw[fermionnoarrow,  thin] (v1) -- (v5);
\draw[fermionnoarrow,  ultra thick] (v2) -- (v5);
\draw[fermionnoarrow,  ultra thick] (v5) -- (v6);
\draw[fermionnoarrow,  ultra thick] (v6) -- (v1);
\draw[fermionnoarrow, ultra thick] (v1) -- (p1);
\draw[fermionnoarrow, ultra thick] (v2) -- (p2);
\draw[fermionnoarrow, thin] (v5) -- (p3);
\draw[fermionnoarrow, ultra thick, blue] (v6) -- (p4);
\node  at (-0.9,-0.75) {\phantom{$p_1$}}; 
\node  at (2.9,-0.75) {\phantom{$p_4$}}; 
\end{tikzpicture}
}\\
\subfloat[Sector $\{1,1,0,0,1,1,1,0,0\}$ of PL2: 6 MIs]{\label{fig:PL2_110011100}
\begin{tikzpicture}
\coordinate (v1) at (0,0);
\coordinate (v2) at (0,1.5);
\coordinate (v5) at (1.5,1.5);
\coordinate (v6) at (1.5,0.0);
\coordinate (v3) at (0.75,1.5);
\coordinate (v4) at (0.75,0);
\coordinate (p1) at (-1,-0.5);
\coordinate (p2) at (-1,2);
\coordinate (p3) at (2.5,2);
\coordinate (p4) at (2.5,-0.5);
\draw[fermionnoarrow,  ultra thick] (v1) -- (v2);
\draw[fermionnoarrow,  ultra thick] (v2) -- (v6);
\draw[fermionnoarrow,  thin] (v2) -- (v5);
\draw[fermionnoarrow,  ultra thick] (v5) -- (v6);
\draw[fermionnoarrow,  thin] (v6) -- (v1);
\draw[fermionnoarrow, thin] (v2) -- (p2);
\draw[fermionnoarrow, ultra thick] (v1) -- (p1);
\draw[fermionnoarrow, ultra thick] (v5) -- (p3);
\draw[fermionnoarrow, ultra thick, blue] (v6) -- (p4);
\node  at (-0.9,-0.75) {\phantom{$p_1$}}; 
\node  at (2.9,-0.75) {\phantom{$p_4$}}; 
\end{tikzpicture}
}\quad
\subfloat[Sector $\{0,1,1,1,0,1,1,0,0\}$ of PL2: 5 MIs]{\label{fig:PL2_011101100}
\begin{tikzpicture}
\coordinate (v1) at (0,0);
\coordinate (v2) at (0,1.5);
\coordinate (v5) at (1.5,1.5);
\coordinate (v6) at (1.5,0.0);
\coordinate (v3) at (0.75,1.5);
\coordinate (v4) at (0.75,0);
\coordinate (p1) at (-1,-0.5);
\coordinate (p2) at (-1,2);
\coordinate (p3) at (2.5,2);
\coordinate (p4) at (2.5,-0.5);
\draw[fermionnoarrow,  ultra thick] (v1) -- (v2);
\draw[fermionnoarrow,  ultra thick] (v1) -- (v5);
\draw[fermionnoarrow,  ultra thick] (v2) -- (v5);
\draw[fermionnoarrow,  ultra thick] (v5) -- (v6);
\draw[fermionnoarrow,  ultra thick] (v6) -- (v1);
\draw[fermionnoarrow, thin] (v2) -- (p2);
\draw[fermionnoarrow, ultra thick] (v1) -- (p1);
\draw[fermionnoarrow, ultra thick] (v5) -- (p3);
\draw[fermionnoarrow, ultra thick, blue] (v6) -- (p4);
\node  at (-0.9,-0.75) {\phantom{$p_1$}}; 
\node  at (2.9,-0.75) {\phantom{$p_4$}}; 
\end{tikzpicture}
}\\
\subfloat[Sector $\{1,1,0,0,1,1,1,0,0\}$ of PL3: 4 MIs]{\label{fig:PL3_110011100}
\begin{tikzpicture}
\coordinate (v1) at (0,0);
\coordinate (v2) at (0,1.5);
\coordinate (v5) at (1.5,1.5);
\coordinate (v6) at (1.5,0.0);
\coordinate (v3) at (0.75,1.5);
\coordinate (v4) at (0.75,0);
\coordinate (p1) at (-1,-0.5);
\coordinate (p2) at (-1,2);
\coordinate (p3) at (2.5,2);
\coordinate (p4) at (2.5,-0.5);
\draw[fermionnoarrow,  ultra thick] (v1) -- (v2);
\draw[fermionnoarrow,  ultra thick] (v2) -- (v6);
\draw[fermionnoarrow,  ultra thick] (v2) -- (v5);
\draw[fermionnoarrow,  ultra thick] (v5) -- (v6);
\draw[fermionnoarrow,  thin] (v6) -- (v1);
\draw[fermionnoarrow, ultra thick] (v1) -- (p1);
\draw[fermionnoarrow, ultra thick] (v2) -- (p2);
\draw[fermionnoarrow, thin] (v5) -- (p3);
\draw[fermionnoarrow, ultra thick, blue] (v6) -- (p4);
\node  at (-0.9,-0.75) {\phantom{$p_1$}}; 
\node  at (2.9,-0.75) {\phantom{$p_4$}}; 
\end{tikzpicture}
}\quad
\subfloat[Sector $\{0,1,1,1,0,1,1,0,0\}$ of PL3: 7 MIs]{\label{fig:PL3_011101100}
\begin{tikzpicture}
\coordinate (v1) at (0,0);
\coordinate (v2) at (0,1.5);
\coordinate (v5) at (1.5,1.5);
\coordinate (v6) at (1.5,0.0);
\coordinate (v3) at (0.75,1.5);
\coordinate (v4) at (0.75,0);
\coordinate (p1) at (-1,-0.5);
\coordinate (p2) at (-1,2);
\coordinate (p3) at (2.5,2);
\coordinate (p4) at (2.5,-0.5);
\draw[fermionnoarrow,  ultra thick] (v1) -- (v2);
\draw[fermionnoarrow,  ultra thick] (v1) -- (v5);
\draw[fermionnoarrow,  thin] (v2) -- (v5);
\draw[fermionnoarrow,  ultra thick] (v5) -- (v6);
\draw[fermionnoarrow,  ultra thick] (v6) -- (v1);
\draw[fermionnoarrow, ultra thick] (v1) -- (p1);
\draw[fermionnoarrow, ultra thick] (v2) -- (p2);
\draw[fermionnoarrow, thin] (v5) -- (p3);
\draw[fermionnoarrow, ultra thick, blue] (v6) -- (p4);
\node  at (-0.9,-0.75) {\phantom{$p_1$}}; 
\node  at (2.9,-0.75) {\phantom{$p_4$}}; 
\end{tikzpicture}
}
\caption{Four-point kite sectors appearing in the families \TOP{PL1}\,, \TOP{PL2} and \TOP{PL3}}
\label{fig:4pt_kite}
\end{figure}

The most complicated sectors appearing in the planar families \TOP{PL1}\,, \TOP{PL2} and \TOP{PL3} are the four-point kite sectors shown in \cref{fig:4pt_kite}. In~\cite{Bargiela:2025vwl} it was shown that, for general mass configurations, one can always compute an algebraic leading singularity for the corner integral of these sectors. This happens because one of the Baikov polynomials (in $D=4$) appears with a vanishing exponent in the integrand of the corner integral:
\begin{equation}
    \mathcal{I}_{\mathrm{4pt-kite}} \bigl|^{\mathrm{MC}}_{\varepsilon=0} \propto \int \frac{\mathrm{d} z_1 \mathrm{d} z_2}{\sqrt{\mathcal{E}_1}\mathcal{B}_1^0 \sqrt{\mathcal{B}_2}}\,.
    \label{eq:maxcut_4ptkite}
\end{equation}
Indeed, the sector in \cref{fig:PL1_110011100} is associated with a logarithmic geometry. We thus choose the basis
\begin{equation}
    \begin{split}
        \mathcal{I}_{31}^{\TOP{PL1}} & =\varepsilon^{3} \sqrt{\lambda_{\text{K}}\left(s,m^2,m^{2}\right)} \left(m^2-t\right) I_{110011200}^{(\TOP{PL1})}\,,\\
        \mathcal{I}_{32}^{\TOP{PL1}} & =\varepsilon^{3} m^2 \left(m^2-u\right) \left[I_{110012100}^{(\TOP{PL1})}+I_{110021100}^{(\TOP{PL1})}\right]\,,\\
        \mathcal{I}_{33}^{\TOP{PL1}} & =\varepsilon^{3} m^2 \left(m^2-t\right) I_{210011100}^{(\TOP{PL1})}\,,\\
        \mathcal{I}_{34}^{\TOP{PL1}} & =\varepsilon^{4} \left(m^2-u\right) I_{110011100}^{(\TOP{PL1})}\,,
    \end{split}
    \label{eq:basis_PL1_110011100}
\end{equation}
and the corresponding DEs are canonical. 

\subsubsection*{The nested square root}

As already observed in the literature for similar four-point kite sectors~\cite{FebresCordero:2023pww,Li:2026emp}, the sector in \cref{fig:PL1_011101100} is instead associated with a nested square root. As mentioned in \cref{sec:introduction}, it is not clear whether the corresponding canonical DEs can be expressed in terms of dlog one-forms. While this is the case for the sector studied in~\cite{Becchetti:2025oyb}, the one presented in~\cite{FebresCordero:2023pww} involves also differential forms associated with elliptic geometries. Here we limit ourselves to show how the DEs can be put in $\varepsilon$-factorised form, without discussing the analytic structure of the connection matrix. We start by choosing the basis
\begin{equation}
\begin{split}
\mathcal{I}_{21}^{\TOP{PL1}} & =\varepsilon^{4}\sqrt{\lambda_{\text{K}}\left(q^{2},u,m^{2}\right)}I_{011101100}^{(\TOP{PL1})}\,,\\
\mathcal{I}_{22}^{\TOP{PL1}} & =\varepsilon^{3}\sqrt{\lambda_{\text{K}}\left(s,m^{2},m^{2}\right)}\left(m^{2}-t\right)I_{011101200}^{(\TOP{PL1})}\,,\\
\mathcal{I}_{23}^{\TOP{PL1}}& =\varepsilon^{2} m^{4}\left(m^{2}-t\right)I_{012201100}^{(\TOP{PL1})}-\varepsilon^3 m^{2}\left(m^{2}+t -q^{2}\right)I_{012101100}^{(\TOP{PL1})}\\
&-\varepsilon^3 \frac{\left(m^{2}-t\right)\left(s-2m^{2}\right)}{2}I_{011101200}^{(\TOP{PL1})} + \varepsilon^3 m^2 \left( 2 m^2 -s \right) I_{011102100}^{(\TOP{PL1})}\\
&+\varepsilon^3 \left( 3 m^4 + q^2 s -m^2 (2 q^2 + 2 s +t) \right) I_{011201100}^{(\TOP{PL1})}\,,\\
\mathcal{I}_{24}^{\TOP{PL1}} & =\varepsilon^{3}m^{2}\sqrt{\lambda_{\text{K}}\left(q^{2},t,m^{2}\right)}I_{012101100}^{(\TOP{PL1})}\,,\\
\mathcal{I}_{25}^{\TOP{PL1}} & =\varepsilon^{3}m^{4} \left[I_{011102100}^{(\TOP{PL1})}+I_{011201100}^{(\TOP{PL1})}\right]\,,\\
\mathcal{I}_{26}^{\TOP{PL1}} & =\varepsilon^{3}m^{4} \left[I_{011102100}^{(\TOP{PL1})}-I_{011201100}^{(\TOP{PL1})}\right]\,.
\end{split}
\label{eq:basis_PL1_011101100}
\end{equation}
The first four integrals are chosen so that their normalisation can be computed from the DEs, by requiring that the corresponding diagonal entry is $\varepsilon$-factorised. We obtain the linear combination that defines $\mathcal{I}_{23}^{\TOP{PL1}}$ by imposing that the third row of the differential equation matrices is $\varepsilon$-factorised. For the four-by-four block associated with the first four MIs, it suffices to integrate the $\varepsilon^0$ term of the DEs to obtain the coefficients of $I_{012101100}^{(\TOP{PL1})}$ and $I_{011101200}^{(\TOP{PL1})}$ in the linear combination. For the last two terms, we resort instead to an ansatz for the form of the coefficients, whose form we determine by solving algebraic equations. 

The basis in \cref{eq:basis_PL1_011101100} satisfies DEs that are mostly $\varepsilon$-factorised, with the exception of the block coupling integrals $\mathcal{I}_{25}^{\TOP{PL1}}$ and $\mathcal{I}_{26}^{\TOP{PL1}}$. The presence of a coupled two-by-two block does not necessarily imply the presence of a nested square root, since there might be some transformation that decouples the integrals without introducing this kind of analytic structures. 
We remove this ambiguity by analysing the Picard-Fuchs operator~\cite{Muller-Stach:2012tgj,Adams:2017tga} associated with these two integrals. We work on the maximal cut, setting $\varepsilon=0$ and on a numerical slice
\begin{equation}
    \vec{x} = \vec{a} \lambda + \vec{b}, \quad \vec{a}, \vec{b} \in \mathbb{Z}^4 \,,
    \label{eq:slice_PF}
\end{equation}
where $\vec{a}, \vec{b}$ are chosen randomly, with the only constraint that no denominator of the DEs vanishes. This reduces the problem to one dimension. The Picard-Fuchs operator $L$ associated with the integral $J$ is then defined as
\begin{equation}
    L = \sum_{i=1}^N c_i(\lambda) \frac{\mathrm{d}^i}{\mathrm{d} \lambda^i}, \quad L \ J\vert_{\varepsilon=0}^{\mathrm{MC}} = 0\,,
    \label{eq:PF_operator}
\end{equation}
where the coefficients $c_i(\lambda)$ are some algebraic functions of $\lambda$. It is known~\cite{Adams:2017tga} that if the Picard-Fuchs operator factorises in linear factors, then it is possible to put the differential equation in $\varepsilon$-factorised form through a transformation that involves only rational functions and simple square roots. If instead the operator in \cref{eq:PF_operator} contains irreducible quadratic factors, the integral is part of a $2\times 2$ block of the DEs that can only be decoupled by introducing more complicated functions. This can be due to the presence of an elliptic curve~\cite{Adams:2017tga} or of a nested square root~\cite{Badger:2024fgb}.

Indeed, the Picard-Fuchs operators associated with the integrals $\mathcal{I}_{25}^{\TOP{PL1}}$ and $\mathcal{I}_{26}^{\TOP{PL1}}$ are quadratic and irreducible. In order to prove that the sector is associated with a nested square root and not with an elliptic curve, we construct explicitly a transformation that puts the DEs in $\varepsilon$-factorised form. Similarly to what was observed in~\cite{FebresCordero:2023pww}, rescaling $\mathcal{I}_{25}^{\TOP{PL1}}$ and $\mathcal{I}_{26}^{\TOP{PL1}}$ as
\begin{equation}
    \begin{split}
        \mathcal{I}_{25}^{\TOP{PL1}} &\longrightarrow \frac{\sqrt{q^2-4 m^2} \sqrt[4]{Q_6 (\vec{x})}}{m^4} \mathcal{I}_{25}^{\TOP{PL1}}\,,\\
        \mathcal{I}_{26}^{\TOP{PL1}} &\longrightarrow \frac{\sqrt{q^2} \sqrt[4]{Q_6 (\vec{x})}}{m^4} \mathcal{I}_{26}^{\TOP{PL1}}\,,\\
    \end{split}
    \label{eq:PL1_011101100_basis_rescaling}
\end{equation}
leads to DEs that are $\varepsilon$-factorised on the diagonal. The polynomial $Q_6(\vec{x})$ is irreducible, has degree six, and appears in the denominator of the DEs. It is also found by \textsc{Sofia}~\cite{Caron-Huot:2024brh,Correia:2025wtb}, both as a rational singularity and as the argument of a square root. We refer to the ancillary files~\cite{zenodo} for its exact expression.

The presence of the quartic root in \cref{eq:PL1_011101100_basis_rescaling} hints to the fact that the two integrals are indeed associated with a nested square root, where the internal square root has the argument $Q_6(\vec{x})$. We thus look for a matrix $T_{\mathrm{NR}}$ similar to that of~\cite{FebresCordero:2023pww} such that the transformation
\begin{equation}
    \begin{pmatrix}
        \mathcal{I}_{25}^{\TOP{PL1}}\\
        \mathcal{I}_{26}^{\TOP{PL1}}
    \end{pmatrix} \longrightarrow \begin{pmatrix}
        \tilde{\mathcal{I}}_{25}^{\TOP{PL1}}\\
        \tilde{\mathcal{I}}_{26}^{\TOP{PL1}}
    \end{pmatrix} =  T_{\mathrm{NR}} \cdot \begin{pmatrix}
        \mathcal{I}_{25}^{\TOP{PL1}}\\
        \mathcal{I}_{26}^{\TOP{PL1}}
    \end{pmatrix}\,,
    \label{eq:nested_root_rotation}
\end{equation}
makes the DEs $\varepsilon$-factorised for this sector. From the above discussion, we expect the transformation to involve nested square roots $\sqrt{N_\pm}$ with the general form
\begin{equation}
    N_\pm (\vec{x})= Q_a(\vec{x}) \pm Q_b(\vec{x}) \sqrt{Q_6(\vec{x})}\,,
    \label{eq:nested_root_general}
\end{equation}
which are related to each other by the transformation $\sqrt{Q_6(\vec{x})} \to -\sqrt{Q_6(\vec{x})}$. In the language of \cite{Becchetti:2025oyb}, the nested square roots $\sqrt{N_\pm}$ form a \textit{duplet} under this sign flip. We require that the canonical integrals $\tilde{\mathcal{I}}_{25}^{\TOP{PL1}}$ and $\tilde{\mathcal{I}}_{26}^{\TOP{PL1}}$ to have the same behaviour under this transformation, imposing some constraints on the structure of $T_{\mathrm{NR}}$. The structure of the $\varepsilon^0$-term of the DEs, together with dimensional arguments, further constrains the form of $T_{\mathrm{NR}}$ to
\begin{equation}
    T_{\mathrm{NR}} =\frac{1}{2 m^4} \begin{pmatrix}
        \frac{\sqrt{\lambda_{\text{K}}\left(q^{2},m^2,m^{2}\right)}}{q^2} \sqrt{N_-} & \sqrt{N_+}\\
        \frac{\sqrt{\lambda_{\text{K}}\left(q^{2},m^2,m^{2}\right)}}{q^2} \sqrt{N_+} & \sqrt{N_-}
    \end{pmatrix}\,.
    \label{eq:nested_root_transformation_general}
\end{equation}

By dimensional arguments, $Q_b$ is a polynomial of odd degree, while $\deg(Q_a)=\deg(Q_b)+3$. It suffices to fix their coefficients to fully determine the transformation in \cref{eq:nested_root_transformation_general}. By requiring that the diagonal entries of the DEs are $\varepsilon$-factorised, we determine that
\begin{equation}
    Q_b(\vec{x}) = q^2\,.
    \label{eq:QB}
\end{equation}
When considering the off-diagonal entries, it is useful to think about the different square roots that appear in the DEs after applying the transformation in \cref{eq:nested_root_transformation_general}. These are $\sqrt{\lambda_{\text{K}}\left(q^{2},m^2,m^{2}\right)}, \sqrt{N_\pm}, \sqrt{Q_6}$ and their products $\sqrt{\lambda_{\text{K}}\left(q^{2},m^2,m^{2}\right)} \sqrt{N_\pm}$,\\
$\sqrt{\lambda_{\text{K}}\left(q^{2},m^2,m^{2}\right)} \sqrt{Q_6}$ and $\sqrt{N_+}\sqrt{N_-}$. If these roots are all independent, their coefficients in the $\varepsilon^0$-term all have to vanish individually. However, working under this assumption we are left with uncancelled terms proportional to $\sqrt{\lambda_{\text{K}}\left(q^{2},m^2,m^{2}\right)}$ and $\sqrt{N_+}\sqrt{N_-}$. We thus impose the additional constraint
\begin{equation}
\begin{split}
    \sqrt{N_+ (\vec{x})}\sqrt{N_- (\vec{x})} &= \sqrt{\lambda_{\text{K}}\left(q^{2},m^2,m^{2}\right)} Q_c(\vec{x})\,,\\
    Q_a^2(\vec{x}) -Q_b^2 (\vec{x}) Q_6 (\vec{x}) &= \lambda_{\text{K}}\left(q^{2},m^2,m^{2}\right)Q_c^2 (\vec{x})\,,
\end{split}
    \label{eq:Qc_definition}
\end{equation}
where $Q_c$ is a polynomial of degree three. Again, we determine its coefficients by imposing that the off-diagonal entries of the DEs are $\varepsilon$-factorised. This fixes
\begin{equation}
    Q_c(\vec{x}) = \frac{1}{2} q^2 \left( m^4 + 4 m^2 s -s^2 -2 m^2 t + t^2 \right),
    \label{eq:QC}
\end{equation}
which in turn fixes the form of $Q_a$ through \cref{eq:Qc_definition}. We then verify that the transformation in \cref{eq:nested_root_transformation_general} yields $\varepsilon$-factorised DEs for this sector, both on the maximal cut and beyond. Nevertheless, for our scope of integrating the DEs numerically, we prefer to avoid introducing nested square roots in the DEs, hence we content ourselves with the MIs in \cref{eq:basis_PL1_011101100}.

We remark that we could have arrived at \cref{eq:Qc_definition} also by symmetry arguments. In general, we expect that MIs have a definite parity, even or odd, under the flip of the sign of the square roots that appear in their normalisation. In the case of nested square roots, this picture was generalised in~\cite{Becchetti:2025oyb}: the MIs are even/odd under $\sqrt{N_\pm} \to -\sqrt{N_\pm}$, and they form a duplet under $\sqrt{Q_6(\vec{x})} \to -\sqrt{Q_6(\vec{x})}$. Clearly, while \cref{eq:nested_root_transformation_general} was constructed to account for the duplet structure, it does not yield MIs with a definite behaviour under the flip of the sign of the external square roots, nor under $\sqrt{\lambda_{\text{K}}\left(q^{2},m^2,m^{2}\right)} \to -\sqrt{\lambda_{\text{K}}\left(q^{2},m^2,m^{2}\right)}$. This suggests that the two nested square roots and $\sqrt{\lambda_{\text{K}}\left(q^{2},m^2,m^{2}\right)}$ are not multiplicatively independent, and in fact replacing the relation \cref{eq:Qc_definition} in \cref{eq:nested_root_transformation_general} yields MIs that have manifestly the correct transformation properties.

\subsubsection*{Elliptic geometries}

The four-point kite sectors, in general, are not constrained to be $\dlog$. For instance, one of the sectors appearing in~\cite{Adams:2018kez} exhibits an elliptic geometry. This cannot be seen from the integrand of the corner integral because of the vanishing exponent of the polynomial $\mathcal{B}_1$ in \cref{eq:maxcut_4ptkite}\,, but it becomes manifest in integrals involving dots (or auxiliary propagators). This is precisely the situation encountered in the sectors of \TOP{PL2} and \TOP{PL3}. 
Ideally, we would follow the same strategy as for the simpler elliptic sectors, analysing the maximal cut in the loop-by-loop Baikov representation and identifying the integral associated with the differential of the first kind. However, integrals involving dots generally contain double poles, and thus vanish when trying to localise the integral on the maximal cut. 

To tackle the elliptic sectors in \cref{fig:4pt_kite}\,, we thus found and employed an alternative strategy. Inspired by~\cite{Dlapa:2021qsl}\,, where it was shown that one can exploit reducible super-sectors to find $\dlog$ integrals, we study the six-propagators sectors of \TOP{PL2} and \TOP{PL3} that reduce to the four-point kite and box-bubble sectors. The simplest case is that of sector $\{ 1,1,0,1,1,1,1,0,0 \}$ of \TOP{PL3}\,, whose corner integral is associated with an elliptic curve, isomorphic to that in \cref{eq:maxcut_PL1_011111100}. Crucially, the loop-by-loop Baikov representation obtained by integrating out first the $k_1$-loop involves only one ISP, and thus we immediately identify three MIs for this sector, analogously to what we did for the ones discussed in \cref{sec:simple_sectors_PL1,sec:simple_sectors}. If we now release the cut on the fourth propagator, we verify that this sector reduces to that in \cref{fig:PL3_110011100}. 

As we mentioned at the beginning of this section, the corner integral of the four-point kite sector admits an algebraic leading singularity and, in this case, it is linearly independent from the MIs coming from the reducible sector. We can thus choose the four MIs for the sector in \cref{fig:PL3_110011100} to be
\begin{equation}
    \begin{split}
        \mathcal{I}_{35}^{\TOP{PL3}} & =\varepsilon^{4} \left(u-m^2\right) I_{110011100}^{(\TOP{PL3})}\,,\\
        \mathcal{I}_{36}^{\TOP{PL3}} & =\varepsilon^{4} m^2 \left(q^2-s\right) I_{110111100}^{(\TOP{PL3})}\,,\\
        \mathcal{I}_{37}^{\TOP{PL3}} & =\varepsilon^{4} m^4 \partial_t \left( \left(q^2-s\right) I_{110111100}^{(\TOP{PL3})} \right)\,,\\
        \mathcal{I}_{38}^{\TOP{PL3}} & =\varepsilon^{4} \left(q^2-s\right) \left[I_{11011110-1}^{(\TOP{PL3})}- I_{110011100}^{(\TOP{PL3})} \right]\,.
    \end{split}
    \label{eq:basis_PL3_110011100}
\end{equation}
On the maximal cut, this set of MIs satisfies DEs of the form
\begin{equation}
   \mathrm{d} \begin{pmatrix}
       \mathcal{I}^{\TOP{PL3}}_{35}\\
       \mathcal{I}^{\TOP{PL3}}_{36}\\
       \mathcal{I}^{\TOP{PL3}}_{37}\\
       \mathcal{I}^{\TOP{PL3}}_{38}
   \end{pmatrix} = \begin{pmatrix}
        * \varepsilon & * \varepsilon & 0 & * \varepsilon\\
        * \varepsilon & * + * \varepsilon & * & * \varepsilon\\
        * \varepsilon + * \varepsilon^2 & * + * \varepsilon+* \varepsilon^2 & *+* \varepsilon & * \varepsilon + * \varepsilon^2\\
        * \varepsilon & * + * \varepsilon & * & * \varepsilon
    \end{pmatrix} \cdot \begin{pmatrix}
       \mathcal{I}^{\TOP{PL3}}_{35}\\
       \mathcal{I}^{\TOP{PL3}}_{36}\\
       \mathcal{I}^{\TOP{PL3}}_{37}\\
       \mathcal{I}^{\TOP{PL3}}_{38}
   \end{pmatrix}.
   \label{eq:PL3_110011100_maxcut_DEs}
\end{equation}
An interesting observation is that the corner integral indeed behaves like a $\dlog$ integral. By this we mean not only that the corresponding entries of the DEs are $\varepsilon$-factorised, but that they can actually be expressed in terms of logarithmic one-forms, with the exception of the couplings to integrals associated with elliptic differentials of the first kind.

The sector in \cref{fig:PL3_011101100} can be treated analogously, starting from its reducible elliptic super-sector $\{ 0,1,1,1,1,1,1,0,0\}$. Since the sector has 7 MIs, apart from the corner integrals and the three integrals coming from the super-sector we need to choose other three. Employing a loop-by-loop approach, we can construct an additional MI as a product of a triangle in $D=4$ and a bubble in $D=2$. Finally, we choose two integrals with a dot to close the system, such that the corresponding entries of the DEs are at worst linear in $\varepsilon$.

The basis for the sector is then
\begin{equation}
    \begin{split}
        \mathcal{I}_{28}^{\TOP{PL3}} & =\varepsilon^{4} \sqrt{\lambda_{\text{K}}\left(q^{2},u,m^{2}\right)} I_{011101100}^{(\TOP{PL3})}\,,\\
        \mathcal{I}_{29}^{\TOP{PL3}} & =\varepsilon^{3} m^4 \left[ I_{011201100}^{(\TOP{PL3})}+I_{011102100}^{(\TOP{PL3})} \right]\,,\\
        \mathcal{I}_{30}^{\TOP{PL3}} & =\varepsilon^{3} m^2 \left[I_{-121101100}^{(\TOP{PL3})}+I_{-111101200}^{(\TOP{PL3})}-I_{0211011-10}^{(\TOP{PL3})}-I_{0111012-10}^{(\TOP{PL3})} \right] + \dots\,,\\
        \mathcal{I}_{31}^{\TOP{PL3}} & =\varepsilon^{3} m^4 I_{011201100}^{(\TOP{PL3})}\,,\\
        \mathcal{I}_{32}^{\TOP{PL3}} & =\varepsilon^{4} m^2 \left(q^2-s\right) I_{011111100}^{(\TOP{PL3})}\,,\\
        \mathcal{I}_{33}^{\TOP{PL3}} & =\varepsilon^{4} m^4 \partial_t \left( \left(q^2-s\right) I_{011111100}^{(\TOP{PL3})} \right)\,,\\
        \mathcal{I}_{34}^{\TOP{PL3}} & =\varepsilon^{4} \left(q^2-s\right) \left[I_{01011110-1}^{(\TOP{PL3})}-I_{011101100}^{(\TOP{PL3})} \right]\,,
    \end{split}
    \label{eq:basis_PL3_011101100}
\end{equation}
where the ellipsis refers to contributions from $\mathcal{I}_{28}^{\TOP{PL3}}$ and $\mathcal{I}_{34}^{\TOP{PL3}}$, which ensure that the differential equation is $\varepsilon$-factorised in the entries coupling $\mathcal{I}_{30}^{\TOP{PL3}}$ to these integrals, as shown below; their explicit form is omitted for brevity.

On the maximal cut, the DEs for the basis in \cref{eq:basis_PL3_011101100} take the form
\begin{equation}
   \mathrm{d} \begin{pmatrix}
       \mathcal{I}^{\TOP{PL3}}_{28}\\
       \mathcal{I}^{\TOP{PL3}}_{29}\\
       \mathcal{I}^{\TOP{PL3}}_{30}\\
       \mathcal{I}^{\TOP{PL3}}_{31}\\
       \mathcal{I}^{\TOP{PL3}}_{32}\\
       \mathcal{I}^{\TOP{PL3}}_{33}\\
       \mathcal{I}^{\TOP{PL3}}_{34}
   \end{pmatrix} = \begin{pmatrix}
        * \varepsilon & * \varepsilon & 0 & * \varepsilon & * \varepsilon & 0 & * \varepsilon\\
        * \varepsilon & * + * \varepsilon & * \varepsilon & * + * \varepsilon & * + * \varepsilon & * & * \varepsilon\\
        * \varepsilon & * + * \varepsilon & * \varepsilon & * + * \varepsilon & * + * \varepsilon & * & * \varepsilon\\
        * \varepsilon & * + * \varepsilon & * \varepsilon & * + * \varepsilon & * + * \varepsilon & * & * \varepsilon\\
        0 & * \varepsilon & * \varepsilon & * \varepsilon & * + * \varepsilon & * & * \varepsilon\\
        * \varepsilon^2 & * \varepsilon + * \varepsilon^2 & * \varepsilon + * \varepsilon^2 & * \varepsilon + * \varepsilon^2 & * + * \varepsilon + * \varepsilon^2 & * + * \varepsilon & * \varepsilon + * \varepsilon^2\\
        * \varepsilon & * \varepsilon & * \varepsilon & * \varepsilon & * + * \varepsilon & * & * \varepsilon
    \end{pmatrix} \cdot \begin{pmatrix}
       \mathcal{I}^{\TOP{PL3}}_{28}\\
       \mathcal{I}^{\TOP{PL3}}_{29}\\
       \mathcal{I}^{\TOP{PL3}}_{30}\\
       \mathcal{I}^{\TOP{PL3}}_{31}\\
       \mathcal{I}^{\TOP{PL3}}_{32}\\
       \mathcal{I}^{\TOP{PL3}}_{33}\\
       \mathcal{I}^{\TOP{PL3}}_{34}
   \end{pmatrix}.
   \label{eq:PL3_011101100_maxcut_DEs}
\end{equation}
Analogously to the other four-point kite sector of \TOP{PL3}, the corner integral in this case also exhibits a $\dlog$ representation. Beyond the maximal cut, however, this sector couples to the box–bubble sector shown in \cref{fig:PL3_010111100}. This coupling arises because the sector $\{0,1,1,1,1,1,1,0,0\}$ is also a super-sector of the box–bubble topology.
As a result, both sectors are associated with the same elliptic curve, and the construction naturally leads to a system of three coupled sectors governed by a common elliptic geometry.

We treat the sector in \cref{fig:PL2_110011100} analogously. This sector has two reducible super-sectors, both of which are associated with an elliptic curve isomorphic to that in \cref{eq:elliptic_curve_010111100_PL2}. Since the sector has six MIs, apart from the corner we have to select other two. We select the basis
\begin{equation}
    \begin{split}
        \mathcal{I}_{23}^{\TOP{PL2}} & =\varepsilon^{4} \sqrt{\lambda_{\text{K}}\left(s,m^2,m^2\right)} I_{110011100}^{(\TOP{PL2})}\,,\\
        \mathcal{I}_{24}^{\TOP{PL2}} & =\varepsilon^{3} m^2 \sqrt{\lambda_{\text{K}}\left(q^2,u,m^2\right)} I_{120011100}^{(\TOP{PL2})}\,,\\
        \mathcal{I}_{25}^{\TOP{PL2}} & =\varepsilon^{2} m^2 I_{12-1012100}^{(\TOP{PL2})} + \dots\,,\\
        \mathcal{I}_{26}^{\TOP{PL2}} & =\varepsilon^{4} m^2 \left(u-m^2 \right) I_{111011100}^{(\TOP{PL2})}\,,\\
        \mathcal{I}_{27}^{\TOP{PL2}} & =\varepsilon^{4} m^4 \partial_{q^2} \left( \left(u-m^2 \right) I_{111011100}^{(\TOP{PL2})} \right)\,,\\
        \mathcal{I}_{28}^{\TOP{PL2}} & =\varepsilon^{4} \left(u-m^2 \right) \left[I_{1110111-10}^{(\TOP{PL2})}-I_{110011100}^{(\TOP{PL2})}\right]\,,
    \end{split}
    \label{eq:basis_PL2_110011100}
\end{equation}
where the ellipsis refers to contributions from $\mathcal{I}_{23}^{\TOP{PL2}}$, $\mathcal{I}_{24}^{\TOP{PL2}}$ and $\mathcal{I}_{28}^{\TOP{PL2}}$, as well as from sub-sectors.
On the maximal cut, this basis satisfies DEs of the form
\begin{equation}
   \mathrm{d} \begin{pmatrix}
       \mathcal{I}^{\TOP{PL2}}_{23}\\
       \mathcal{I}^{\TOP{PL2}}_{24}\\
       \mathcal{I}^{\TOP{PL2}}_{25}\\
       \mathcal{I}^{\TOP{PL2}}_{26}\\
       \mathcal{I}^{\TOP{PL2}}_{27}\\
       \mathcal{I}^{\TOP{PL2}}_{28}
   \end{pmatrix} = \begin{pmatrix}
        * \varepsilon & * \varepsilon & * \varepsilon & * \varepsilon & 0 & * \varepsilon\\
        * \varepsilon & * \varepsilon & * \varepsilon & * + * \varepsilon & * & * \varepsilon\\
        * \varepsilon & * \varepsilon & * \varepsilon & * + * \varepsilon & * & * \varepsilon\\
        0 & * \varepsilon & * \varepsilon & * + * \varepsilon & * & 0\\
        * \varepsilon^2 & * \varepsilon + * \varepsilon^2 & * \varepsilon + * \varepsilon^2 & * + * \varepsilon + * \varepsilon^2 & * + * \varepsilon & * \varepsilon^2\\
        * \varepsilon & * \varepsilon & * \varepsilon & * + * \varepsilon & * & * \varepsilon,
    \end{pmatrix} \cdot \begin{pmatrix}
       \mathcal{I}^{\TOP{PL2}}_{23}\\
       \mathcal{I}^{\TOP{PL2}}_{24}\\
       \mathcal{I}^{\TOP{PL2}}_{25}\\
       \mathcal{I}^{\TOP{PL2}}_{26}\\
       \mathcal{I}^{\TOP{PL2}}_{27}\\
       \mathcal{I}^{\TOP{PL2}}_{28}
   \end{pmatrix}.
   \label{eq:PL2_110011100_maxcut_DEs}
\end{equation}
We remark that, although the entries involving integrals $\mathcal{I}^{\TOP{PL2}}_{24}$, $\mathcal{I}^{\TOP{PL2}}_{25}$ and $\mathcal{I}^{\TOP{PL2}}_{28}$ are $\epsilon$-factorised, whenever they do not couple to $\mathcal{I}^{\TOP{PL2}}_{26}$ and $\mathcal{I}^{\TOP{PL2}}_{27}$, in general we could not express them in terms of logarithmic one-forms. On the contrary, the corner integral again exhibits a $\dlog$ behaviour.

Finally, we turn our attention to the sector in \cref{fig:PL2_011101100}. This sector does not have reducible super-sectors, hence we cannot apply the same strategy to identify its geometry. 
Instead, we analyse the Picard-Fuchs operator of integrals with dotted propagators, to investigate whether they contain irreducible quadratic factors. Indeed, the Picard-Fuchs operator of the integral $I_{011101200}^{(\TOP{PL2})}$ is quadratic and irreducible, hinting to the fact that it is associated with an elliptic curve. 
Since our goal is to obtain DEs in the form of \cref{eq:deq}, and not to find a canonical basis, we do not attempt to prove explicitly that this integral is associated with an elliptic geometry. Instead, we construct a basis under the assumption that it is elliptic, and we verify that the resulting DEs take the expected form. Assuming that the integral $I_{011101200}^{(\TOP{PL2})}$ is related to the differential of the first kind, we then take its derivative as second MI and we need an integral with an additional pole to complete the elliptic block with a third kind differential. It is natural to try to take an integral with an ISP. We verify that the integral $I_{01110120-1}^{(\TOP{PL2})}$ leads to DEs that are at worst linear in $\varepsilon$. Moreover, beyond the maximal cut this integral receives contributions from $I_{001101200}^{(\TOP{PL2})}$, which we can compute integrating out the DEs. This is an additional confirmation of the fact that this is the correct choice for the third MI.

Together with the corner integral, this gives us four MIs out of the five we need. The last MI can be constructed as a product of a four-dimensional triangle and a two-dimensional bubble ($\mathcal{I}^{\TOP{PL2}}_{30}$ in \cref{eq:basis_PL2_011101100} below), as we did for the sector in \cref{fig:PL3_011101100}. This leads us to the basis
\begin{equation}
    \begin{split}
        \mathcal{I}_{29}^{\TOP{PL2}} & =\varepsilon^{4} \left(q^2-s \right) I_{011101100}^{(\TOP{PL2})}\,,\\
        \mathcal{I}_{30}^{\TOP{PL2}} & =\varepsilon^{3} m^2 \sqrt{\lambda_{\text{K}}\left(q^2,m^2,m^2\right)} \left[ I_{01120110-1}^{(\TOP{PL2})}-I_{0112-11100}^{(\TOP{PL2})}\right]\,,\\
        \mathcal{I}_{31}^{\TOP{PL2}} & =\varepsilon^{3} m^2 \left(q^2-s \right) I_{011101200}^{(\TOP{PL2})}\,,\\
        \mathcal{I}_{32}^{\TOP{PL2}} & =\varepsilon^{3} m^4 \partial_t \left( \left(q^2-s \right) I_{011101200}^{(\TOP{PL2})} \right)\,,\\
        \mathcal{I}_{33}^{\TOP{PL2}} & =\varepsilon^{3} \left(q^2-s \right) \left[I_{01110120-1}^{(\TOP{PL2})}-2 I_{001101200}^{(\TOP{PL2})}\right]\,,
    \end{split}
    \label{eq:basis_PL2_011101100}
\end{equation}
which, on the maximal cut, satisfies equations of the form
\begin{equation}
   \mathrm{d} \begin{pmatrix}
       \mathcal{I}^{\TOP{PL2}}_{29}\\
       \mathcal{I}^{\TOP{PL2}}_{30}\\
       \mathcal{I}^{\TOP{PL2}}_{31}\\
       \mathcal{I}^{\TOP{PL2}}_{32}\\
       \mathcal{I}^{\TOP{PL2}}_{33}
   \end{pmatrix} = \begin{pmatrix}
        * \varepsilon & * \varepsilon & * \varepsilon & 0 & * \varepsilon\\
        0 & * \varepsilon & 0 & 0 & 0\\
        0 & 0 & * + * \varepsilon & * & * \varepsilon\\
        0 & 0 & * + * \varepsilon + * \varepsilon^2 & * + * \varepsilon & * \varepsilon + * \varepsilon^2\\
        0 & 0 & * + * \varepsilon & * & * \varepsilon
    \end{pmatrix} \cdot \begin{pmatrix}
       \mathcal{I}^{\TOP{PL2}}_{29}\\
       \mathcal{I}^{\TOP{PL2}}_{30}\\
       \mathcal{I}^{\TOP{PL2}}_{31}\\
       \mathcal{I}^{\TOP{PL2}}_{32}\\
       \mathcal{I}^{\TOP{PL2}}_{33}
   \end{pmatrix},
   \label{eq:PL2_011101100_maxcut_DEs}
\end{equation}
where both the first and the second integral behave as $\dlog$ integrals. In fact, the second integral decouples completely from the others on the maximal cut. Beyond the maximal cut, the same integral receives contributions from sub-sectors, which we determine by integrating the $\varepsilon^0$-term of the differential equation.

~

This concludes our discussion of the four-point kite integrals. We thus move to the only missing sectors of the planar families, the top sectors of the families in \cref{fig:PL2,fig:PL3}.

\subsection{The top sectors of \TOP{PL2} and \TOP{PL3}}
\label{sec:top_sectors}

The top sectors of \TOP{PL2} and \TOP{PL3}\,, depicted in \cref{fig:PL2,fig:PL3}\,, are both associated with an elliptic curve and each contains five MIs, making them more involved than the sectors discussed in \cref{sec:simple_sectors_PL1,sec:simple_sectors}. In the following, we discuss in detail how we obtained a basis for \TOP{PL2}; the same strategy applies to \TOP{PL3}.

Following the strategy used for the previously discussed sectors, we start by investigating the loop-by-loop Baikov representation of the corner integral. If we start from the $k_2$-loop, the only residual denominator is $D_8$, and the maximal cut of the integral is
\begin{equation}
    I_{111111100}^{(\TOP{PL2})} \bigl|^{\rm MC}_{\varepsilon=0} \propto \int \frac{\mathrm{d} z}{(u-m^2)^2\sqrt{\mathcal{P}_{4}(z)}}\,,
    \label{eq:maxcut_PL2_111111100_D8}
\end{equation}
where the elliptic curve $\mathcal{P}_4(z)$ turns out to be isomorphic to that of \cref{eq:elliptic_curve_010111100_PL2}. 
This immediately gives us three MIs for this sector, associated with elliptic differentials of the first, second, and third kind respectively:
\begin{equation}
\begin{split}
    \mathcal{I}^{\TOP{PL2}}_1 &= \varepsilon^4 m^2 (u-m^2)^2 I_{111111100}^{(\TOP{PL2})}\,,\\
    \mathcal{I}^{\TOP{PL2}}_2 &= \varepsilon^4 (u-m^2)^2 I_{1111111-10}^{(\TOP{PL2})}\,,\\
    \mathcal{I}^{\TOP{PL2}}_5 &= \varepsilon^4 m^4 \partial_{q^2} \bigg( (u-m^2)^2 I_{111111100}^{(\TOP{PL2})} \bigg).
\end{split}
\label{eq:basis_PL2_top_maxcut_D8}
\end{equation}
If instead we analyse the loop-by-loop Baikov representation starting from the $k_1$-loop, the residual denominator is $D_9$, and the integral is again proportional to an elliptic differential of the first kind
\begin{equation}
    I_{111111100}^{(\TOP{PL2})} \bigl|^{\rm MC}_{\varepsilon=0} \propto \int \frac{\mathrm{d} z}{(u-m^2) \sqrt{\lambda_{\text{K}}\left(q^2,u,m^{2}\right)} \sqrt{ \mathcal{P}_{4}(z)}}\,,
    \label{eq:maxcut_PL2_111111100_D9}
\end{equation}
where, unsurprisingly, the elliptic curve is isomorphic to those in \cref{eq:elliptic_curve_010111100_PL2,eq:maxcut_PL2_111111100_D8}. We identify the fourth MI as
\begin{equation}
    \mathcal{I}^{\TOP{PL2}}_3 = \varepsilon^4 (u-m^2) \sqrt{\lambda_{\text{K}}\left(q^2,u,m^{2}\right)} I_{11111110-1}^{(\TOP{PL2})}\,.
    \label{eq:basis_PL2_top_maxcut_D9}
\end{equation}
We find the missing MI by studying the global Baikov representation. We make the \emph{ansatz}:
\begin{equation}
    \mathcal{I}^{\TOP{PL2}}_4 = [c_1 +c_2 D_8 + c_3 D_4 + c_4 D_8 D_9] I_{111111100}^{(\TOP{PL2})}
    \label{eq:ansatz_top_PL2}
\end{equation}
where the square brackets indicate that the numerator has to be inserted under the integral sign and the coefficients $c_1, c_2, c_3$, and $c_4$ can depend on the kinematic invariants. On the maximal cut, at $\varepsilon=0$, we have
\begin{equation}
     \mathcal{I}^{\TOP{PL2}}_4 \bigl|^{\rm MC}_{\varepsilon=0} \propto \int \mathrm{d} D_8\, \mathrm{d} D_9 \frac{a_1 +a_2 D_8 + a_3 D_4 + a_4 D_8 D_9}{B(D_8,D_9; s, t, m^2,q^2)}\,,
    \label{eq:maxcut_PL2_111111100_full}
\end{equation}
where $B(D_8,D_9; s, t, m^2,q^2)$ is an algebraic function. 
We carry out the integration over $D_8$ in \cref{eq:maxcut_PL2_111111100_full}\,, and we arrive at the expression
\begin{equation}
   \mathcal{I}^{\TOP{PL2}}_4 \bigl|^{\rm MC}_{\varepsilon=0} \propto \int \mathrm{d} D_9 \left( \tilde{B}_1(D_9)\log(\alpha_1(D_8,D_9) + \tilde{B}_2(D_9)\tanh^{-1}(\alpha_2(D_8,D_9) \right)\,,
    \label{eq:maxcut_PL2_111111100_full_intD8}
\end{equation}
where $\alpha_1$, $\alpha_2$ and $\tilde{B}_2$ are algebraic functions, while $\tilde{B}_1$ is a rational function:
\begin{equation}
    \tilde{B}_1(D_9; s, t, m^2,q^2; c_2, c_4) = \frac{c_2 + c_4 D_9}{4 D_9 \left( m^4 + u (q^2-s-t+D_9) \right)}\,.
    \label{eq:B1tilde_PL2}
\end{equation}
It is clear that the first term term in the integrand of \cref{eq:maxcut_PL2_111111100_full_intD8} is, if not $\dlog$\,, at least free of any dependence on the elliptic curve. Meanwhile, the denominator of $\tilde{B}_2$ contains the square root of a degree four polynomial in $D_9$, a clear indication that this term is associated with an elliptic geometry.
Guided by the analysis of the four-point kite integrals of \cref{sec:fourpt_kite}\,, where we observed that the corner integral behaves like a $\dlog$ integral, we look for a choice of the coefficients of the \emph{ansatz} in \cref{eq:ansatz_top_PL2} such that $\tilde{B}_2 = 0$. A solution exists, and it fixes all the coefficients but one. Keeping $c_2$ as free parameter, \cref{eq:B1tilde_PL2} becomes
\begin{equation}
    \tilde{B}_1(D_9; s, t, m^2,q^2; c_2) = \frac{c_2}{4 (u-m^2)^2 D_9}\,,
    \label{eq:B1tilde_PL2_solc2}
\end{equation}
which, combined with \cref{eq:maxcut_PL2_111111100_full_intD8} and setting $c_2=4(u-m^2)^2$, gives an integral whose integrand is in $\dlog$-form with unit leading singularity.
This integral is independent of the ones in \cref{eq:basis_PL2_top_maxcut_D8,eq:basis_PL2_top_maxcut_D9}, and we can choose it as our last MI:
\begin{equation}
\begin{split}\mathcal{I}_{3}^{\TOP{PL2}}  =&\phantom{x}\varepsilon^{4}\left[2uI_{1111111-1-1}^{(\TOP{PL2})}+2\left(u-m^{2}\right)^{2}I_{1111111-10}^{(\TOP{PL2})}\right.\\
 & +\left(5m^{4}-(q^{2}-s-t)(s+t)+m^{2}(q^{2}-4(s+t))\right)I_{11111110-1}^{(\TOP{PL2})}\\
 & \left.+\left(u-m^{2}\right)\left(m^{4}+m^{2}(s-q^{2}-2t)+t(s+t-q^{2})\right)I_{111111100}^{(\TOP{PL2})}\right]\,.
\end{split}
\end{equation}
On the maximal cut, this set of MIs satisfies a differential equation of the form
\begin{equation}
   \mathrm{d} \begin{pmatrix}
       \mathcal{I}^{\TOP{PL2}}_1\\
       \mathcal{I}^{\TOP{PL2}}_2\\
       \mathcal{I}^{\TOP{PL2}}_3\\
       \mathcal{I}^{\TOP{PL2}}_4\\
       \mathcal{I}^{\TOP{PL2}}_5
   \end{pmatrix} = \begin{pmatrix}
        *+*\varepsilon & * \varepsilon & * \varepsilon & * \varepsilon & *\\
        * + * \varepsilon & * \varepsilon & * \varepsilon & * \varepsilon & *\\
        * + * \varepsilon & * \varepsilon & * \varepsilon & * \varepsilon & *\\
        0 & 0 & 0 & * \varepsilon & 0\\
        *+* \varepsilon+* \varepsilon^2 & * \varepsilon+* \varepsilon^2 & * \varepsilon+* \varepsilon^2 & * \varepsilon+* \varepsilon^2 & *+* \varepsilon
    \end{pmatrix} \cdot  \begin{pmatrix}
       \mathcal{I}^{\TOP{PL2}}_1\\
       \mathcal{I}^{\TOP{PL2}}_2\\
       \mathcal{I}^{\TOP{PL2}}_3\\
       \mathcal{I}^{\TOP{PL2}}_4\\
       \mathcal{I}^{\TOP{PL2}}_5
   \end{pmatrix}\,,
   \label{eq:PL2_topsector_maxcut_DEs}
\end{equation}
where we see that the fourth integral, the $\dlog$ MI, decouples from the other MIs. Beyond the maximal cut the DEs follow the same pattern, i.e.~they are at most quadratic in $\varepsilon$ in the entries coupling $\mathcal{I}^{\TOP{PL2}}_5$ to the sub-sectors and linear in $\varepsilon$ otherwise. Integrating the $\varepsilon^0$ term in the DEs we determine additional sub-sector contributions to $\mathcal{I}^{\TOP{PL2}}_4$, such that the entries coupling this integral to its $\dlog$ sub-sectors are $\varepsilon$-factorised.

The top sector of \TOP{PL3} is also associated with an elliptic curve, which is isomorphic to the one in \cref{eq:maxcut_PL1_011111100}. In order to select the MIs, 
we adopt the same strategy as for \TOP{PL2}: choosing the four integrals associated with the differentials of the first, second and third kind, and a $\dlog$ integral that we obtain from analysing the standard Baikov representation. The basis for this sector is then given by:
\begin{equation}
\begin{split}
    \mathcal{I}^{\TOP{PL3}}_1 &= \varepsilon^4 s m^2 (q^2 -s) I_{111111100}^{(\TOP{PL3})}\,,\\
    \mathcal{I}^{\TOP{PL3}}_2 &= \varepsilon^4 s (q^2 -s) I_{11111110-1}^{(\TOP{PL3})}\,,\\
    \mathcal{I}^{\TOP{PL3}}_3 &= \text{$\dlog$-integral} ,\\
    \mathcal{I}^{\TOP{PL3}}_4 &= \varepsilon^4 s (s-4 m^2) I_{1111111-10}^{(\TOP{PL3})}\,,\\
    \mathcal{I}^{\TOP{PL3}}_5 &= \varepsilon^4 m^4 \partial_{t} \bigg( s (q^2 -s) I_{111111100}^{(\TOP{PL3})} \bigg),
\end{split}
\label{eq:basis_PL3_top_elliptic}
\end{equation}
where we refer to the ancillary files~\cite{zenodo} for the explicit expression of $\mathcal{I}^{\TOP{PL3}}_3$. On the maximal cut, this basis satisfies DEs of the same form as \cref{eq:PL2_topsector_maxcut_DEs}. As we did for \TOP{PL2}\,, we refine it by adding sub-sector contributions to $\mathcal{I}^{\TOP{PL3}}_3$ that make its couplings to $\dlog$ sub-sectors $\varepsilon$-factorised.

\section{Representation of the differential equations and numerical evaluation}
\label{sec:benchmarks}
\begin{table}[t]
\centering
    \begin{tabular}{cc ccc cc cccc cccc cc}
        \toprule
    \multirow{2}{*}{Family} &&\multicolumn{3}{c}{\# Letters}  && \multicolumn{3}{c}{\# One-forms} \\
        \cmidrule(lr){2-5}  \cmidrule(lr){7-9}
        &&Even{}  &&Algebraic{}        &&Rational{}  &&Algebraic{} \\
        \midrule
		 \TOP{PL1} &&	17 && 15 && 48 && 34\\
		 \TOP{PL2} &&	25&& 15&& 138 && 127\\
		 \TOP{PL3} &&	20&& 13&& 123 && 96 \\
      \bottomrule
    \end{tabular}
\caption{
Number of letters and one-forms appearing in the differential equations for the integral families \TOP{PL1}, \TOP{PL2}, and \TOP{PL3}, classified according to their rational or algebraic structure.
}\label{tab:letters}
\end{table}

For the chosen basis of master integrals, we construct DEs of the form of \cref{eq:deq}, displaying a polynomial dependence on $\varepsilon$ and an algebraic dependence on the kinematic variables. We determine the letters of the kinematic alphabet $W_\alpha(\vec{x})$ with the aid of \textsc{BaikovLetter}~\cite{Jiang:2024eaj}. The alphabet for the three planar families includes 53 letters, of which 32 are rational and 21 are algebraic. For the non-logarithmic one-forms $\omega_\beta(\vec{x})$ we follow the strategy of~\cite{Badger:2024fgb,Becchetti:2025qlu}: we use the entries of the DEs as an \emph{ansatz} for the one-forms and we determine a minimal independent set. 
We summarise the number of differential forms of each kind appearing in the DEs of the planar families in \cref{tab:letters}. Since the non-logarithmic one-forms are defined starting from the DEs of the individual families, we refrain from constructing an independent basis across all three families, in order to keep their relation to the DEs manifest. Let us also stress that the entries associated exclusively with $\dlog$ integrals do not contain non-logarithmic one-forms $\omega_\beta(\vec{x})$. These only appear in sectors involving elliptic integrals or the nested square root structure of \TOP{PL1}.
The complete alphabet and set of one-forms for each integral family are provided in~\cite{zenodo}.

\subsection{Structure of DEs}

To numerically integrate the DEs associated with \TOP{PL1}, \TOP{PL2}, and \TOP{PL3}, we first expand the basis of master integrals as,
\begin{align}
\vec{\mathcal{I}}(\vec{x};\varepsilon)&=\sum_{w=0}^{w_{\mathrm{max}}}\varepsilon^w\,\vec{\mathcal{I}}^{(w)}(\vec{x})\,,
\label{eq:eps_exp}
\end{align}
and we insert this expansion in \cref{eq:deq}. Both sides of the equation are polynomial in $\varepsilon$, and thus the coefficient of each power of $\varepsilon$ yields a differential equation for the coefficients $\vec{\mathcal{I}}^{(w)}(\vec{x})$. This leads us to a linear system of coupled differential equations  for the vector $\vec{u} = \vec{\mathcal{I}}^{(0)}\cup\hdots\cup\vec{\mathcal{I}}^{(4)}$ of all functions appearing up to $\mathcal{O}(\epsilon^4)$. This system only depends on the kinematics, and we can solve it numerically.

A few comments regarding the upper limit $w_{\mathrm{max}}$ in the expansion of \cref{eq:eps_exp} are in order. For canonical DEs, it is conjectured that only terms up to $\mathcal{O}(\varepsilon^4)$ appear in two-loop finite remainders. Since the basis we use is not canonical, higher terms in the expansion in \cref{eq:eps_exp} might be needed in this case. However, preliminary studies of the amplitude for the process in \cref{eq:ampl}, in particular of the $n_F$ contribution corresponding to the gauge-invariant subset of diagrams containing a closed fermion loop, show that only terms up to $\mathcal{O}(\varepsilon^4)$ appear up to the finite part. We therefore truncate the expansion in \cref{eq:eps_exp} to $w_{\mathrm{max}}=4$.

We then parametrise the DEs along a path connecting initial (boundary) and final (target) point, resulting in a system that only depends on the path parameter $y$ and takes the form
\begin{align}
\frac{\mathrm{d}}{\mathrm{d}y}\vec{u}\left(y;\epsilon\right) & =\left[\sum_{i=1}^{M}c_{i}\left(y\right)\,A_{i}\right]\vec{u}\left(y;\epsilon\right)\,,
\label{eq:de-system}
\end{align}
where $M$ denotes the number of letters and one-forms appearing in the DE for a given integral family (see \cref{tab:letters}). For the $\dlog$ forms, $c_i(y)=\mathrm{d}\log a_i(y)/\mathrm{d}y$, while for the one-forms,
\begin{align}
c_{i}\left(y\right)=\sum_{x\in\left\{ s,t,m^{2},q^{2}\right\} }\frac{\mathrm{d}\omega_{i}\left(\vec{x}\right)}{\mathrm{d}x}\frac{\mathrm{d}x}{\mathrm{d}y}\,.
\end{align}
In every family the $w=0$ block is entirely trivial, as the corresponding functions are constants; the genuinely non-trivial numerical integration is for the four blocks involving functions of higher order.

\subsection{Path and deformation}
For the integration path connecting the initial and final points, we use a three-segment contour for all three families. The target point $(s_f,t_f,q_f^2)$ is reached from a boundary point $(s_i,t_i,q_i^2)$ by first lifting $t$ off the real axis, performing the bulk of the displacement with $t$ held off-axis, and finally returning $t$ to the real axis.
Throughout the calculation, we retain the full electron-mass dependence, fixing $m^2=(0.00051099895,\mathrm)^2$.
The imaginary part is kept negative throughout, in accordance with the Feynman $+\mathrm{i} 0$ prescription. We choose the deformation as:
\begin{align}
\delta\;=\;-\,\mathrm{i}\,\max\!\Bigl(10^{-3}\,\bigl|t\bigr|_{{\rm larger}},\;6\times10^{-8}\,|s|\Bigr),
\label{eq:delta}
\end{align}
where $|t|_{\rm larger}$ denotes the maximum of $|t_i|$ and $|t_f|$.

The same prescription is used for all three families, with the second term providing a lower bound proportional to $s$. This floor is necessary because the boundary points of all families carry the hierarchy $t/s\simeq0$, causing the deformation to become excessively small and bringing the integration path close to the branch point at $t=0$. We find that introducing the floor substantially reduces the computational cost while maintaining, or slightly improving, the numerical accuracy. The precise value of the floor is not critical within a reasonable range; however, making the deformation significantly smaller leads to a rapid increase in the integration time, while increasing it beyond the values in Eq. \cref{eq:delta} tends to degrade the numerical accuracy. We therefore adopt \cref{eq:delta} as a common prescription for all three integral families. In particular, reducing the deformation below the value in \cref{eq:delta} can increase the computational cost by more than an order of magnitude, with no corresponding gain in numerical accuracy.

\subsection{Boundary points}

In the first version of our implementation, we 
determine the boundary conditions
 by evaluating the complete basis of master integrals with \textsc{AMFlow}~\cite{Liu:2017jxz,Liu:2021wks,Liu:2022chg} at the phase-space point:
\begin{align}
\vec{x}_{0}=
\left\{ s_0,t_0,q^{2}_0,m^{2}_0\right\}
=
\left\{ 16\,,-\frac{708703}{136301126}\,,\frac{67727737}{149466910}\,,\frac{27238}{104312222829}\right\} \,.
\end{align}
We choose this boundary point to mimic the kinematic configurations relevant for radiative-return experiments. 
 In these low-energy processes, with typical centre-of-mass energies $\sqrt{s}\sim1-10\,\text{GeV}$, retaining the full electron-mass dependence generates a pronounced hierarchy of scales, $m_e^2/s\sim10^{-7}-10^{-8}$, which makes the numerical evaluation of the master integrals challenging. Therefore, the choice of the base point is crucial to make the numerical integrations of the DEs more stable. 
In particular, $\vec{x}_0$ corresponds to a forward, hard-photon configuration, characterised by nearly collinear photon emission and small $|t|$, with most of the available energy carried by the radiated photon. This region is especially relevant for radiative-return measurements, where the radiative cross section is largest.

\begin{figure}[t]
\centering
\includegraphics[scale=0.7]{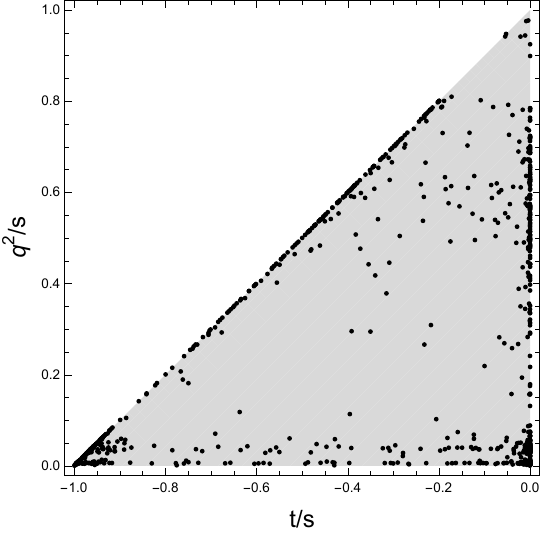}
\caption{
Distribution of 1000 phase-space points generated with {\sc Phokhara} for $s\in\{1.0404,\, 16,\, 100\}\,{\rm GeV}^2$. 
The points are shown in the $(t/s,q^2/s)$ plane within the physical region \cref{eq:physical_region}.}
\label{fig:physical_region}
\end{figure}

Moreover, $\vec{x}_0$ lies near a corner of the physical region \cref{eq:physical_region}, as illustrated in \cref{fig:physical_region}. Although this choice is well suited to the forward region, it introduces two limitations when integrating the DEs across the full phase space. First, reaching a target point near the opposite corner requires traversing almost the entire physical region, increasing the computational cost and reducing the numerical accuracy. Second, for \TOP{PL2}, whose natural kinematic variables are $(t,u,q^2)$ rather than $(s,t,q^2)$, the original boundary point is particularly far from the region where the numerical integration becomes most challenging.
To overcome these limitations, we extend our numerical implementation by replacing the single boundary point with a pool of boundary points distributed throughout the physical region. For each target, the integration starts from the nearest available boundary point, reducing the length of the integration path and improving numerical stability. In particular, for each integral family, the pool includes points close to the $t=0$ boundary ($t/s\sim10^{-6}$), covering the forward region; points near the opposite boundary ($|t/s|\gtrsim0.7$, corresponding to $|u/s|\lesssim0.3$); and intermediate configurations with $q^2/s$ of order a few tenths, covering the region between these two limits.

The pool is constructed from boundary points evaluated with \textsc{AMFlow}, supplemented by additional phase-space points distributed over the plane, whose values have already been independently validated using \texttt{DiffExp} (follow discussion in \cref{sec:performance}). This allows us to enlarge the pool without requiring additional \textsc{AMFlow} evaluations.

\subsection{Numerical implementation}

We solve the DEs numerically within the {\sc Julia} framework 
In this proof-of-concept implementation, all calculations are carried out in double precision ({\tt Float64} in {\sc Julia}).
We code \cref{eq:de-system} as a single in-place r.h.s. function that loops over the $M$
(matrix, integration kernels) pairs, evaluating each kernel (logarithm or one-form) at the current path
parameter and accumulating $c_i(x)\, A_i\, \vec{u}$ via sparse
matrix-vector products. The vector $\vec{u}$, its constant matrices,
and the integration kernels are the only topology-specific inputs, so the same driver
code applies unchanged to other integral families with a DE of the same
form.

Because the deformation~\eqref{eq:delta} keeps the r.h.s. analytic along the entire integration path, high-order explicit integrators are, in principle, well suited to this problem, as their higher order can reduce the number of steps required to achieve a given accuracy. We solve (in our first implementation) the DE using the non-stiff integrators \texttt{Tsit5()}~\cite{Tsitouras2011} and \texttt{Vern()}~\cite{Verner2010}, as implemented in \texttt{OrdinaryDiffEq.jl}~\cite{Rackauckas2020,Rackauckas2019}, using double precision with $\texttt{reltol}=\texttt{abstol}=10^{-12}$. To assess the dependence on the choice of integrator, we performed a dedicated benchmark for the \TOP{PL3} family, comparing \texttt{Tsit5()} with the higher-order Verner methods of orders 6--9 at matched tolerances. All methods give mutually consistent results for the phase-space points where they converge, but their computational costs differ substantially. In particular, \texttt{Vern9()} can require up to two to three orders of magnitude more computational time than \texttt{Tsit5()} for individual kinematic configurations, without providing a corresponding improvement in numerical accuracy. This behaviour is particularly pronounced close to physical thresholds, where the adaptive step-size control becomes more demanding. 

Based on this comparison, we adopt \texttt{Tsit5()} as the default integrator for all subsequent numerical calculations. 
Each target is integrated twice, first at a loose tolerance and then at a tighter one, and the two results are
compared. If they agree to the requested number of digits, which in our case is set to 2, the \emph{tighter}
of the two is kept and the looser one discarded. Otherwise the next tighter
tolerance is tried and compared against the one before it, and so on. The
looser run is therefore never the answer: it serves only as a yardstick
against which the tighter one is judged. Comparing two solutions in this way
needs no reference value, so the check is available at any phase-space point.

In addition to the fallback strategy, we apply two guards to every family: a single segment is limited to $20\,000$
integrator steps, and a whole target to $180\,$s. They exist because a few
\TOP{PL2} targets near $u=0$ converge at no tolerance we tried and would
otherwise hold the run indefinitely; such a target is reported as unconverged
rather than silently consuming the run. No target in the runs reported here
reached either limit.

\subsection{Performance \& Benchmarks}
\label{sec:performance}

After implementing DEs and boundary conditions in {\sc Julia}, we investigate the numerical performance throughout the physical region \cref{eq:physical_region}. 
To this end, we generate a sample of 1000 realistic phase-space points with the Monte Carlo event generator {\sc Phokhara}~\cite{Campanario:2019mjh}, by considering the centre-of-mass energies $s\in\{1.0404,\, 16,\, 100\}\,{\rm GeV}^2$. 
We show in \cref{fig:physical_region} the distribution of these phase-space points within the physical region 
\eqref{eq:physical_region}. 
For visualisation purposes, we display the rescaled kinematic variables $t/s$ and $q^2/s$, while neglecting the electron-mass dependence by taking $m^2/s\to0$.

\begin{table}[h]
\centering
\begin{tabular}{clccc}
\toprule
step & \texttt{reltol} / \texttt{abstol} & \TOP{PL1} & \TOP{PL2} & \TOP{PL3} \\
\midrule
1 & $3\times10^{-3}\,/\,3\times10^{-6}$ & ---                 & yardstick           & --- \\
2 & $1\times10^{-3}\,/\,1\times10^{-6}$ & yardstick           & $99.90\,/\,99.90$   & yardstick \\
3 & $1\times10^{-4}\,/\,1\times10^{-7}$ & $100.00\,/\,100.00$ & $0.10\,/\,100.00$   & $99.80\,/\,99.80$ \\
4 & $1\times10^{-6}\,/\,1\times10^{-9}$ & ---                 & ---                 & $0.10\,/\,99.90$ \\
5 & $1\times10^{-8}\,/\,1\times10^{-11}$ & ---                & ---                 & $0.10\,/\,100.00$ \\
6 & $1\times10^{-10}\,/\,1\times10^{-12}$ & ---               & ---                 & --- \\
\bottomrule
\end{tabular}
\caption{%
The tolerance sequence ordered from loose to tight. Each entry gives
the percentage of integrated targets accepted at that step and the cumulative percentage accepted by then. 
}
\label{tab:tol}
\end{table}

For \TOP{PL1} and \TOP{PL3}, we observe a typical target is integrated at relative/absolute tolerances of 
$10^{-3}/10^{-6}$ and then at $10^{-4}/10^{-7}$, and the second result is
returned; for \TOP{PL2} the two are $3\times10^{-3}/3\times10^{-6}$ and
$10^{-3}/10^{-6}$, and again it is the second that is returned.
The tolerances are drawn from the fixed sequence of \cref{tab:tol}. 
Notice that \TOP{PL2} is integrated one step more loosely than the other two because the
cost of its runs grows very steeply: the same target costs about $5\,$s at
$10^{-2}$, $15\,$s at $10^{-3}$ and $110\,$s at $10^{-4}$. 

In \cref{tab:tol}, we showcase that the numerical integration of realistic phase-space points is stable and internally consistent, with the results converging to the requested number of digits as the tolerance is tightened. This, however, does not by itself establish the correctness of the result. In particular, since the physical kinematics are approached by introducing a small negative imaginary part, we must verify that the chosen contour and the associated analytic continuation reproduce the correct physical result. An independent validation against a reliable reference calculation is therefore required to establish the absolute accuracy of the numerical integration.

To validate our implementation, we compare our results against independent evaluations from \textsc{AMFlow} at the physical points:
\begin{equation}
\begin{split}
\vec{x}_{1}= & \left\{ 16\,,-\frac{446241}{15096592942}\,,\frac{430523521}{700501982}\,,\frac{27238}{104312222829}\right\} \,,\\
\vec{x}_{2}= & \left\{ \frac{2601}{2500}\,,-\frac{392013}{62926006274}\,,\frac{86179753}{146223505}\,,\frac{27238}{104312222829}\right\} \,,\\
\vec{x}_{3}= & \left\{ 100\,,-\frac{477365}{9555073077}\,,\frac{62422497}{101504386}\,,\frac{27238}{104312222829}\right\} \,.
\end{split}
\end{equation}
This comparison yields at least eight significant figures of agreement across these phase-space points.
Although this level of agreement already provides a non-trivial validation, we further benchmark our implementation against {\sc DiffExp}~\cite{Hidding:2020ytt}, which implements the method of generalised power series expansions~\cite{Pozzorini:2005ff,Moriello:2019yhu}. We choose the parameters of {\sc DiffExp} such as to achieve at least eight significant figures.\footnote{More precisely, we used \texttt{ExpansionOrder$\to$35}, \texttt{WorkingPrecision$\to$250} and \texttt{ChopPrecision$\to$200}.} Since we do not optimise the parameters to make the numerical evaluation more efficient, the execution times are not directly comparable between the two frameworks, and we focus exclusively on numerical agreement. 

\begin{table}[t]
\centering
\begin{tabular}{lrrrr|rrrr}
\toprule
 & \multicolumn{4}{c|}{Integrated} & \multicolumn{4}{c}{Scored against \texttt{DiffExp}} \\
Family & Median & Mean & Max & Total & Median digits & $\geq2$ & $\geq4$ & $\geq6$ \\
\midrule
\TOP{PL1} & $0.18\,$s & $0.22\,$s & $0.9\,$s   & $3.7\,$min  & $5.27$ & $100\%$ & $98.6\%$ & $26.4\%$ \\
\TOP{PL2} & $7.36\,$s & $8.40\,$s & $142.3\,$s & $2.3\,$h    & $8.56$ & $100\%$ & $99.8\%$ & $86.6\%$ \\
\TOP{PL3} & $0.76\,$s & $1.43\,$s & $34.3\,$s  & $23.8\,$min & $6.53$ & $100\%$ & $97.8\%$ & $65.1\%$ \\
\bottomrule
\end{tabular}
\caption{%
Evaluation of the $1000$ phase-space points. 
The two halves have different denominators. \emph{Integrated} covers the targets actually propagated along a
contour. 
\emph{Scored} covers the comparison against \texttt{DiffExp}. 
``Median digits'' is the worst of the four non-trivial $\epsilon$ orders. No target reached the time budget.}
\label{tab:perf}
\end{table}

\begin{figure}[t]
\centering
\includegraphics[scale=0.34]{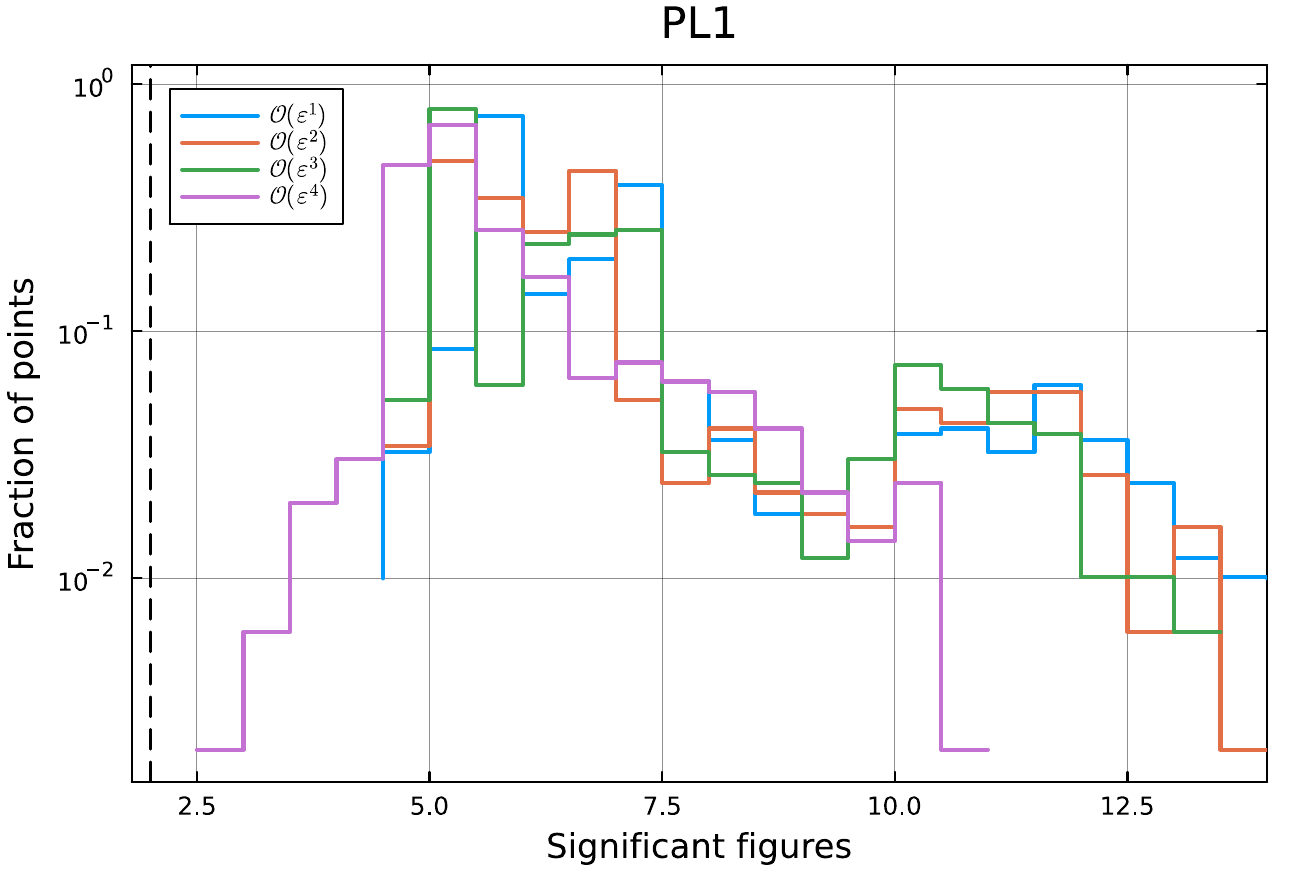}
\includegraphics[scale=0.34]{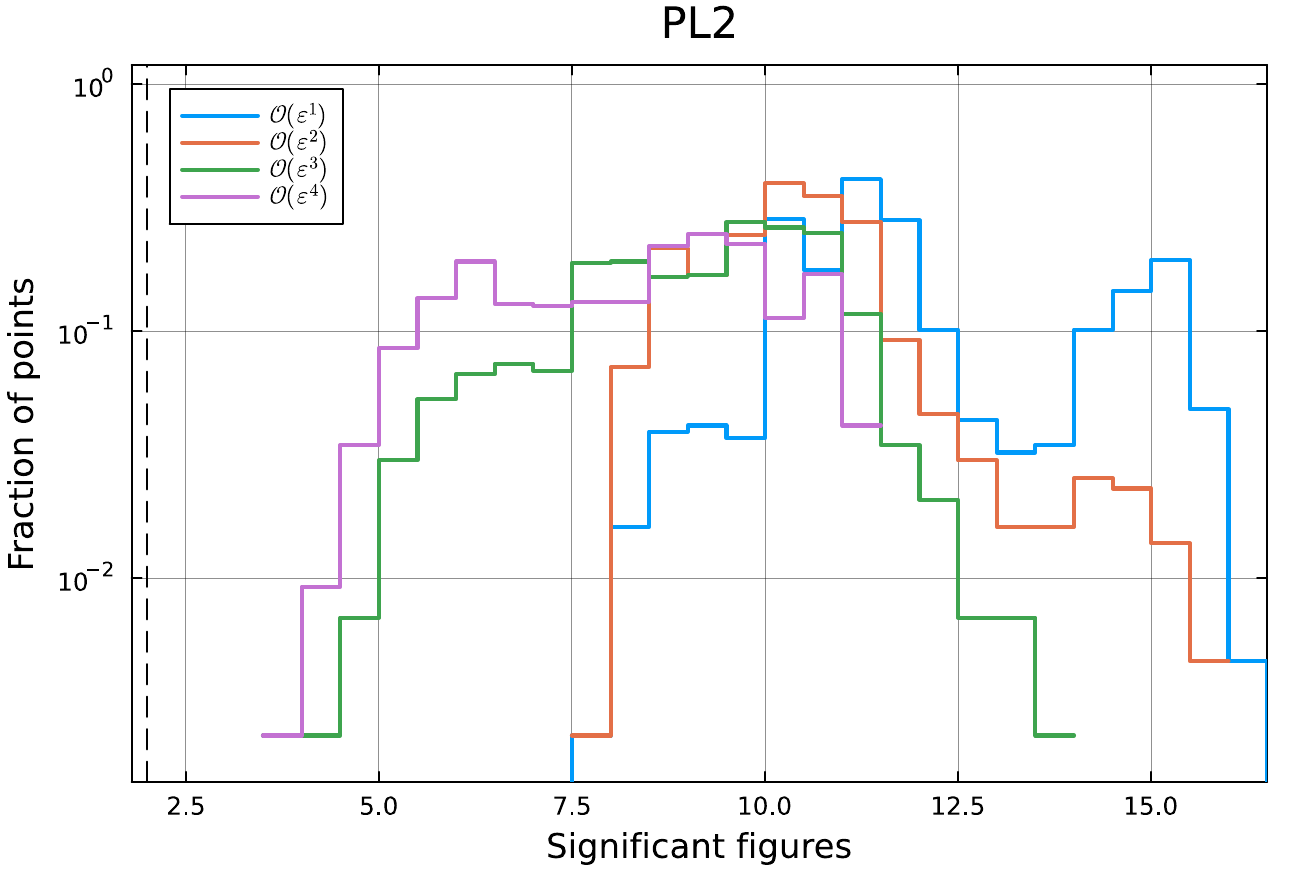}
\includegraphics[scale=0.34]{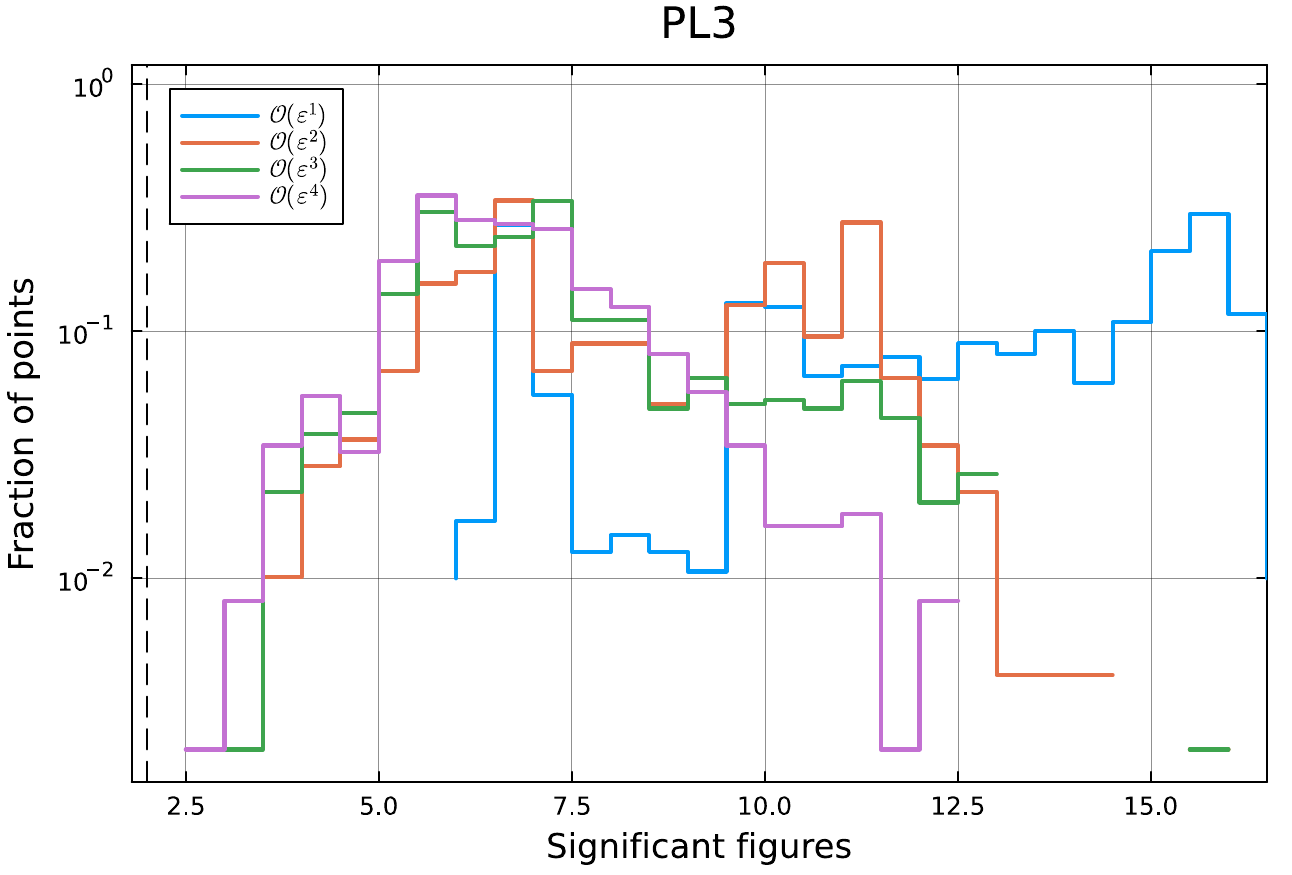}
\caption{Significant figures reached by the numerical integration as
compared with {\sc DiffExp} for all master integrals at
different orders in $\varepsilon$ over 1000 phase-space points.
The dashed line marks the two-digit target.}
\label{fig:significant_figures}
\end{figure}

In \cref{tab:perf} and \cref{fig:significant_figures}, we summarise the comparison family by family and report the worst agreement at each phase-space point. As already suggested in \cref{tab:letters}, the different integral families exhibit different levels of numerical complexity. In particular, we classify \TOP{PL1} as moderately difficult, while \TOP{PL2} and \TOP{PL3} are significantly more challenging. 
Per order, the medians of digits of agreement are $6.50/6.33/6.22/5.27$ for \TOP{PL1},
$11.48/10.38/9.43/8.58$ for \TOP{PL2} and $12.84/8.13/6.99/6.61$ for
\TOP{PL3} at $\mathcal{O}(\varepsilon^{1\hdots4})$. Every target in every family
reaches at least two significant figures, the minima being $2.88$, $3.83$ and
$2.97$. It is worth noting that \TOP{PL2}, integrated at the loosest
tolerance of the three and by far the most expensive family, is also the most
accurate: its difficulty is cost rather than precision, which is why
integrating it more tightly buys nothing.

In this proof-of-concept implementation, we observe that the numerical integration becomes particularly challenging for phase-space points close to singularities and physical thresholds, especially for functions that start appearing at $\mathcal{O}(\varepsilon^3)$. This behaviour is expected, as the higher terms in the $\varepsilon$-expansion generally contain more singularities. This correlates with the presence of the non-logarithmic one-forms. If needed, the performance could then be improved by constructing a more compact representation for these one-forms. 
Moreover, since the poles of the amplitude are determined by lower loops, it might be possible to ensure that non-logarithmic integrals appear only at $\mathcal{O}(\varepsilon^4)$. This could be achieved for example by choosing finite non-logarithmic integrals, as was done in~\cite{Badger:2024dxo,Becchetti:2026yxl}. If this were not possible, one might investigate whether performing a grading of functions~\cite{Chicherin:2020oor,Chicherin:2021dyp,Gehrmann:2024tds} could help achieving the same goal, e.g.~by studying the linear relations between coefficients $\mathcal{I}^{(w)}$ associated with the same elliptic differentials and making the cancellations between these functions explicit.
We postpone the investigation of these optimisations for future work.

\section{Conclusions}
\label{sec:conclusion}
Achieving NNLO precision for radiative return processes requires the evaluation of two-loop scattering amplitudes with multiple kinematic scales, including internal masses. Initial-state radiation contributions provide a natural starting point for this programme, reducing the problem to the study of genuine four-point two-loop amplitudes. In this work, we began the calculation of the corresponding planar integral families and adopted a systematic framework for the construction and fully numerical integration of their differential equations.

By building on canonical differential equations, we constructed systems that remain polynomial in the dimensional regulator $\varepsilon$, while isolating the sectors associated with elliptic geometries. Rather than introducing explicit elliptic functions into the differential equations, we exploited their relation to elliptic differentials of the first, second, and third kind at the level of the integrand of the master integrals. This step proved particularly challenging for the integrals belonging to four-point kite sectors, since in those cases the presence of an elliptic geometry is not manifest for the corner integral. 
To overcome this obstacle, we analysed the Picard-Fuchs operators corresponding to these integrals, and, when possible, we exploited the presence of reducible super-sectors.
Additionally, one of these four-point kite sectors was associated with a duplet of nested square roots. While we managed to construct a basis satisfying $\varepsilon$-factorised DEs for the corresponding integrals, for the purpose of the numerical evaluation we preferred to work with differential equations linear in $\varepsilon$, but involving only rational functions and simple square roots.

Beyond the analytic construction of the differential equations, we investigated their numerical integration in the physical production region relevant for low-energy radiative-return experiments. By implementing the differential equations in {\sc Julia} and employing non-stiff integrators, we demonstrated stable numerical evaluations throughout the physical region and systematically benchmarked them against independent calculations from \textsc{AMFlow} and \textsc{DiffExp}. We obtained evaluation times ranging from $\mathcal{O}(40-800\,\mathrm{ms})$.
Future optimisation of the Monte Carlo implementation may benefit from grids of boundary points, allowing shorter integration paths and improved numerical performance. The benchmarked phase-space points generated in this work constitute a natural starting point for developing such strategies.

The framework presented in this work constitutes a first step towards the construction of NNLO amplitudes for radiative return processes. The extension to the calculation of non-planar Feynman integrals and the assembly of the complete scattering amplitudes are currently under investigation and will be presented in forthcoming publications. 

\section*{Acknowledgments}

We are indebted to Thomas Dave and Pau Petit Ros\`as  for numerous checks performed at different stages of the project.
We are also grateful to Sara Maggio for enlightening discussions on the top sector of $\TOP{PL3}$.
We would like to thank Federico Coro, Dhimiter Canko and Simone Zoia for useful discussions,  
Antonela Matijasic, Dmytro Melnichenko and Stefan Weinzierl for collaboration on closely related projects, 
and Pau Petit Ros\`as, Tiziano Peraro and Simone Zoia for comments on the draft. 
M.P. thanks the University of Liverpool for hospitality while carrying out this project.
Parts of the numerical implementation and its performance study were carried out with the assistance of AI agents, specifically Claude Opus 5 via Claude Code.
This work was supported by the European Research Council (ERC) under the European Union’s Horizon Europe research and innovation program grant agreement 101040760, \textit{High-precision multi-leg Higgs and top physics with finite fields} (ERC Starting Grant FFHiggsTop), the Leverhulme Trust, LIP-2021-014, and through the Ramón y Cajal contract RYC2024-050272-I, funded by MICIU/AEI/10.13039/501100011033 and FSE+.

\appendix

\bibliographystyle{JHEP}
\bibliography{Biblio}

@article{Rackauckas2019,
	author = {Rackauckas, Christopher and Nie, Qing},
	doi = {10.1016/j.advengsoft.2019.03.009},
	journal = {Advances in Engineering Software},
	pages = {1--6},
	title = {Confederated Modular Differential Equation APIs for Accelerated Algorithm Development and Benchmarking},
	volume = {132},
	year = {2019}}

@article{Campanario:2019mjh,
	archiveprefix = {arXiv},
	author = {Campanario, Francisco and Czy{\.z}, Henryk and Gluza, Janusz and Jeli{\'n}ski, Tomasz and Rodrigo, Germ{\'a}n and Tracz, Szymon and Zhuridov, Dmitry},
	doi = {10.1103/PhysRevD.100.076004},
	eprint = {1903.10197},
	journal = {Phys. Rev. D},
	number = {7},
	pages = {076004},
	primaryclass = {hep-ph},
	title = {{Standard model radiative corrections in the pion form factor measurements do not explain the $a_\mu$ anomaly}},
	volume = {100},
	year = {2019}}

@misc{zenodo,
	author = {Pozzoli, Mattia and Torres Bobadilla, William J.},
	doi = {\href{https://doi.org/10.5281/zenodo.20826750}{10.5281/zenodo.20826750}},
	month = jul,
	publisher = {Zenodo},
	title = {{Ancillary files for ``First look at the evaluation of two-loop Feynman integrals for radiative return processes''}},
	url = {https://doi.org/10.5281/zenodo.20826750},
	year = 2026}

@article{Gehrmann:2024tds,
	archiveprefix = {arXiv},
	author = {Gehrmann, Thomas and Henn, Johannes and Jakub{\v{c}}{\'\i}k, Petr and Lim, Jungwon and Mella, Cesare Carlo and Syrrakos, Nikolaos and Tancredi, Lorenzo and Torres Bobadilla, William J.},
	doi = {10.1007/JHEP12(2024)215},
	eprint = {2410.19088},
	journal = {JHEP},
	pages = {215},
	primaryclass = {hep-th},
	reportnumber = {TUM-HEP 1529/24, MPP-2024-193, ZU-TH 52/24},
	title = {{Graded transcendental functions: an application to four-point amplitudes with one off-shell leg}},
	volume = {12},
	year = {2024}}

@article{Chicherin:2021dyp,
	archiveprefix = {arXiv},
	author = {Chicherin, Dmitry and Sotnikov, Vasily and Zoia, Simone},
	doi = {10.1007/JHEP01(2022)096},
	eprint = {2110.10111},
	journal = {JHEP},
	pages = {096},
	primaryclass = {hep-ph},
	reportnumber = {LAPTH-041/21, MPP-2021-182},
	title = {{Pentagon functions for one-mass planar scattering amplitudes}},
	volume = {01},
	year = {2022}}

@article{Rackauckas2020,
	author = {Christopher Rackauckas and Qing Nie},
	doi = {10.5334/jors.151},
	journal = {Journal of Open Research Software},
	number = {1},
	pages = {15},
	title = {DifferentialEquations.jl -- A Performant and Feature-Rich Ecosystem for Solving Differential Equations in Julia},
	volume = {5},
	year = {2017}}

@article{Verner2010,
	author = {J. H. Verner},
	doi = {10.1007/s11075-009-9290-3},
	journal = {Numerical Algorithms},
	number = {2--3},
	pages = {383--396},
	title = {Numerically optimal Runge--Kutta pairs with interpolants},
	volume = {53},
	year = {2010}}

@article{Tsitouras2011,
	author = {Ch. Tsitouras},
	doi = {10.1016/j.camwa.2011.06.002},
	journal = {Computers \& Mathematics with Applications},
	number = {2},
	pages = {770--775},
	title = {Runge--Kutta pairs of order 5(4) satisfying only the first column simplifying assumption},
	volume = {62},
	year = {2011}}

@article{Pak:2011xt,
	archiveprefix = {arXiv},
	author = {Pak, Alexey},
	doi = {10.1088/1742-6596/368/1/012049},
	editor = {Teodorescu, Liliana and Britton, David and Glover, Nigel and Heinrich, Gudrun and Lauret, Jerome and Naumann, Axel and Speer, Thomas and Teixeira-Dias, Pedro},
	eprint = {1111.0868},
	journal = {J. Phys. Conf. Ser.},
	pages = {012049},
	primaryclass = {hep-ph},
	title = {{The toolbox of modern multi-loop calculations: novel analytic and semi-analytic techniques}},
	volume = {368},
	year = {2012}}

@article{Laporta:2000dsw,
	archiveprefix = {arXiv},
	author = {Laporta, S.},
	doi = {10.1142/S0217751X00002159},
	eprint = {hep-ph/0102033},
	journal = {Int. J. Mod. Phys. A},
	pages = {5087--5159},
	title = {{High-precision calculation of multiloop Feynman integrals by difference equations}},
	volume = {15},
	year = {2000}}

@article{Liu:2017jxz,
	archiveprefix = {arXiv},
	author = {Liu, Xiao and Ma, Yan-Qing and Wang, Chen-Yu},
	doi = {10.1016/j.physletb.2018.02.026},
	eprint = {1711.09572},
	journal = {Phys. Lett. B},
	pages = {353--357},
	primaryclass = {hep-ph},
	title = {{A Systematic and Efficient Method to Compute Multi-loop Master Integrals}},
	volume = {779},
	year = {2018}}

@article{Liu:2021wks,
	archiveprefix = {arXiv},
	author = {Liu, Xiao and Ma, Yan-Qing},
	doi = {10.1103/PhysRevD.105.L051503},
	eprint = {2107.01864},
	journal = {Phys. Rev. D},
	number = {5},
	pages = {L051503},
	primaryclass = {hep-ph},
	title = {{Multiloop corrections for collider processes using auxiliary mass flow}},
	volume = {105},
	year = {2022}}

@inproceedings{Bourjaily:2022bwx,
	archiveprefix = {arXiv},
	author = {Bourjaily, Jacob L. and others},
	booktitle = {{Snowmass 2021}},
	eprint = {2203.07088},
	month = {3},
	primaryclass = {hep-ph},
	reportnumber = {BONN-TH-2022-05, UUITP-11/22, CERN-TH-2022-029, TUM-HEP-1391/22, HU-EP-22/08, MITP-22-022},
	title = {{Functions Beyond Multiple Polylogarithms for Precision Collider Physics}},
	year = {2022}}

@article{Liu:2022chg,
	archiveprefix = {arXiv},
	author = {Liu, Xiao and Ma, Yan-Qing},
	doi = {10.1016/j.cpc.2022.108565},
	eprint = {2201.11669},
	journal = {Comput. Phys. Commun.},
	pages = {108565},
	primaryclass = {hep-ph},
	title = {{AMFlow: A Mathematica package for Feynman integrals computation via auxiliary mass flow}},
	volume = {283},
	year = {2023}}

@article{Delto:2023kqv,
	archiveprefix = {arXiv},
	author = {Delto, Maximilian and Duhr, Claude and Tancredi, Lorenzo and Zhu, Yu Jiao},
	doi = {10.1103/PhysRevLett.132.231904},
	eprint = {2311.06385},
	journal = {Phys. Rev. Lett.},
	number = {23},
	pages = {231904},
	primaryclass = {hep-ph},
	reportnumber = {BONN-TH-2023-13, TUM-HEP-1479/23},
	title = {{Two-Loop QED Corrections to the Scattering of Four Massive Leptons}},
	volume = {132},
	year = {2024}}

@article{Becchetti:2025oyb,
	archiveprefix = {arXiv},
	author = {Becchetti, Matteo and Dlapa, Christoph and Zoia, Simone},
	doi = {10.1103/zt4w-c1jk},
	eprint = {2503.03603},
	journal = {Phys. Rev. D},
	number = {3},
	pages = {L031501},
	primaryclass = {hep-th},
	reportnumber = {DESY 25-029, ZU-TH 13/25},
	title = {{Canonical differential equations for the elliptic two-loop five-point integral family relevant to tt{\textasciimacron}+jet production at leading color}},
	volume = {112},
	year = {2025}}

@article{Becchetti:2025rrz,
	archiveprefix = {arXiv},
	author = {Becchetti, Matteo and Coro, Federico and Nega, Christoph and Tancredi, Lorenzo and Wagner, Fabian J.},
	doi = {10.1007/JHEP06(2025)033},
	eprint = {2502.00118},
	journal = {JHEP},
	pages = {033},
	primaryclass = {hep-ph},
	reportnumber = {TUM-HEP 1554/25},
	title = {{Analytic two-loop amplitudes for $ q\overline{q}\to \gamma \gamma $ and gg {\textrightarrow} {\ensuremath{\gamma}}{\ensuremath{\gamma}} mediated by a heavy-quark loop}},
	volume = {06},
	year = {2025}}

@article{Coro:2025vgn,
	archiveprefix = {arXiv},
	author = {Coro, Federico and Nega, Christoph and Tancredi, Lorenzo and Wagner, Fabian J.},
	doi = {10.1007/JHEP01(2026)090},
	eprint = {2509.15315},
	journal = {JHEP},
	pages = {090},
	primaryclass = {hep-ph},
	reportnumber = {TUM-HEP 1570/25},
	title = {{Analytic two-loop amplitudes for di-jet and {\ensuremath{\gamma}}+jet production mediated by a heavy-quark loop}},
	volume = {01},
	year = {2026}}

@article{Aliaj:2026iny,
	archiveprefix = {arXiv},
	author = {Aliaj, Rigers and Dian, Garbriele and Papathanasiou, Georgios},
	eprint = {2603.16743},
	month = {3},
	primaryclass = {hep-th},
	title = {{Novel cluster-algebraic letters for 5- and 6-point QCD processes}},
	year = {2026}}

@article{Boughezal:2007ny,
	archiveprefix = {arXiv},
	author = {Boughezal, R. and Czakon, M. and Schutzmeier, T.},
	doi = {10.1088/1126-6708/2007/09/072},
	eprint = {0707.3090},
	journal = {JHEP},
	pages = {072},
	primaryclass = {hep-ph},
	title = {{NNLO fermionic corrections to the charm quark mass dependent matrix elements in $\bar B \to X_s \gamma$}},
	volume = {09},
	year = {2007}}

@article{Czakon:2008zk,
	archiveprefix = {arXiv},
	author = {Czakon, M.},
	doi = {10.1016/j.physletb.2008.05.028},
	eprint = {0803.1400},
	journal = {Phys. Lett. B},
	pages = {307--314},
	primaryclass = {hep-ph},
	title = {{Tops from Light Quarks: Full Mass Dependence at Two-Loops in QCD}},
	volume = {664},
	year = {2008}}

@article{Mandal:2018cdj,
	archiveprefix = {arXiv},
	author = {Mandal, Manoj K. and Zhao, Xiaoran},
	doi = {10.1007/JHEP03(2019)190},
	eprint = {1812.03060},
	journal = {JHEP},
	pages = {190},
	primaryclass = {hep-ph},
	reportnumber = {CP3-18-71, MCNET-18-32},
	title = {{Evaluating multi-loop Feynman integrals numerically through differential equations}},
	volume = {03},
	year = {2019}}

@article{Czakon:2020vql,
	archiveprefix = {arXiv},
	author = {Czakon, Micha L. and Niggetiedt, Marco},
	doi = {10.1007/JHEP05(2020)149},
	eprint = {2001.03008},
	journal = {JHEP},
	pages = {149},
	primaryclass = {hep-ph},
	reportnumber = {P3H-20-01, TTK-20-01},
	title = {{Exact quark-mass dependence of the Higgs-gluon form factor at three loops in QCD}},
	volume = {05},
	year = {2020}}

@article{Czakon:2021yub,
	archiveprefix = {arXiv},
	author = {Czakon, M. and Harlander, R. V. and Klappert, J. and Niggetiedt, M.},
	doi = {10.1103/PhysRevLett.127.162002},
	eprint = {2105.04436},
	journal = {Phys. Rev. Lett.},
	note = {[Erratum: Phys.Rev.Lett. 131, 179901 (2023)]},
	number = {16},
	pages = {162002},
	primaryclass = {hep-ph},
	reportnumber = {TTK-21-17, P3H-21-031},
	title = {{Exact Top-Quark Mass Dependence in Hadronic Higgs Production}},
	volume = {127},
	year = {2021}}

@article{Haisch:2024nzv,
	archiveprefix = {arXiv},
	author = {Haisch, Ulrich and Niggetiedt, Marco},
	doi = {10.1007/JHEP10(2024)236},
	eprint = {2408.13186},
	journal = {JHEP},
	pages = {236},
	primaryclass = {hep-ph},
	reportnumber = {MPP-2024-197},
	title = {{Exact two-loop amplitudes for Higgs plus jet production with a cubic Higgs self-coupling}},
	volume = {10},
	year = {2024}}

@article{Badger:2025ljy,
	archiveprefix = {arXiv},
	author = {Badger, Simon and Becchetti, Matteo and Brancaccio, Colomba and Czakon, Micha{\l} and Hartanto, Heribertus Bayu and Poncelet, Rene and Zoia, Simone},
	doi = {10.1007/JHEP05(2026)044},
	eprint = {2511.11424},
	journal = {JHEP},
	pages = {044},
	primaryclass = {hep-ph},
	reportnumber = {P3H-25-092, TTK-25-38, ZU-TH 74/25, IFJPAN-IV-2025-23},
	title = {{Double virtual QCD corrections to $ t\overline{t} $+jet production at the LHC}},
	volume = {05},
	year = {2026}}

@article{Bargiela:2025vwl,
	archiveprefix = {arXiv},
	author = {Bargiela, Piotr and Frellesvig, Hjalte and Marzucca, Robin and Morales, Roger and Seefeld, Florian and Wilhelm, Matthias and Yang, Tong-Zhi},
	doi = {10.1007/JHEP05(2026)057},
	eprint = {2512.13794},
	journal = {JHEP},
	pages = {057},
	primaryclass = {hep-th},
	reportnumber = {LITP-25-15, ZU-TH 81/25},
	title = {{The spectrum of Feynman-integral geometries at two loops}},
	volume = {05},
	year = {2026}}

@article{Fadin:2023phc,
	archiveprefix = {arXiv},
	author = {Fadin, V. S. and Lee, R. N.},
	doi = {10.1007/JHEP11(2023)148},
	eprint = {2308.09479},
	journal = {JHEP},
	pages = {148},
	primaryclass = {hep-ph},
	title = {{Two-loop radiative corrections to $e^+e^-\to\gamma\gamma*$ cross section}},
	volume = {11},
	year = {2023}}

@article{Badger:2023xtl,
	archiveprefix = {arXiv},
	author = {Badger, Simon and Kry{\'s}, Jakub and Moodie, Ryan and Zoia, Simone},
	doi = {10.1007/JHEP11(2023)041},
	eprint = {2307.03098},
	journal = {JHEP},
	pages = {041},
	primaryclass = {hep-ph},
	title = {{Lepton-pair scattering with an off-shell and an on-shell photon at two loops in massless QED}},
	volume = {11},
	year = {2023}}

@article{Budassi:2026lmr,
	archiveprefix = {arXiv},
	author = {Budassi, Ettore and Carloni Calame, Carlo M. and Ghilardi, Marco and Gurgone, Andrea and Montagna, Guido and Moretti, Mauro and Nicrosini, Oreste and Piccinini, Fulvio and Ucci, Francesco P.},
	eprint = {2601.19530},
	month = {1},
	primaryclass = {hep-ph},
	title = {{Radiative return at NLOPS accuracy}},
	year = {2026}}

@article{CarloniCalame:2026hhy,
	archiveprefix = {arXiv},
	author = {Carloni Calame, Carlo M. and Ghilardi, Marco and Gurgone, Andrea and Montagna, Guido and Moretti, Mauro and Nicrosini, Oreste and Piccinini, Fulvio and Ucci, Francesco P.},
	eprint = {2603.28621},
	month = {3},
	primaryclass = {hep-ph},
	title = {{Structure-dependent radiative corrections to $e^+ e^- \to \pi^+ \pi^- \gamma$ in the GVMD approach}},
	year = {2026}}

@article{PetitRosas:2026iuq,
	archiveprefix = {arXiv},
	author = {Petit Ros{\`a}s, Pau and Shekhovtsova, Olga and Torres Bobadilla, William J.},
	eprint = {2603.13171},
	month = {3},
	primaryclass = {hep-ph},
	title = {{Radiative return meets GVMD}},
	year = {2026}}

@article{Aliberti:2024fpq,
	archiveprefix = {arXiv},
	author = {Aliberti, Riccardo and others},
	doi = {10.21468/SciPostPhysCommRep.9},
	eprint = {2410.22882},
	month = {10},
	primaryclass = {hep-ph},
	title = {{Radiative corrections and Monte Carlo tools for low-energy hadronic cross sections in $e^+ e^-$ collisions}},
	year = {2024}}

@article{Dave:2026pvq,
	archiveprefix = {arXiv},
	author = {Dave, Thomas and Paltrinieri, J{\'e}r{\'e}my and Petit Ros{\`a}s, Pau and Torres Bobadilla, William J.},
	eprint = {2604.16251},
	month = {4},
	primaryclass = {hep-ph},
	title = {{Tensor decomposition of $e^+e^-\to\pi^+\pi^-\gamma$ to higher orders in the dimensional regulator}},
	year = {2026}}

@article{Aliberti:2025beg,
	archiveprefix = {arXiv},
	author = {Aliberti, R. and others},
	doi = {10.1016/j.physrep.2025.08.002},
	eprint = {2505.21476},
	journal = {Phys. Rept.},
	pages = {1--158},
	primaryclass = {hep-ph},
	reportnumber = {CERN-TH-2025-101, FERMILAB-PUB-25-0344-T, INT-PUB-25-015, IPARCOS-UCM-25-029, KEK Preprint 2025-22, LTH 1403, MITP-25-037, UWThPh 2025-15, UWThPh 2025-15, ZU-TH 37/25, IPARCOS-UCM-25-029},
	title = {{The anomalous magnetic moment of the muon in the Standard Model: an update}},
	volume = {1143},
	year = {2025}}

@article{Gorges:2023zgv,
	archiveprefix = {arXiv},
	author = {G{\"o}rges, Lennard and Nega, Christoph and Tancredi, Lorenzo and Wagner, Fabian J.},
	doi = {10.1007/JHEP07(2023)206},
	eprint = {2305.14090},
	journal = {JHEP},
	pages = {206},
	primaryclass = {hep-th},
	title = {{On a procedure to derive {\ensuremath{\epsilon}}-factorised differential equations beyond polylogarithms}},
	volume = {07},
	year = {2023}}

@article{Forner:2026vby,
	archiveprefix = {arXiv},
	author = {Forner, Felix and Mella, Cesare Carlo and Nega, Christoph and Tancredi, Lorenzo and Wagner, Fabian J.},
	eprint = {2604.25270},
	month = {4},
	primaryclass = {hep-th},
	reportnumber = {OUTP-26-03P, TUM-HEP-1601/26},
	title = {{Integrand Analysis, Leading Singularities and Canonical Bases beyond Polylogarithms}},
	year = {2026}}

@article{Chaubey:2025adn,
	archiveprefix = {arXiv},
	author = {Chaubey, Ekta and Sotnikov, Vasily},
	doi = {10.1103/4fjc-lfnx},
	eprint = {2504.20897},
	journal = {Phys. Rev. Lett.},
	number = {10},
	pages = {101903},
	primaryclass = {hep-th},
	reportnumber = {BONN-TH-2025-18, ZU-TH 29/25},
	title = {{Elliptic Leading Singularities and Canonical Integrands}},
	volume = {135},
	year = {2025}}

@article{Adams:2018yfj,
	archiveprefix = {arXiv},
	author = {Adams, Luise and Weinzierl, Stefan},
	doi = {10.1016/j.physletb.2018.04.002},
	eprint = {1802.05020},
	journal = {Phys. Lett. B},
	pages = {270--278},
	primaryclass = {hep-ph},
	reportnumber = {MITP-18-011},
	title = {{The $\varepsilon$-form of the differential equations for Feynman integrals in the elliptic case}},
	volume = {781},
	year = {2018}}

@article{Duhr:2025lbz,
	archiveprefix = {arXiv},
	author = {Duhr, Claude and Maggio, Sara and Nega, Christoph and Sauer, Benjamin and Tancredi, Lorenzo and Wagner, Fabian J.},
	doi = {10.1007/JHEP06(2025)128},
	eprint = {2503.20655},
	journal = {JHEP},
	pages = {128},
	primaryclass = {hep-th},
	reportnumber = {BONN-TH-2025-11, TUM-HEP 1559/25, HU-EP-25/13-RTG},
	title = {{Aspects of canonical differential equations for Calabi-Yau geometries and beyond}},
	volume = {06},
	year = {2025}}

@article{Adams:2017tga,
	archiveprefix = {arXiv},
	author = {Adams, Luise and Chaubey, Ekta and Weinzierl, Stefan},
	doi = {10.1103/PhysRevLett.118.141602},
	eprint = {1702.04279},
	journal = {Phys. Rev. Lett.},
	number = {14},
	pages = {141602},
	primaryclass = {hep-ph},
	title = {{Simplifying Differential Equations for Multiscale Feynman Integrals beyond Multiple Polylogarithms}},
	volume = {118},
	year = {2017}}

@article{Frellesvig:2021hkr,
	archiveprefix = {arXiv},
	author = {Frellesvig, Hjalte},
	doi = {10.1007/JHEP03(2022)079},
	eprint = {2110.07968},
	journal = {JHEP},
	pages = {079},
	primaryclass = {hep-th},
	title = {{On epsilon factorized differential equations for elliptic Feynman integrals}},
	volume = {03},
	year = {2022}}

@article{Dlapa:2022wdu,
	archiveprefix = {arXiv},
	author = {Dlapa, Christoph and Henn, Johannes M. and Wagner, Fabian J.},
	doi = {10.1007/JHEP08(2023)120},
	eprint = {2211.16357},
	journal = {JHEP},
	pages = {120},
	primaryclass = {hep-ph},
	reportnumber = {DESY 22-189, MPP-2022-138, TUM-HEP-1427/22},
	title = {{An algorithmic approach to finding canonical differential equations for elliptic Feynman integrals}},
	volume = {08},
	year = {2023}}

@article{Pogel:2022ken,
	archiveprefix = {arXiv},
	author = {P{\"o}gel, Sebastian and Wang, Xing and Weinzierl, Stefan},
	doi = {10.1103/PhysRevLett.130.101601},
	eprint = {2211.04292},
	journal = {Phys. Rev. Lett.},
	number = {10},
	pages = {101601},
	primaryclass = {hep-th},
	reportnumber = {MITP/22-094, TUM-HEP-1431/22},
	title = {{Taming Calabi-Yau Feynman Integrals: The Four-Loop Equal-Mass Banana Integral}},
	volume = {130},
	year = {2023}}

@article{Pogel:2022vat,
	archiveprefix = {arXiv},
	author = {P{\"o}gel, Sebastian and Wang, Xing and Weinzierl, Stefan},
	doi = {10.1007/JHEP04(2023)117},
	eprint = {2212.08908},
	journal = {JHEP},
	pages = {117},
	primaryclass = {hep-th},
	title = {{Bananas of equal mass: any loop, any order in the dimensional regularisation parameter}},
	volume = {04},
	year = {2023}}

@article{Frellesvig:2023iwr,
	archiveprefix = {arXiv},
	author = {Frellesvig, Hjalte and Weinzierl, Stefan},
	doi = {10.21468/SciPostPhys.16.6.150},
	eprint = {2301.02264},
	journal = {SciPost Phys.},
	number = {6},
	pages = {150},
	primaryclass = {hep-th},
	reportnumber = {MITP/23-001},
	title = {{On $\varepsilon$-factorised bases and pure Feynman integrals}},
	volume = {16},
	year = {2024}}

@article{Driesse:2024feo,
	archiveprefix = {arXiv},
	author = {Driesse, Mathias and Jakobsen, Gustav Uhre and Klemm, Albrecht and Mogull, Gustav and Nega, Christoph and Plefka, Jan and Sauer, Benjamin and Usovitsch, Johann},
	doi = {10.1038/s41586-025-08984-2},
	eprint = {2411.11846},
	journal = {Nature},
	number = {8063},
	pages = {603--607},
	primaryclass = {hep-th},
	reportnumber = {HU-EP-24/32-RTG, QMUL-PH-24-26, BONN-TH-2024-15, TUM-HEP-1532/24},
	title = {{Emergence of Calabi{\textendash}Yau manifolds in high-precision black-hole scattering}},
	volume = {641},
	year = {2025}}

@article{Duhr:2024uid,
	archiveprefix = {arXiv},
	author = {Duhr, Claude and Porkert, Franziska and Stawinski, Sven F.},
	doi = {10.1007/JHEP02(2025)014},
	eprint = {2412.02300},
	journal = {JHEP},
	pages = {014},
	primaryclass = {hep-th},
	reportnumber = {BONN-TH-2024-17},
	title = {{Canonical differential equations beyond genus one}},
	volume = {02},
	year = {2025}}

@article{Chen:2025hzq,
	archiveprefix = {arXiv},
	author = {Chen, Jiaqi and Yang, Li Lin and Zhang, Yiyang},
	doi = {10.1007/JHEP04(2026)077},
	eprint = {2503.23720},
	journal = {JHEP},
	pages = {077},
	primaryclass = {hep-th},
	title = {{On an approach to canonicalizing elliptic Feynman integrals}},
	volume = {04},
	year = {2026}}

@article{e-collaboration:2025frv,
	archiveprefix = {arXiv},
	author = {Bree, Iris and others},
	collaboration = {{\ensuremath{\varepsilon}}-collaboration},
	doi = {10.1103/pyt8-d7rt},
	eprint = {2506.09124},
	journal = {Phys. Rev. Lett.},
	number = {24},
	pages = {241602},
	primaryclass = {hep-th},
	title = {{Geometric Bookkeeping Guide to Feynman Integral Reduction and {\ensuremath{\epsilon}}-Factorized Differential Equations}},
	volume = {136},
	year = {2026}}

@article{Bree:2025tug,
	archiveprefix = {arXiv},
	author = {Bree, Iris and others},
	collaboration = {{\ensuremath{\epsilon}}},
	doi = {10.1103/mjpn-61yv},
	eprint = {2511.15381},
	journal = {Phys. Rev. D},
	number = {11},
	pages = {116019},
	primaryclass = {hep-th},
	title = {{New algorithms for Feynman integral reduction and epsilon-factorized differential equations}},
	volume = {113},
	year = {2026}}

@article{Henn:2020lye,
	archiveprefix = {arXiv},
	author = {Henn, Johannes and Mistlberger, Bernhard and Smirnov, Vladimir A. and Wasser, Pascal},
	doi = {10.1007/JHEP04(2020)167},
	eprint = {2002.09492},
	journal = {JHEP},
	pages = {167},
	primaryclass = {hep-ph},
	reportnumber = {MPP-2020-13, MITP/20-006, MIT-CTP/5176},
	title = {{Constructing d-log integrands and computing master integrals for three-loop four-particle scattering}},
	volume = {04},
	year = {2020}}

@article{Meyer:2017joq,
	archiveprefix = {arXiv},
	author = {Meyer, Christoph},
	doi = {10.1016/j.cpc.2017.09.014},
	eprint = {1705.06252},
	journal = {Comput. Phys. Commun.},
	pages = {295--312},
	primaryclass = {hep-ph},
	reportnumber = {HU-EP-17-10},
	title = {{Algorithmic transformation of multi-loop master integrals to a canonical basis with CANONICA}},
	volume = {222},
	year = {2018}}

@article{Frellesvig:2017aai,
	archiveprefix = {arXiv},
	author = {Frellesvig, Hjalte and Papadopoulos, Costas G.},
	doi = {10.1007/JHEP04(2017)083},
	eprint = {1701.07356},
	journal = {JHEP},
	pages = {083},
	primaryclass = {hep-ph},
	title = {{Cuts of Feynman Integrals in Baikov representation}},
	volume = {04},
	year = {2017}}

@article{Baikov:1996iu,
	archiveprefix = {arXiv},
	author = {Baikov, P. A.},
	doi = {10.1016/S0168-9002(97)00126-5},
	editor = {Werlen, M. and Perret-Gallix, D.},
	eprint = {hep-ph/9611449},
	journal = {Nucl. Instrum. Meth. A},
	pages = {347--349},
	reportnumber = {INP-96-42-449},
	title = {{Explicit solutions of the multiloop integral recurrence relations and its application}},
	volume = {389},
	year = {1997}}

@article{Baikov:1996rk,
	archiveprefix = {arXiv},
	author = {Baikov, P. A.},
	doi = {10.1016/0370-2693(96)00835-0},
	eprint = {hep-ph/9603267},
	journal = {Phys. Lett. B},
	pages = {404--410},
	reportnumber = {INP-96-10-417},
	title = {{Explicit solutions of the three loop vacuum integral recurrence relations}},
	volume = {385},
	year = {1996}}

@book{Lang1987-zb,
	address = {New York, NY},
	author = {Lang, Serge},
	edition = 2,
	language = {en},
	month = may,
	publisher = {Springer},
	series = {Graduate texts in mathematics},
	title = {Elliptic functions},
	year = 1987}

@article{Argeri:2014qva,
	archiveprefix = {arXiv},
	author = {Argeri, Mario and Di Vita, Stefano and Mastrolia, Pierpaolo and Mirabella, Edoardo and Schlenk, Johannes and Schubert, Ulrich and Tancredi, Lorenzo},
	doi = {10.1007/JHEP03(2014)082},
	eprint = {1401.2979},
	journal = {JHEP},
	pages = {082},
	primaryclass = {hep-ph},
	title = {{Magnus and Dyson Series for Master Integrals}},
	volume = {03},
	year = {2014}}

@article{Dlapa:2021qsl,
	archiveprefix = {arXiv},
	author = {Dlapa, Christoph and Li, Xiaodi and Zhang, Yang},
	doi = {10.1007/JHEP07(2021)227},
	eprint = {2103.04638},
	journal = {JHEP},
	pages = {227},
	primaryclass = {hep-th},
	reportnumber = {MPP-2021-23, PCFT-21-10, USTC-ICTS-21-10},
	title = {{Leading singularities in Baikov representation and Feynman integrals with uniform transcendental weight}},
	volume = {07},
	year = {2021}}

@article{Flieger:2022xyq,
	archiveprefix = {arXiv},
	author = {Flieger, Wojciech and Torres Bobadilla, William J.},
	doi = {10.1140/epjp/s13360-024-05796-7},
	eprint = {2210.09872},
	journal = {Eur. Phys. J. Plus},
	number = {11},
	pages = {1022},
	primaryclass = {hep-th},
	reportnumber = {MPP-2022-129},
	title = {{Landau and leading singularities in arbitrary space-time dimensions}},
	volume = {139},
	year = {2024}}

@article{PetitRosas:2025xhm,
	archiveprefix = {arXiv},
	author = {Petit Ros{\`a}s, Pau and Torres Bobadilla, William J.},
	doi = {10.1007/JHEP09(2025)210},
	eprint = {2507.12548},
	journal = {JHEP},
	pages = {210},
	primaryclass = {hep-ph},
	title = {{Fast evaluation of Feynman integrals for Monte Carlo generators}},
	volume = {09},
	year = {2025}}

@article{Adams:2018bsn,
	archiveprefix = {arXiv},
	author = {Adams, Luise and Chaubey, Ekta and Weinzierl, Stefan},
	doi = {10.1103/PhysRevLett.121.142001},
	eprint = {1804.11144},
	journal = {Phys. Rev. Lett.},
	number = {14},
	pages = {142001},
	primaryclass = {hep-ph},
	title = {{Planar Double Box Integral for Top Pair Production with a Closed Top Loop to all orders in the Dimensional Regularization Parameter}},
	volume = {121},
	year = {2018}}

@article{Adams:2018kez,
	archiveprefix = {arXiv},
	author = {Adams, Luise and Chaubey, Ekta and Weinzierl, Stefan},
	doi = {10.1007/JHEP10(2018)206},
	eprint = {1806.04981},
	journal = {JHEP},
	pages = {206},
	primaryclass = {hep-ph},
	reportnumber = {MITP-18-045},
	title = {{Analytic results for the planar double box integral relevant to top-pair production with a closed top loop}},
	volume = {10},
	year = {2018}}

@article{Frellesvig:2024ymq,
	archiveprefix = {arXiv},
	author = {Frellesvig, Hjalte},
	doi = {10.1007/JHEP04(2025)111},
	eprint = {2412.01804},
	journal = {JHEP},
	pages = {111},
	primaryclass = {hep-th},
	title = {{The loop-by-loop Baikov representation {\textemdash} Strategies and implementation}},
	volume = {04},
	year = {2025}}

@article{Badger:2024fgb,
	archiveprefix = {arXiv},
	author = {Badger, Simon and Becchetti, Matteo and Giraudo, Nicol{\`o} and Zoia, Simone},
	doi = {10.1007/JHEP07(2024)073},
	eprint = {2404.12325},
	journal = {JHEP},
	pages = {073},
	primaryclass = {hep-ph},
	reportnumber = {CERN-TH-2024-048, ZU-TH 22/24},
	title = {{Two-loop integrals for $ t\overline{t} $+jet production at hadron colliders in the leading colour approximation}},
	volume = {07},
	year = {2024}}

@article{Becchetti:2025qlu,
	archiveprefix = {arXiv},
	author = {Becchetti, Matteo and Canko, Dhimiter and Chestnov, Vsevolod and Peraro, Tiziano and Pozzoli, Mattia and Zoia, Simone},
	doi = {10.1007/JHEP07(2025)001},
	eprint = {2504.13011},
	journal = {JHEP},
	pages = {001},
	primaryclass = {hep-ph},
	reportnumber = {ZU-TH 28/25},
	title = {{Two-loop Feynman integrals for leading colour $ t\overline{t}W $ production at hadron colliders}},
	volume = {07},
	year = {2025}}

@article{Jiang:2024eaj,
	archiveprefix = {arXiv},
	author = {Jiang, Xuhang and Liu, Jiahao and Xu, Xiaofeng and Yang, Li Lin},
	doi = {10.1016/j.physletb.2025.139443},
	eprint = {2401.07632},
	journal = {Phys. Lett. B},
	pages = {139443},
	primaryclass = {hep-ph},
	title = {{Symbol letters of Feynman integrals from Gram determinants}},
	volume = {864},
	year = {2025}}

@article{Peraro:2016wsq,
	archiveprefix = {arXiv},
	author = {Peraro, Tiziano},
	doi = {10.1007/JHEP12(2016)030},
	eprint = {1608.01902},
	journal = {JHEP},
	pages = {030},
	primaryclass = {hep-ph},
	reportnumber = {EDINBURGH-2016-14},
	title = {{Scattering amplitudes over finite fields and multivariate functional reconstruction}},
	volume = {12},
	year = {2016}}

@article{vonManteuffel:2014ixa,
	archiveprefix = {arXiv},
	author = {von Manteuffel, Andreas and Schabinger, Robert M.},
	doi = {10.1016/j.physletb.2015.03.029},
	eprint = {1406.4513},
	journal = {Phys. Lett. B},
	pages = {101--104},
	primaryclass = {hep-ph},
	reportnumber = {MITP-14-009},
	title = {{A novel approach to integration by parts reduction}},
	volume = {744},
	year = {2015}}

@article{Chetyrkin:1981qh,
	author = {Chetyrkin, K. G. and Tkachov, F. V.},
	doi = {10.1016/0550-3213(81)90199-1},
	journal = {Nucl. Phys. B},
	pages = {159--204},
	title = {{Integration by Parts: The Algorithm to Calculate beta Functions in 4 Loops}},
	volume = {192},
	year = {1981}}

@article{Tkachov:1981wb,
	author = {Tkachov, F. V.},
	doi = {10.1016/0370-2693(81)90288-4},
	journal = {Phys. Lett.},
	pages = {65-68},
	slaccitation = {%%CITATION = PHLTA,100B,65;%%},
	title = {{A Theorem on Analytical Calculability of Four Loop Renormalization Group Functions}},
	volume = {100B},
	year = {1981}}

@article{Barucchi:1973zm,
	author = {Barucchi, G. and Ponzano, G.},
	doi = {10.1063/1.1666327},
	journal = {J. Math. Phys.},
	pages = {396--401},
	title = {{Differential equations for one-loop generalized Feynman integrals}},
	volume = {14},
	year = {1973}}

@article{Kotikov:1991hm,
	author = {Kotikov, A. V.},
	doi = {10.1016/0370-2693(91)90834-D},
	journal = {Phys. Lett. B},
	pages = {314--322},
	title = {{Differential equations method: The Calculation of vertex type Feynman diagrams}},
	volume = {259},
	year = {1991}}

@article{Bern:1993kr,
	archiveprefix = {arXiv},
	author = {Bern, Zvi and Dixon, Lance J. and Kosower, David A.},
	doi = {10.1016/0550-3213(94)90398-0},
	eprint = {hep-ph/9306240},
	journal = {Nucl. Phys. B},
	pages = {751--816},
	reportnumber = {SLAC-PUB-5947, SACLAY-SPH-T-92-048, UCLA-92-043},
	title = {{Dimensionally regulated pentagon integrals}},
	volume = {412},
	year = {1994}}

@article{Gehrmann:1999as,
	archiveprefix = {arXiv},
	author = {Gehrmann, T. and Remiddi, E.},
	doi = {10.1016/S0550-3213(00)00223-6},
	eprint = {hep-ph/9912329},
	journal = {Nucl. Phys. B},
	pages = {485--518},
	reportnumber = {TTP-99-49},
	title = {{Differential equations for two loop four point functions}},
	volume = {580},
	year = {2000}}

@article{Kotikov:1990kg,
	author = {Kotikov, A. V.},
	doi = {10.1016/0370-2693(91)90413-K},
	journal = {Phys. Lett. B},
	pages = {158--164},
	reportnumber = {ITF-90-31E},
	title = {{Differential equations method: New technique for massive Feynman diagrams calculation}},
	volume = {254},
	year = {1991}}

@article{Henn:2013pwa,
	archiveprefix = {arXiv},
	author = {Henn, Johannes M.},
	doi = {10.1103/PhysRevLett.110.251601},
	eprint = {1304.1806},
	journal = {Phys. Rev. Lett.},
	pages = {251601},
	primaryclass = {hep-th},
	title = {{Multiloop integrals in dimensional regularization made simple}},
	volume = {110},
	year = {2013}}

@article{Peraro:2019svx,
	archiveprefix = {arXiv},
	author = {Peraro, Tiziano},
	doi = {10.1007/JHEP07(2019)031},
	eprint = {1905.08019},
	journal = {JHEP},
	pages = {031},
	primaryclass = {hep-ph},
	reportnumber = {ZU-TH 24/19},
	title = {{$\text{FiniteFlow}$: multivariate functional reconstruction using finite fields and dataflow graphs}},
	volume = {07},
	year = {2019}}

@article{Lee:2012cn,
	archiveprefix = {arXiv},
	author = {Lee, R. N.},
	eprint = {1212.2685},
	month = {12},
	primaryclass = {hep-ph},
	title = {{Presenting LiteRed: a tool for the Loop InTEgrals REDuction}},
	year = {2012}}

@article{Wu:2023upw,
	archiveprefix = {arXiv},
	author = {Wu, Zihao and Boehm, Janko and Ma, Rourou and Xu, Hefeng and Zhang, Yang},
	doi = {10.1016/j.cpc.2023.108999},
	eprint = {2305.08783},
	journal = {Comput. Phys. Commun.},
	pages = {108999},
	primaryclass = {hep-ph},
	reportnumber = {USTC-ICTS/PCFT-23-15},
	title = {{NeatIBP 1.0, a package generating small-size integration-by-parts relations for Feynman integrals}},
	volume = {295},
	year = {2024}}

@article{Lange:2025fba,
	archiveprefix = {arXiv},
	author = {Lange, Fabian and Usovitsch, Johann and Wu, Zihao},
	doi = {10.1016/j.cpc.2025.109999},
	eprint = {2505.20197},
	journal = {Comput. Phys. Commun.},
	pages = {109999},
	primaryclass = {hep-ph},
	reportnumber = {ZU-TH 39/25, HU-EP-25/17-RTG},
	title = {{Kira 3: integral reduction with efficient seeding and optimized equation selection}},
	volume = {322},
	year = {2026}}

@article{BaBar:2012bdw,
	archiveprefix = {arXiv},
	author = {Lees, J. P. and others},
	collaboration = {BaBar},
	doi = {10.1103/PhysRevD.86.032013},
	eprint = {1205.2228},
	journal = {Phys. Rev. D},
	pages = {032013},
	primaryclass = {hep-ex},
	reportnumber = {BABAR-PUB-12-003},
	title = {{Precise Measurement of the $e^+ e^- \to \pi^+\pi^- (\gamma)$ Cross Section with the Initial-State Radiation Method at BABAR}},
	volume = {86},
	year = {2012}}

@article{Belle:2007ebm,
	archiveprefix = {arXiv},
	author = {Mori, T. and others},
	collaboration = {Belle},
	doi = {10.1143/JPSJ.76.074102},
	eprint = {0704.3538},
	journal = {J. Phys. Soc. Jap.},
	pages = {074102},
	primaryclass = {hep-ex},
	title = {{High statistics measurement of the cross-sections of gamma gamma ---\ensuremath{>} pi+ pi- production}},
	volume = {76},
	year = {2007}}

@article{KLOE:2008fmq,
	archiveprefix = {arXiv},
	author = {Ambrosino, F. and others},
	collaboration = {KLOE},
	doi = {10.1016/j.physletb.2008.10.060},
	eprint = {0809.3950},
	journal = {Phys. Lett. B},
	pages = {285--291},
	primaryclass = {hep-ex},
	title = {{Measurement of $\sigma(e^+ e^- \to \pi^+ \pi^- \gamma(\gamma)$ and the dipion contribution to the muon anomaly with the KLOE detector}},
	volume = {670},
	year = {2009}}

@article{KLOE:2010qei,
	archiveprefix = {arXiv},
	author = {Ambrosino, F. and others},
	collaboration = {KLOE},
	doi = {10.1016/j.physletb.2011.04.055},
	eprint = {1006.5313},
	journal = {Phys. Lett. B},
	pages = {102--110},
	primaryclass = {hep-ex},
	title = {{Measurement of $\sigma(e^+ e^- \to \pi^+ \pi^-)$ from threshold to 0.85 GeV$^2$ using Initial State Radiation with the KLOE detector}},
	volume = {700},
	year = {2011}}

@article{KLOE:2012anl,
	archiveprefix = {arXiv},
	author = {Babusci, D. and others},
	collaboration = {KLOE},
	doi = {10.1016/j.physletb.2013.02.029},
	eprint = {1212.4524},
	journal = {Phys. Lett. B},
	pages = {336--343},
	primaryclass = {hep-ex},
	title = {{Precision measurement of $\sigma(e^+e^-\rightarrow \pi^+\pi^-\gamma)/ \sigma(e^+e^-\rightarrow \mu^+\mu^-\gamma)$ and determination of the $\pi^+\pi^-$ contribution to the muon anomaly with the KLOE detector}},
	volume = {720},
	year = {2013}}

@article{KLOE-2:2017fda,
	archiveprefix = {arXiv},
	author = {Anastasi, A. and others},
	collaboration = {KLOE-2},
	doi = {10.1007/JHEP03(2018)173},
	eprint = {1711.03085},
	journal = {JHEP},
	pages = {173},
	primaryclass = {hep-ex},
	title = {{Combination of KLOE $\sigma\big(e^+e^-\rightarrow\pi^+\pi^-\gamma(\gamma)\big)$ measurements and determination of $a_{\mu}^{\pi^+\pi^-}$ in the energy range $0.10 < s < 0.95$ GeV$^2$}},
	volume = {03},
	year = {2018}}

@article{BESIII:2015equ,
	archiveprefix = {arXiv},
	author = {Ablikim, M. and others},
	collaboration = {BESIII},
	doi = {10.1016/j.physletb.2015.11.043},
	eprint = {1507.08188},
	journal = {Phys. Lett. B},
	note = {[Erratum: Phys.Lett.B 812, 135982 (2021)]},
	pages = {629--638},
	primaryclass = {hep-ex},
	title = {{Measurement of the $e^+e^-\to\pi^+\pi^-$ cross section between 600 and 900 MeV using initial state radiation}},
	volume = {753},
	year = {2016}}

@article{Badger:2025ilt,
	archiveprefix = {arXiv},
	author = {Badger, Simon and Brancaccio, Colomba and Becchetti, Matteo and Czakon, Michal and Hartanto, Heribertus Bayu and Poncelet, Rene and Zoia, Simone},
	eprint = {2511.11431},
	month = {11},
	primaryclass = {hep-ph},
	reportnumber = {IFJPAN-IV-2025-24, P3H-25-093, TTK-25-39, ZU-TH 75/25},
	title = {{Higher-order QCD corrections to top-quark pair production in association with a jet}},
	year = {2025}}

@article{Pozzorini:2005ff,
	archiveprefix = {arXiv},
	author = {Pozzorini, S. and Remiddi, E.},
	doi = {10.1016/j.cpc.2006.05.005},
	eprint = {hep-ph/0505041},
	journal = {Comput. Phys. Commun.},
	pages = {381--387},
	reportnumber = {TTP05-06, SFB-CPP-05-14},
	title = {{Precise numerical evaluation of the two loop sunrise graph master integrals in the equal mass case}},
	volume = {175},
	year = {2006}}

@article{Moriello:2019yhu,
	archiveprefix = {arXiv},
	author = {Moriello, Francesco},
	doi = {10.1007/JHEP01(2020)150},
	eprint = {1907.13234},
	journal = {JHEP},
	pages = {150},
	primaryclass = {hep-ph},
	title = {{Generalised power series expansions for the elliptic planar families of Higgs + jet production at two loops}},
	volume = {01},
	year = {2020}}

@article{Correia:2025wtb,
	archiveprefix = {arXiv},
	author = {Correia, Miguel and Giroux, Mathieu and Mizera, Sebastian},
	doi = {10.1016/j.cpc.2025.109970},
	eprint = {2503.16601},
	journal = {Comput. Phys. Commun.},
	pages = {109970},
	primaryclass = {hep-th},
	title = {{SOFIA: Singularities of Feynman integrals automatized}},
	volume = {320},
	year = {2026}}

@article{Caron-Huot:2024brh,
	archiveprefix = {arXiv},
	author = {Caron-Huot, Simon and Correia, Miguel and Giroux, Mathieu},
	doi = {10.1103/8rwk-bnph},
	eprint = {2406.05241},
	journal = {Phys. Rev. Lett.},
	number = {13},
	pages = {131603},
	primaryclass = {hep-th},
	title = {{Recursive Landau Analysis}},
	volume = {135},
	year = {2025}}

@article{Laporta:2004rb,
	archiveprefix = {arXiv},
	author = {Laporta, S. and Remiddi, E.},
	doi = {10.1016/j.nuclphysb.2004.10.044},
	eprint = {hep-ph/0406160},
	journal = {Nucl. Phys. B},
	pages = {349--386},
	reportnumber = {CERN-PH-TH-2004-089},
	title = {{Analytic treatment of the two loop equal mass sunrise graph}},
	volume = {704},
	year = {2005}}

@article{Remiddi:2003ci,
	archiveprefix = {arXiv},
	author = {Remiddi, E.},
	editor = {Awramik, M. and Czakon, M. and Gluza, J.},
	eprint = {hep-ph/0310332},
	journal = {Acta Phys. Polon. B},
	pages = {5311--5322},
	reportnumber = {TTP-0326},
	title = {{Differential equations for the two loop equal mass sunrise}},
	volume = {34},
	year = {2003}}

@article{Hidding:2020ytt,
	archiveprefix = {arXiv},
	author = {Hidding, Martijn},
	doi = {10.1016/j.cpc.2021.108125},
	eprint = {2006.05510},
	journal = {Comput. Phys. Commun.},
	pages = {108125},
	primaryclass = {hep-ph},
	title = {{DiffExp, a Mathematica package for computing Feynman integrals in terms of one-dimensional series expansions}},
	volume = {269},
	year = {2021}}

@article{FebresCordero:2023pww,
	archiveprefix = {arXiv},
	author = {Febres Cordero, F. and Figueiredo, G. and Kraus, M. and Page, B. and Reina, L.},
	doi = {10.1007/JHEP07(2024)084},
	eprint = {2312.08131},
	journal = {JHEP},
	pages = {084},
	primaryclass = {hep-ph},
	reportnumber = {CERN-TH-2023-240},
	title = {{Two-loop master integrals for leading-color $ pp\to t\overline{t}H $ amplitudes with a light-quark loop}},
	volume = {07},
	year = {2024}}

@article{Li:2026emp,
	archiveprefix = {arXiv},
	author = {Li, Shu-Xiang and Zhang, Ren-You and Wang, Xiao-Feng and Li, Pan-Feng and Wei, Xiang-Jie and Wang, Yi and Jiang, Yi and Wang, Qing-hai},
	eprint = {2604.27314},
	month = {4},
	primaryclass = {hep-ph},
	title = {{Planar master integrals for two-loop NLO electroweak light-fermion contributions to $g g \rightarrow Z H$}},
	year = {2026}}

@article{Muller-Stach:2012tgj,
	archiveprefix = {arXiv},
	author = {M{\"u}ller-Stach, Stefan and Weinzierl, Stefan and Zayadeh, Raphael},
	doi = {10.1007/s00220-013-1838-3},
	eprint = {1212.4389},
	journal = {Commun. Math. Phys.},
	pages = {237--249},
	primaryclass = {hep-ph},
	title = {{Picard-Fuchs equations for Feynman integrals}},
	volume = {326},
	year = {2014}}

@article{Chicherin:2020oor,
	archiveprefix = {arXiv},
	author = {Chicherin, Dmitry and Sotnikov, Vasily},
	doi = {10.1007/JHEP12(2020)167},
	eprint = {2009.07803},
	journal = {JHEP},
	pages = {167},
	primaryclass = {hep-ph},
	reportnumber = {MPP-2020-171},
	title = {{Pentagon Functions for Scattering of Five Massless Particles}},
	volume = {20},
	year = {2020}}

@article{Badger:2024dxo,
	archiveprefix = {arXiv},
	author = {Badger, Simon and Becchetti, Matteo and Brancaccio, Colomba and Hartanto, Heribertus Bayu and Zoia, Simone},
	doi = {10.1007/JHEP03(2025)070},
	eprint = {2412.13876},
	journal = {JHEP},
	pages = {070},
	primaryclass = {hep-ph},
	reportnumber = {ZU-TH-65/24},
	title = {{Numerical evaluation of two-loop QCD helicity amplitudes for $ gg\to t\overline{t}g $ at leading colour}},
	volume = {03},
	year = {2025}}

@article{Calisto:2023vmm,
	archiveprefix = {arXiv},
	author = {Calisto, Francesco and Moodie, Ryan and Zoia, Simone},
	doi = {10.1007/JHEP07(2024)124},
	eprint = {2312.02067},
	journal = {JHEP},
	pages = {124},
	primaryclass = {hep-ph},
	reportnumber = {CERN-TH-2023-225},
	title = {{Learning Feynman integrals from differential equations with neural networks}},
	volume = {07},
	year = {2024}}

@article{Czakon:2026tog,
	archiveprefix = {arXiv},
	author = {Czakon, Micha{\l} and Tancredi, Lorenzo},
	eprint = {2606.30354},
	month = {6},
	primaryclass = {hep-ph},
	reportnumber = {P3H-26-047,TTK-26-18,TUM-HEP-1607/26},
	title = {{Solution of Canonical Differential Equations for Integrals on Arbitrary Geometries}},
	year = {2026}}

@article{Becchetti:2026yxl,
    author = "Becchetti, Matteo and Canko, Dhimiter and Chen, Xiang and Chestnov, Vsevolod and Delto, Maximilian and Ditsch, Sara and Peraro, Tiziano and Pozzoli, Mattia and Savoini, Chiara and Zoia, Simone",
    title = "{Two-loop QCD amplitudes for $t\bar{t}W$ production at the LHC in the leading-colour approximation}",
    eprint = "2608.03746",
    archivePrefix = "arXiv",
    primaryClass = "hep-ph",
    reportNumber = "ZU-TH 27/26, TUM-HEP-1610/26, MPP-2026-132, CERN-TH-2026-183",
    month = "8",
    year = "2026"
}

\end{document}